\documentclass[11pt,onecolumn]{IEEEtran}
\usepackage{graphicx}
\usepackage{subfigure}
\usepackage{enumerate}
\usepackage{amsmath}
\usepackage{mathrsfs}
\usepackage{amssymb}
\usepackage{slashbox}
\usepackage{algorithm}
\usepackage{algorithmic}
\usepackage{hyperref}
\usepackage{breakurl}
\usepackage{bm}
\usepackage{color}
\usepackage{wrapfigure}

\newtheorem{theorem}{Theorem}
\renewcommand{\baselinestretch}{1.5}

\begin{document}
%
\title{Human Motion Capture Data \\Tailored Transform Coding}
%
%
%


\author{Junhui Hou,
        ~Lap-Pui Chau,
        ~Nadia Magnenat-Thalmann,
        ~and~Ying He
        \thanks{This work carried out at BeingThere Center was supported by the Singapore National
Research Foundation, under its International Research Centre at the
Singapore Funding Initiative, and administered by the IDM Programme
Office.}
\thanks{J. Hou and L.-P. Chau are with the School of Electrical and Electronics Engineering,
Nanyang Technological University, 639798 Singapore (email:
houj0001@ntu.edu.sg, elpchau@ntu.edu.sg).}
\thanks{N. Magnenat-Thalmann is with the Institute
for Media Innovation, Nanyang Technological University, 639798,
Singapore (email: nadiathalmann@ntu.edu.sg).}
\thanks{Y. He is with the School of Computer Engineering, Nanyang Technological University,
639798, Singapore (email:yhe@ntu.edu.sg).}
}

\markboth{}%
{}


\IEEEcompsoctitleabstractindextext{
\begin{abstract}
  Human motion capture (mocap) is a widely used technique for digitalizing human movements. 
  With growing usage, compressing mocap data has received increasing attention, since compact data size enables efficient storage and transmission. 
  Our analysis shows that mocap data have some unique characteristics that distinguish themselves from images and videos.
  Therefore, directly borrowing image or video compression techniques, such as discrete cosine transform, does not work well.
  In this paper, we propose a novel mocap-tailored transform coding algorithm that takes advantage of these features.
  Our algorithm segments the input mocap sequences into clips, which are represented in 2D matrices.
  Then it computes a set of data-dependent orthogonal bases to transform the matrices to frequency domain,
  in which the transform coefficients have significantly less dependency.
  Finally, the compression is obtained by entropy coding of the quantized coefficients and the bases.
  Our method has low computational cost and can be easily extended to compress mocap databases.
  It also requires neither training nor complicated parameter setting.
  Experimental results demonstrate that the proposed scheme significantly outperforms state-of-the-art algorithms in terms of compression performance and speed.
\end{abstract}
}
\maketitle
\IEEEdisplaynotcompsoctitleabstractindextext
\IEEEpeerreviewmaketitle

\section{Introduction}
\label{sec:I}

  Human motion capture (mocap) is the process to digitally record human movement using information (e.g., angle and 3D coordinate) of a set of key points (i.e., markers or
  joints). As a highly successful technique, it has been extensively used in
entertainment, medical, sports and military
  applications \cite{phdthesis}.
  The output of a mocap session is the trajectories of a set of key points.
  An accurate motion capture requires tracking a large number of markers at very high frequency,
  therefore, the generated mocap data is usually bulky, which poses a challenge to both storage and transmission.
  It is highly desirable to develop an efficient method for compressing mocap data.

  The primary goal of data compression is to reduce the redundancy or correlation in the data.
  As an effective tool for data decorrelation, transform coding maps the original data into the transform domain,
  in which the transform coefficients have significantly less dependency than the original data.
  Since most of the energy concentrates on a small portion of the transform domain, compression can be achieved by
  discarding the less-important information (i.e., setting the smallest coefficients to be zeros).
  To recover the data, one simply applies the inverse transformation to the transform coefficients.
  Representative transform coding methods are the Karhunen-Loeve Transform (KLT), Principal Component Analysis (PCA) (or more generally, Singular Value Decomposition (SVD)),
  Discrete Wavelet Transform (DWT), and Discrete Cosine Transform (DCT).
  Among them, the 2D DCT and DWT methods are extremely successful in images/videos compression, e.g., JPEG-2000 and H.264/AVC, due to the locally
  smooth nature in images and/or videos, leading to most transform coefficients occupying low frequencies.
  Since human mocap data can be naturally represented by a 2D matrix, where each row corresponds to the trajectory of a marker,
  one may simply borrow the existing image-based transform coding methods to mocap data.
  Unfortunately, our analysis shows that mocap data have some unique characteristics that distinguish themselves from images and videos.
  Therefore, directly applying DCT or DWT does not work well.

\begin{figure}
\centering
\includegraphics[width=3.2in]{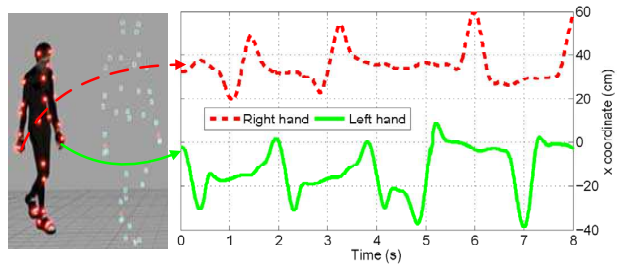}
\caption{Marker-based motion capture device records
the movement of a person by tracking the markers wore on his or her
joints. Each marker trajectory is a smooth curve in $\mathbb{R}^3$.
For better visualization, we only plot the $x$ coordinate. The left
photo is taken from \url{http://nward.com/ict/blender/mocap.htm}}
\label{fig:trajectory}
\end{figure}

This paper presents a novel transform coding method for compressing
human mocap data. Observing that a subset of a long mocap sequence
exhibits stronger data dependency, we segment the input data into a
set of clips, and represent each clip by a matrix, where each row is
the trajectory of a marker. Note that mocap data is smooth in the
dimension of time, since each marker's trajectory is a smooth curve
in $\mathbb{R}^3$. See Fig.~\ref{fig:trajectory}. However, the
columns exhibit much less smoothness due to the complex nature of
human motion. Based on these observations, our transform coding
method adopts a data-dependent left transformation and a
data-independent right transformation, where the former is to reduce
redundancy among the non-smooth columns and the latter is to
decorrelate the smooth rows. Our transform coding is very effective,
since it produces only a small number of coefficients in the
transform domain. Moreover, our method has low computational cost
and can be easily extended for mocap database compression. Unlike
the existing approaches to mocap data compression, our method
requires neither training nor complicated parameter setting.
Experimental results demonstrate that the proposed method
significantly outperforms state-of-the-art algorithms in terms of
speed and compression performance.

 The rest of this paper is organized as follows:
 Section \ref{sec:related} briefly reviews previous work on mocap
 data compression. Section \ref{sec:analysis} analyzes the
 human mocap data and shows its unique features, which distinguish themselves from natural images and videos.
 Then Section \ref{sec:proposed scheme} presents our tailored transform coding,
 which is further generalized to mocap database in Section \ref{sec:extension}.
 Section \ref{sec:experiment} evaluates our method and presents the experimental results.
 Finally,  Section \ref{sec:conclusion} concludes this paper and points out some
 promising future direction as well as the potential of our scheme on other applications.

\section{Previous Work} \label{sec:related}
  Compared to the widely-studied video/image compression, mocap data compression is a relatively young field.
  Liu \emph{et al}. \cite{segmentpca} partitioned a mocap data sequence into subsequences,
  and then applied the PCA to keyframes of each subsequence separately.
  Fitting fixed-length short mocap clips using B\'{e}zier curves, Arikan~\cite{siggraph} performed clustered PCA to reduce their dimensionality.
  Observing that human motion often exhibits repeated patterns, Lin \emph{et al.}~\cite{repeate} applied the PCA to the repeated motion and  interpolated the marker trajectories.
  Tournier \emph{et al}. \cite{Priciple geodesic} proposed a principal geodesic analysis (PGA)-based approach,
  which stores only some poses manifolds, the end-effectors as well as root joint's trajectories and orientations.
  The motion can be  recovered through Inverse Kinematics (IK) in a lossy manner.

  Beaudoin \emph{et al}. \cite{wavelet compression} applied 1D DWT to joint angles,
  and the WT coefficients were optimally selected by taking the visual artifacts into
  account. This method was futher improved by Firouzmanesh et al. \cite{tmm-perceptual}.
  Preda and Preteux \cite{BBA} adopted temporal prediction and spatial
  1D DCT in MPEG-4 bone-based animation (BBA) to remove redundancy.
  Using fuzzy clustering, Chew \emph{et al}. \cite{tmm} reorganized mocap data into matrices
  with strong local correlation, and then adopted
  the JPEG-2000 (employing 2D DWT) to encode them.

  Chattopadhyay \emph{et al}. \cite{Power Aware Approach} proposed a novel model-based indexing method,
  which exploits structural information derived from the skeletal virtual human model.
  Gu \emph{et al}. \cite{pattern} compressed human mocap data by hierarchical structure construction
  and motion pattern indexing.
  They also built a mocap database for each meaningful body part and a sequence of mocap data can be
  efficiently represented as a series of motion pattern indices.
  Taking advantage of the sparsity of mocap data in the action domain and time domain,
  Li \emph{et al}. \cite{sparse} proposed a learning-based scheme to train a set of shift-invariant
  basis functions,
  on which the mocap sequences can be projected, leading to sparse coefficients.
  Lim and Thalmann~\cite{keyframe2}, Xiao \textit{et al.} \cite{cgi} and Kim \textit{et al.}~\cite{keyframe1}  proposed keyframe-based methods,
  which extract a few representative frames, and then represent the whole mocap sequence by interpolating these keyframes.
  Zhu \emph{et al}. \cite{Quaternion} represented human motion data in the quaternion space so that human rotational motion data can be decomposed into the dictionary part
  and the sparse weight part.
  Le \emph{et al}. \cite{marker  optimization} computed optimized marker layouts
  for capturing facial expression as optimization of characteristic control points
  from a set of high-resolution, ground truth facial mesh sequences.
  Hou \emph{et al}. \cite{tensor} proposed a tensor representation, in which mocap clips were assembled
  into a third-order tensor.
  They applied the canonical polyadic decomposition to explore correlation

  Instead of using the standard application-independent transformations,
  our transform coding method is tailored to mocap data by taking advantage
  of their unique properties (see Section 3).
  As a result, our method outperforms the existing techniques in terms of both speed and compression performance.
  Moreover, our method has good scalability so that it works for both a single motion sequence and a
  set of mocap sequences, i.e., mocap database.

\section{Human Mocap Data Analysis}\label{sec:analysis}
As pointed in~\cite{segmentpca,siggraph,repeate}, compressing the
position-based mocap data has advantages over the hierarchical
angle-based representation, which exhibits intrinsic nonlinearity.
Besides, the hierarchical structure results in the accumulation of
errors along the chain of joint angles, resulting in significant
jerkiness~\cite{tmm}. Therefore, we use the position-based human
mocap data in this paper.  We represent a human mocap sequence by a
matrix $\mathbf{M}\in \mathbb{R}^{3n\times f}$, i.e.,
\begin{equation} \nonumber
\centering \mathbf{M}=\left (
\begin{array}{cccc}
x_{1,1} & x_{1,2}&\cdots&x_{1,f} \\
\vdots &\vdots &\ddots&\vdots \\
x_{n,1} & x_{n,2}&\cdots&x_{n,f} \\
y_{1,1} & y_{1,2}&\cdots&y_{1,f} \\
\vdots &\vdots &\ddots&\vdots \\
y_{n,1} & y_{n,2}&\cdots&y_{n,f} \\
z_{1,1} & z_{1,2}&\cdots&z_{1,f} \\
\vdots &\vdots &\ddots&\vdots \\
z_{n,1} & z_{n,2}&\cdots&z_{n,f}
\end{array}
\right),
\end{equation}
where $n$ and $f$ are the numbers of markers and frames,
respectively. $(x_{i,j}$, $y_{i,j}$, $z_{i,j})$ is the coordinate of
the $i$-th marker in the $j$-th frame.

A typical human motion sequence has strong intra-/inter-trajectory
correlation and clip correlation. It also exhibits a strong local
structure. Although some of the properties were also empirically
observed in \cite{segmentpca,wavelet
compression,siggraph,repeate,ICME}, we did not find quantitative
verification in the literature. In this section, we provide concrete
analysis to justify these properties.

\subsection{Trajectory Correlation}
\label{subsec:trajectory corre}
  Each trajectory exhibits coherence (i.e., intra-correlation) since the positions of markers usually vary \textit{smoothly} in time.
  Also, inter-correlation exists among trajectories, due to the highly coordinated and structured nature of human motions.

  To verify the intra-correlation, i.e., small variation or smoothness of each trajectory,
  we compute $\overline{v}=\frac{1}{3nf}\|\nabla \mathbf{M}^\intercal\|_1$,
  where $\nabla\in \mathbb{R}^{(f-1)\times f}$ is
  first order difference matrix, $^\intercal$ is matrix transpose, and $\|\cdot\|_1$ is the sum of absolute value of all
  elements. Using $\overline{\sigma}=\sum_{i=1}^{3n} \sigma_i$ as a reference, where $\sigma_i$ is the standard deviation of the $i$-th row of $\mathbf{M}$, Table \ref{tab:intra-correlation} shows the smoothness measurement on several sequences.
  See Table \ref{tab:sequences details} for details of the test sequences.
  One can clearly see that $\overline{v}\ll\overline{\sigma}$, indicating the small variation within each trajectory.

\begin{figure}[t]
\centering \subfigure[]
{\includegraphics[width=1.7in]{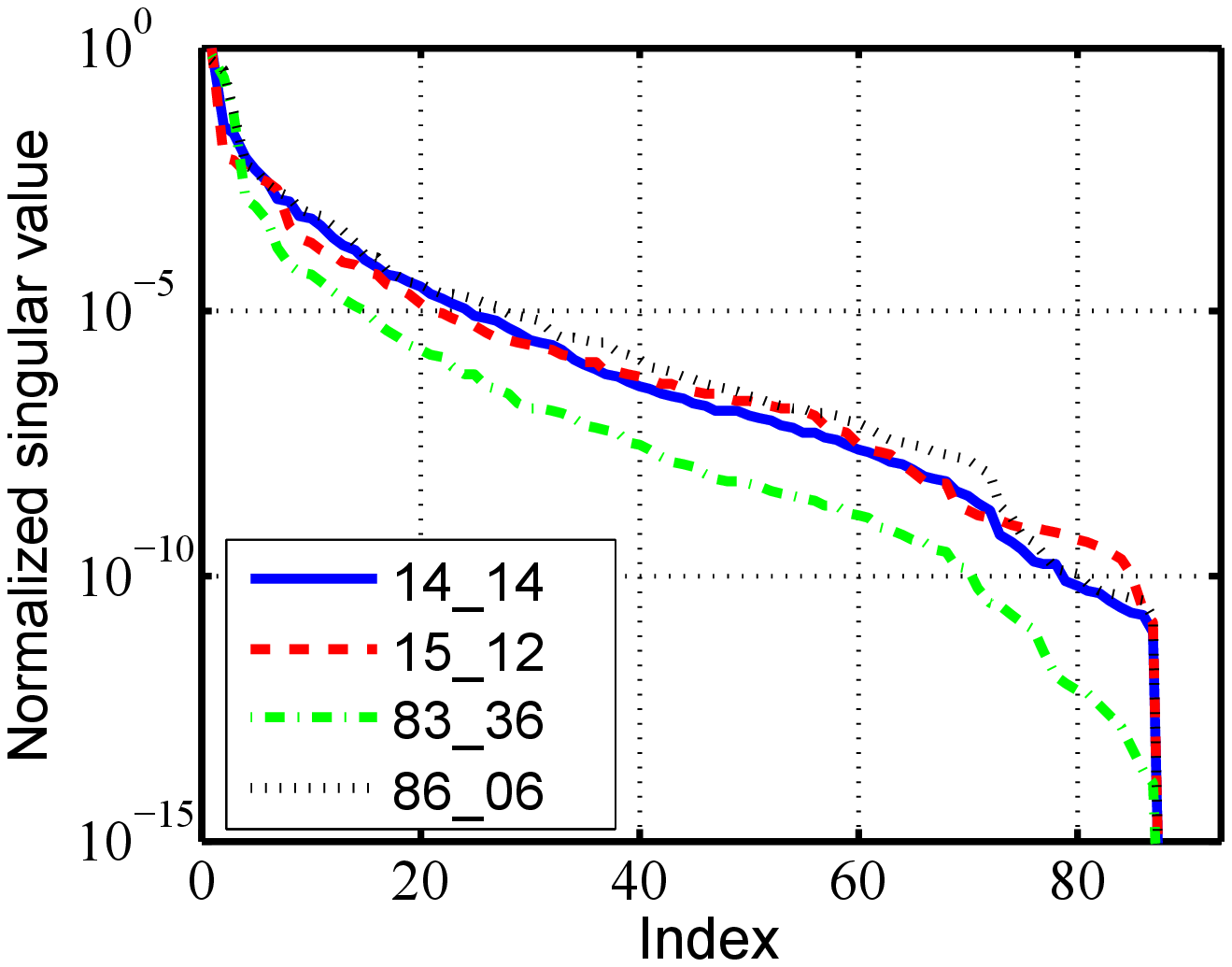}\label{fig:propertyI}}
\subfigure[]
{\label{fig:propertyII}\includegraphics[width=1.7in]{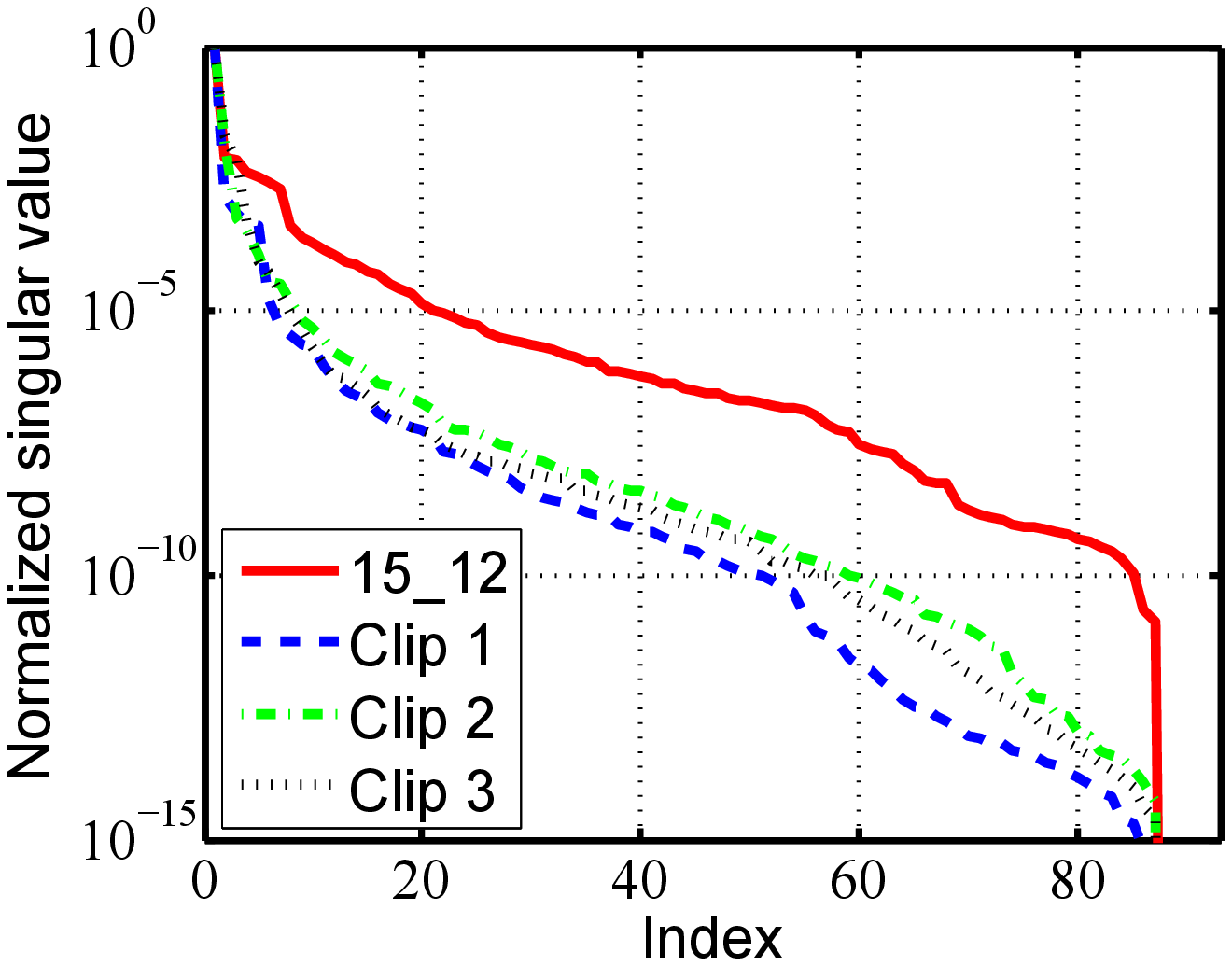}}
\caption{The vertical axis is the normalized singular value and the
horizontal axis is its index. (a) As the normalized singular values
approach zero quickly, matrix $\bf M$ is of low rank and its rows are strongly dependent.
 (b) We randomly extract three clips from a sequence. Each clip has 300 frames. The singular values of the clips approach zero more quickly than that of the whole sequence, indicating that short clips
have stronger inter-correlation than the whole sequence. }
\label{fig:properties verification}
\end{figure}

  To verify the inter-correlation, we apply SVD to $\mathbf{M}$.
  As shown in Fig. \ref{fig:propertyI}, the normalized singular values approach zero quickly, which demonstrates the approximate low-rank nature of $\mathbf{M}$.
  As a result, the rows of $\mathbf{M}$ are strongly dependent.

\begin{table}
\centering \caption{Verification of the intra-trajectory correlation. } \label{tab:intra-correlation}
\begin{tabular}{c|c|c|c|c}
\hline\hline Sequence & 14\_14 &15\_12 & 83\_36& 86\_06\\
\hline $\overline{v}$ &0.0516 & 0.0305&0.0363 &0.0483\\
\hline $\overline{\sigma}$& 2.895&1.683 & 10.217 &8.2831 \\
\hline \hline
\end{tabular}
\end{table}

\subsection{Local v.s. Global}
\label{subsec:local}

Motion in a short period tends to be more correlated than in a long
period, indicating that short clips have stronger inter-correlation
than the whole sequence.

To verify the local property, we randomly extract several clips from
a sequence. Each clip has $F (\ll f)$ frames and is represented by a
matrix $\overline{\mathbf{M}}\in \mathbb{R}^{3n\times F}$.
  Then, we compare their normalized singular values with those of the big matrix $\mathbf{M}$.
  Fig.~\ref{fig:propertyII} shows that the singular values of $\overline{\mathbf{M}}$ approach
  to zero more rapidly than $\mathbf{M}$,
  meaning that the clips have stronger inter-correlation than the whole sequence
  $\mathbf{M}$.

\begin{figure}
\centering \subfigure[Mocap data]
{\label{fig:propertyIII}\includegraphics[width=1.7in]{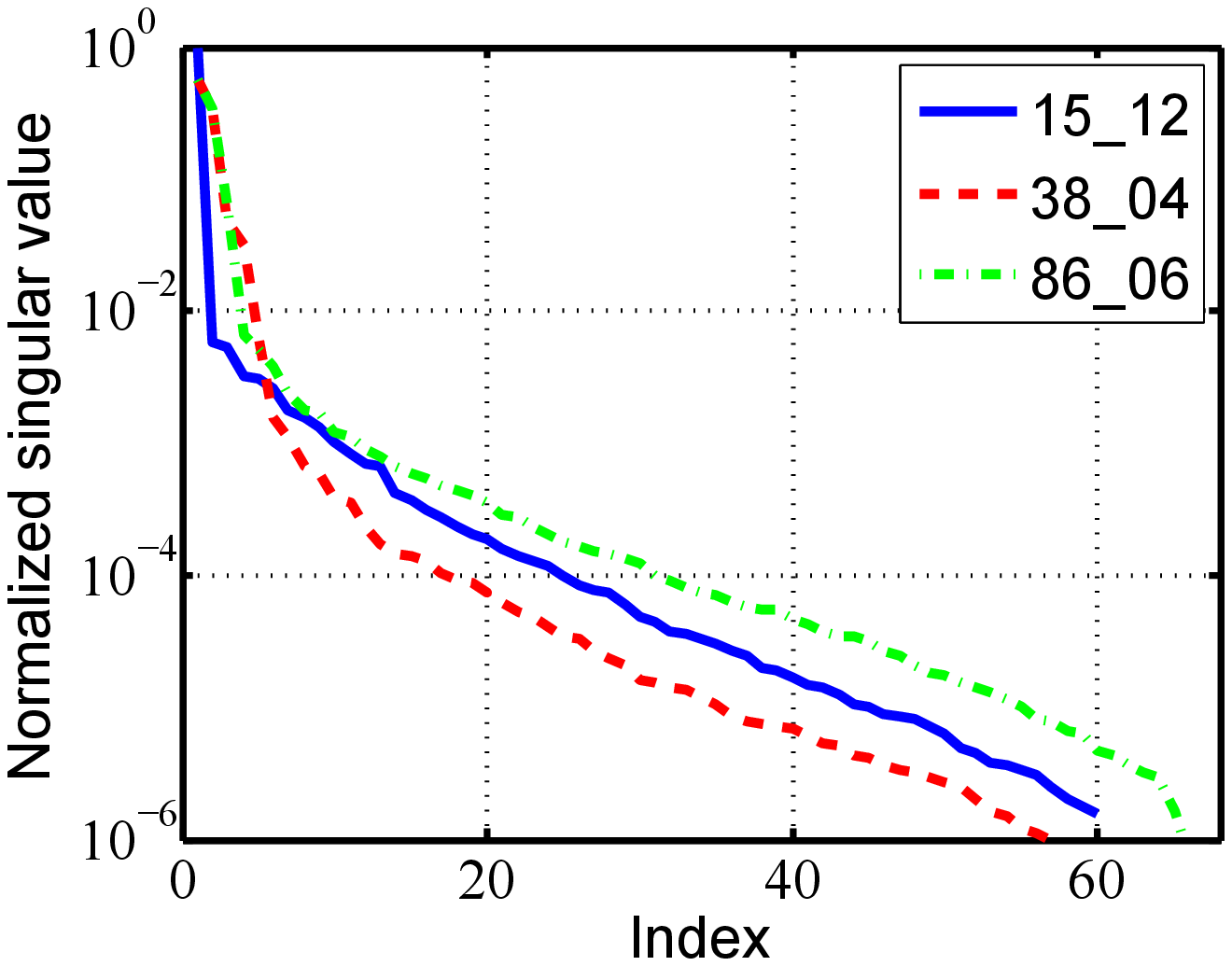}}
\subfigure[Natural videos]
{\label{fig:propertyIIIvideo}\includegraphics[width=1.7in]{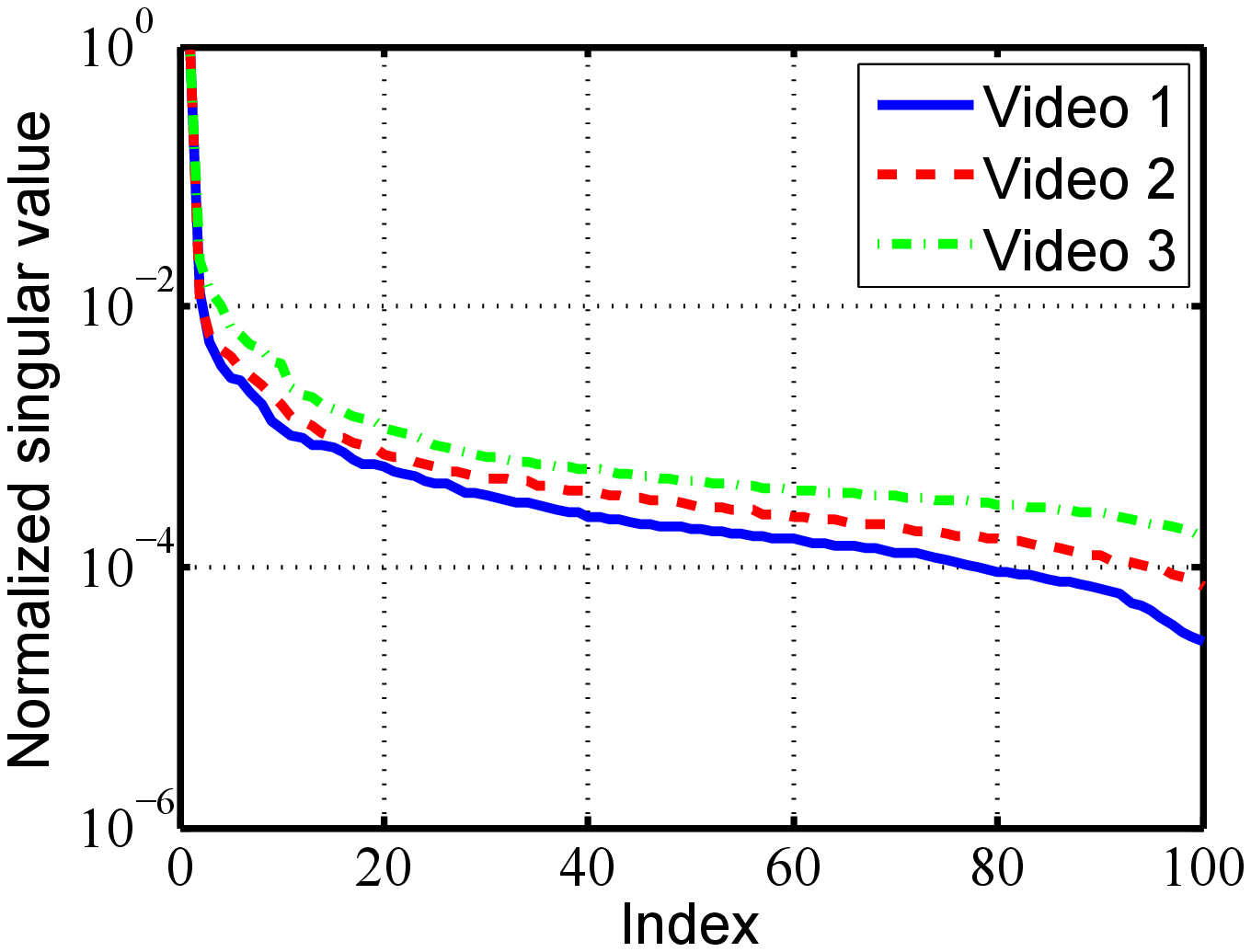}}
\caption{The normalized singular values of an unfolded mocap matrix
approaches zero faster than natural videos. The three natural videos, CREW, SOCCER,
and BUS, are from the standard video compression test sequences [Online]
Available: \url{http://trace.eas.asu.edu/yuv/}.}
\end{figure}

\subsection{Clip Correlation}
\label{subsec:clip corre.}

Intuitive speaking, a long motion sequence (such as walking,
jogging, etc) has repeated motions. Thus, one clip or several
markers' trajectories can be obtained by blending others. Such a
property is more obvious in large mocap database.

Similar to \cite{hosvd}, we  verify the clip correlation by partitioning $\mathbf{M}$ into $J$ non-overlapping
  clips\footnote{Sequences are cropped so that $f$ is a multiple of clip length throughout this paper.}
  represent them as a three dimensional matrix $\mathcal{M}\in \mathbb{R}^{3n\times f/J\times J}$. 
  A new matrix $\mathbf{\widetilde{M}}\in \mathbb{R}^{J\times 3nf/J}$ is formed
  by unfolding the matrices $\mathcal{M}$ along the third dimension. 
  We then factorize $\widetilde{\mathbf{M}}$ by SVD.
   Fig. \ref{fig:propertyIII} shows that the normalized singular values of $\widetilde{\mathbf{M}}$ of all
  test mocap data approach zero rapidly, indicating that mocap clips have strong correlation.
  We also perform the test on natural videos, in which a clip is one video frame.
  As shown in Fig. \ref{fig:propertyIIIvideo}, the clip correlation in natural videos is not as strong as the mocap data.

\section{Our Algorithm}
\label{sec:proposed scheme}

\subsection{Motivation}
As mentioned above, a key issue of compressing mocap data is to
reduce data correlation described in Section \ref{subsec:trajectory
corre}, as much as possible. A straightforward method is to directly
apply 2D transform coding to $\bf M$. However, our analysis in
Section \ref{subsec:local} suggests that compressing subsequences
separately is more effective than that of the sequence as a whole.
Similar observation can be found in \cite{segmentpca},
\cite{siggraph}, \cite{tmm} and image/video compression, where
block-based transform is widely used. Moreover, processing each clip
separately is obviously more memory efficient than that of the whole
sequence. Inspired by these observations, we propose a clip-based
algorithm for compressing
mocap data, in which one mocap sequences is segmented 
into $N$ clips, and the $i$-th clip is denoted by $\mathbf{M}_i\in
\mathbb{R}^{3n\times L_i}$ with $L_i$ stands for the length.

  Within the transform coding framework, one needs to transform the input data into the transform domain,
  in which the transformed coefficients exhibit sparsity so that some smallest ones can be discarded with little information loss.
  A possible solution is to apply SVD to each $\mathbf{M}_i$ and then keep the top $k_i$ important components, i.e.,
  \begin{align}
  &\left[\mathbf{U}_1^\intercal\mathbf{M}_{1}\mathbf{V}_1~~\mathbf{U}_2^\intercal \mathbf{M}_{2}\mathbf{V}_2~~\cdots~~\mathbf{U}_N^\intercal \mathbf{M}_{N}\mathbf{V}_N\right]\nonumber\\
  &=\left[\overline{\mathbf{S}}_1 ~~ \overline{\mathbf{S}}_2~~\cdots~~
  \overline{\mathbf{S}}_N\right], \label{equ:method II}
\end{align}
  where  $\mathbf{U}_i\in \mathbb{R}^{3n\times k_i}$ and $\mathbf{V}_i\in \mathbb{R}^{L_i\times k_i}$ are
  the $k_i$ eigenvectors of $\mathbf{M}_i\mathbf{M}_i^\intercal$ and $\mathbf{M}_i^\intercal\mathbf{M}_i$, corresponding to the $k_i$ largest eigenvalues, respectively,
  and $\overline{\mathbf{S}}_i\in \mathbb{R}^{k_i\times k_i}$ is a diagonal matrix.
  The SVD-based transform coding uses knowledge of the application to choose information to discard.
  However, since one has to store all the transformation matrices $\mathbf{U}_i$ and $\mathbf{V}_i$, $i=1,\cdots,N$,
  it is difficult to obtain high compression performance.

  \begin{figure}
  \centering
  \subfigure[Human motion]{\label{fig:dcts}\includegraphics[width=1.6in]{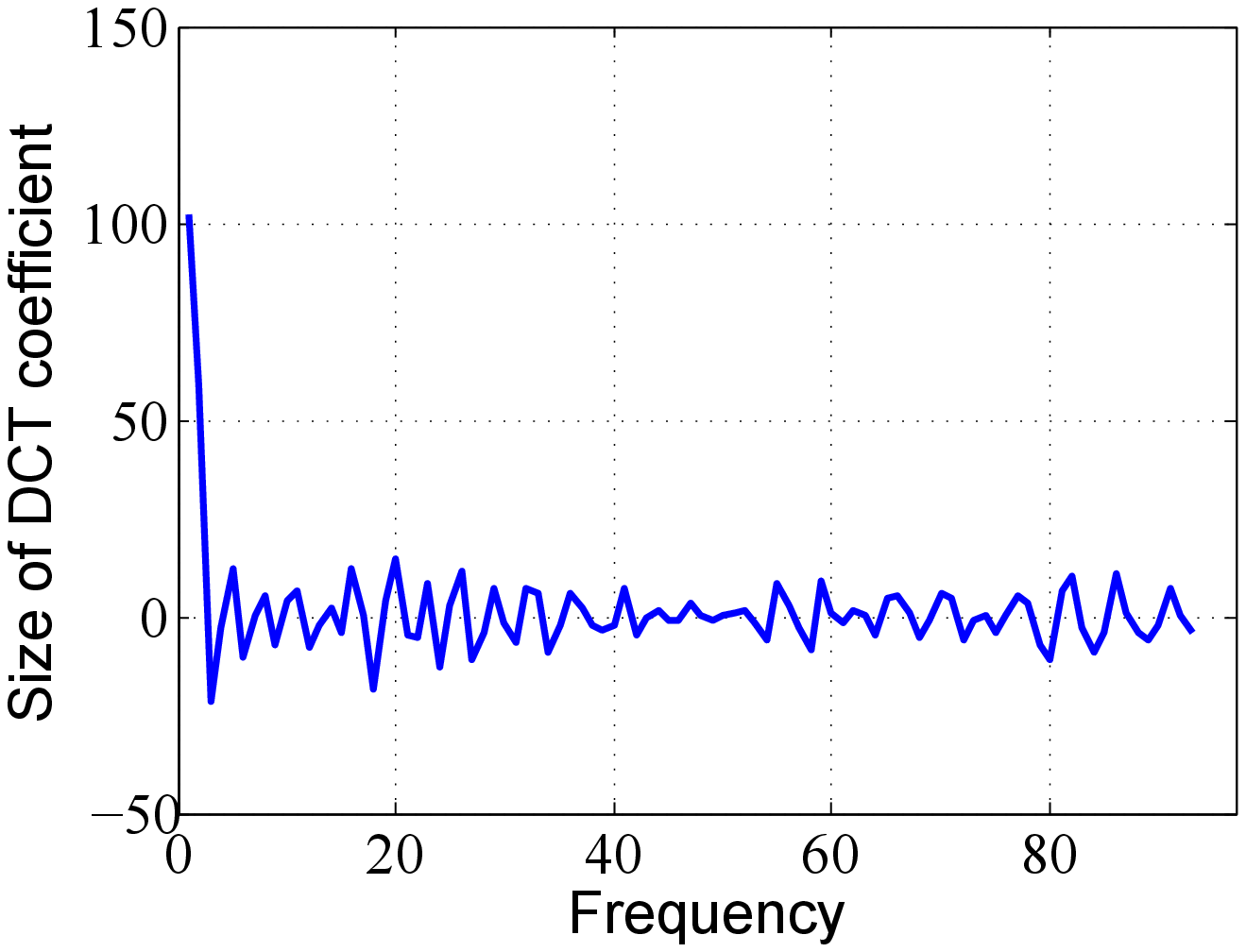}}
  \subfigure[Natural image]{\label{fig:dctt} \includegraphics[width=1.6in]{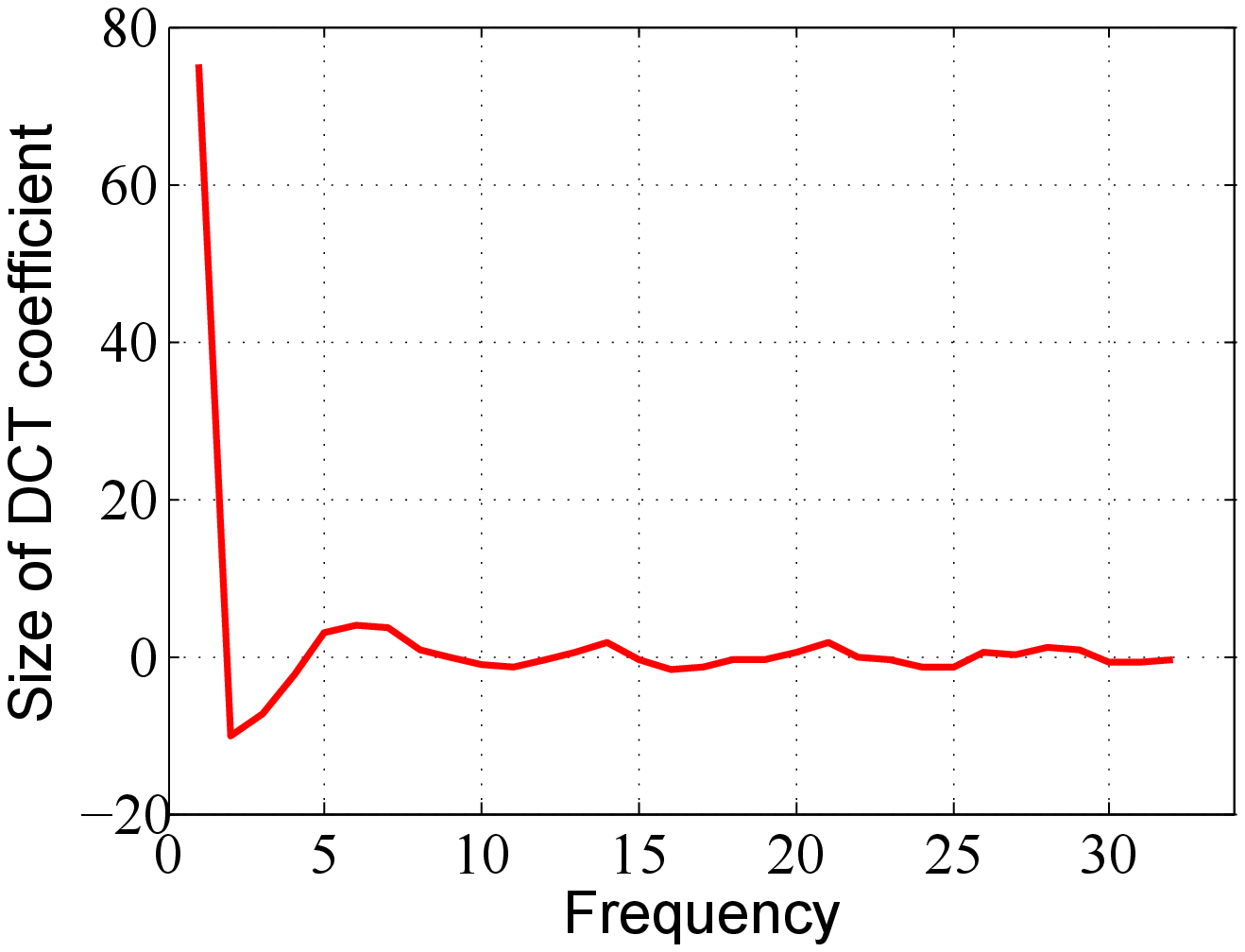}}
  \caption{Applying the 1D DCT to the columns of a natural image, we can see sparsity in the transform domain. However, human motion data do not have such a property due to lack of smoothness in the column direction.}\label{fig:wrongdct}
  \end{figure}

  Alternatively, one may adopt some application-independent transformations, such as 2D DCT or DWT.
  Take 2D DCT as an example.
\begin{align}
&\mathbf{D}_l^\intercal[\mathbf{M}_1 \mathbf{D}_{r_1} ~~
\mathbf{M}_2 \mathbf{D}_{r_2} ~~\cdots~~\mathbf{M}_N
\mathbf{D}_{r_N}] \nonumber \\
&=\left[\widetilde{\mathbf{S}}_1~~\widetilde{\mathbf{S}}_2~~\cdots~~\widetilde{\mathbf{S}}_N\right],
\label{equ:sharedUV}
\end{align}
  where $\mathbf{D}_l\in \mathbb{R}^{3n\times 3n}$ and $\mathbf{D}_{r_i}\in \mathbb{R}^{L_i\times L_i}$ are the standard 1D DCT basis matrices,
  $\widetilde{\mathbf{S}}_i\in \mathbb{R}^{3n\times L_i}$ are the transformed
  coefficients. Compared to the SVD-based transform coding, the DCT-based approach
stores only the transformed matrices  $\widetilde{\mathbf{S}}_i$,
$i=1,\cdots,N$.
  Owning to the locally smooth (or small variation) nature of images, the DCT works extremely well in 2D image and video compression.
  \begin{wrapfigure}{r}{1.9in}
  \centering
  \includegraphics[width=1.9in]{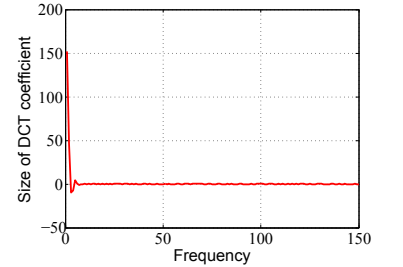}
  \caption{Applying the DCT to $\mathbf{M}_i$'s rows, we see sparsity in the transform domain.}\label{fig:wrongdct2}
  \end{wrapfigure}
  However, as discussed above, the columns in mocap data are not smooth.
  Therefore, it does not make sense to apply the left transform $\mathbf{D}_l$ to mocap data.
  Fig.~\ref{fig:wrongdct}(a) reveals such an issue by applying the 1D DCT to a randomly-chosen column in $\mathbf{M}_i$.
  We can clearly see that the transform coefficients occupy a wide range of frequencies.
  In sharp contrast, applying the 1D DCT to an arbitrary natural image column produces very small range of frequencies in the transformed coefficients.
  See Fig.~\ref{fig:wrongdct}(b).
  On the other hand, rows of a mocap matrix are smooth since each marker has a smooth trajectory.
  Applying the 1D DCT to an arbitrary row of $\mathbf{M}_i$ shows that the DCT coefficients occupy only a small range.
  See Fig.~\ref{fig:wrongdct2}.

\begin{figure}
\centering
\includegraphics[width=3.20in]{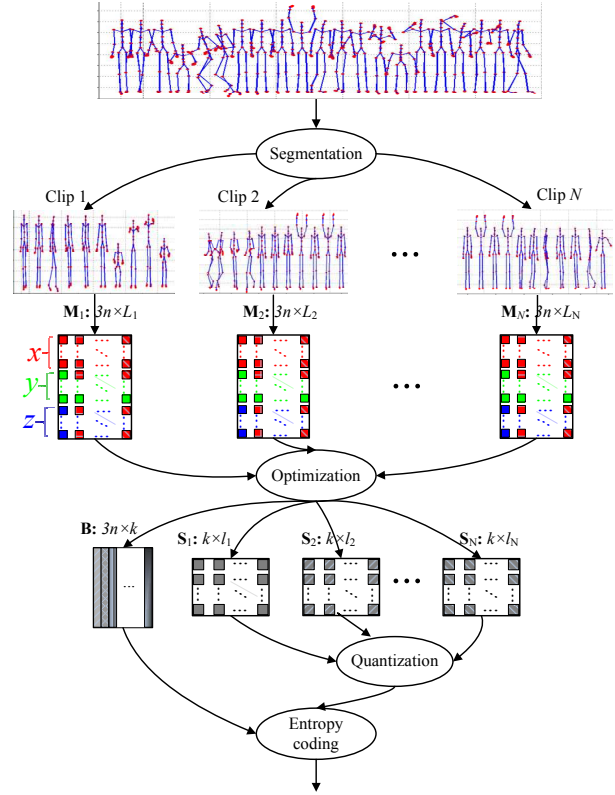}
\caption{The flow chart of our algorithm.} \label{fig:flowchart}
\end{figure}

  Motivated by the above discussions and observations, to develop an effective compression method for mocap sequences,
  one must take knowledge of mocap data into account and also keep the number of transform matrices as small as possible.
  Towards this goal, we propose a novel mocap tailored transform coding method.
  As shown in Fig. \ref{fig:flowchart}, after segmentation, we compute the data-dependent transformation
matrix $\mathbf{B}\in \mathbb{R}^{3n\times k}$,
  \begin{align}
  &\mathbf{B}^\intercal\left[\mathbf{M}_1\widetilde{\mathbf{D}}_{r_1}  ~~ \mathbf{M}_2\widetilde{\mathbf{D}}_{r_2}
  ~~\cdots~~\mathbf{M}_N\widetilde{\mathbf{D}}_{r_N}\right]\nonumber\\
  &= [\mathbf{S}_1~~\mathbf{S}_2~~\cdots~~\mathbf{S}_N],
  \label{equ:shared B}
  \end{align}
  which can reveal the geometric structure of columns of $\mathbf{M}_i$ to lead correlation removal, where $\widetilde{\mathbf{D}}_{r_i}\in \mathbb{R}^{L_i\times l_i}$ is the truncated 1D DCT transform matrix and $\mathbf{S}_i\in \mathbb{R}^{k\times l_i}$ is dense coefficients.

  It is worth noting that our method is able to adopt a single data-dependent left transform due to the strong correlation among mocap clips described in Section \ref{subsec:clip corre.}.
  Our mocap tailored transform coding stores only $k\left(3n+\sum_{i=1}^N l_i\right)$ scalars in the transform domain, whereas the SVD-based method (see Eqn. (\ref{equ:method II})) requires $3n\sum_{i=1}^N k_i+\sum_{i=1}^N (L_il_i+k_i)$ scalars.

\subsection{Mocap Tailored Transform Coding}
  Computing the data-tailored transform matrix $\mathbf{B}$ and coefficient matrices $\mathbf{S}_i$ is formulated as a least-square problem:
  \begin{align}
&\min_{\mathbf{B}, \mathbf{S}_i} \sum_{i=1}^{N} \left
\|\mathbf{M}_i-\mathbf{B}\mathbf{S}_i\widetilde{\mathbf{D}}_{r_i}^\intercal
\right\|_F^2
\nonumber \\
&~~~~~s.t.~~~ \mathbf{B}^\intercal \mathbf{B}
=\mathbf{I}_{k},\label{equ:construct B}
\end{align}
  where $\|\cdot\|_F$ is the Frobenius norm defined as $\|\mathbf{X}\|_F=\sqrt{\sum_i\sum_j X_{i,j}^2}$,
  and $\mathbf{I}_{k}\in \mathbb{R}^{k\times k}$ is the identity
  matrix. To minimize (\ref{equ:construct B}), the partial deviation
  of (\ref{equ:construct B}) with respect to $\mathbf{S}_i$ must be
  equal to zero. So we obtain $\mathbf{S}_i$ must satisfy
  $\mathbf{S}_i=\mathbf{B}^\intercal\mathbf{M}_i\widetilde{\mathbf{D}}_{r_i}$.
 With $\|\mathbf{X}\|_F^2={\rm Tr}(\mathbf{X}\mathbf{X}^T)$ where Tr is
for the trace of matrix, i.e., ${\rm Tr}(\mathbf{X})=\sum_i
X_{i,i}$, we can rewrite the objective function as
\begin{align}
&\sum_{i=1}^{N} \left\|\mathbf{M}_i-\mathbf{B}\mathbf{S}_i\widetilde{\mathbf{D}}_{r_i}^\intercal\right\|_F^2 \nonumber\\
&=\sum_{i=1}^{N}\left\{{\rm
Tr}\left(\mathbf{M}_i\mathbf{M}_i^\intercal\right)+{\rm
Tr}\left(\mathbf{S}_i\mathbf{S}_i^\intercal\right)-2{\rm
Tr}\left(\mathbf{B}\mathbf{S}_i\widetilde{\mathbf{D}}_{r_i}^\intercal\mathbf{M}_i^\intercal\right)\right\}\nonumber\\
&=\sum_{i=1}^N \left\{{\rm
Tr}\left(\mathbf{M}_i\mathbf{M}_i^\intercal\right)+{\rm
Tr}\left(\mathbf{B}^\intercal\mathbf{M}_i\widetilde{\mathbf{D}}_{r_i}\widetilde{\mathbf{D}}_{r_i}^\intercal\mathbf{M}_i^\intercal\mathbf{B}
\right)\nonumber\right. \\
&\left.-2{\rm
Tr}\left(\mathbf{B}\mathbf{B}^\intercal\mathbf{M}_i\widetilde{\mathbf{D}}_{r_i}\widetilde{\mathbf{D}}_{r_i}^\intercal\mathbf{M}_i^\intercal\right)\right\}
\nonumber \\
&=\sum_{i=1}^N {\rm
Tr}\left(\mathbf{M}_i\mathbf{M}_i^\intercal\right)- {\rm
Tr}\left(\mathbf{B}^\intercal\mathbf{C}\mathbf{B}\right),
\label{equ:Jexpansion}
\end{align}
  where $\mathbf{C}=\sum_{i=1}^N \mathbf{M}_i\widetilde{\mathbf{D}}_{r_i}\widetilde{\mathbf{D}}_{r_i}^\intercal\mathbf{M}_i^\intercal$.
  The last equation comes from the fact $\mathrm{Tr}(\mathbf{X}\mathbf{Y})=\mathrm{Tr}(\mathbf{Y}\mathbf{X})$ for compatible matrices $\mathbf{X}$ and $\mathbf{Y}$.

  The first term in Eqn. (\ref{equ:Jexpansion}) is a constant, which is independent of $\mathbf{B}$.
  Therefore, the minimization problem in (\ref{equ:construct B}) is equivalent to the following \textit{maximization} problem, i.e.,
\begin{align}
\max_{\mathbf{B}}~ {\rm
Tr}\left(\mathbf{B}^\intercal\mathbf{C}\mathbf{B}\right)
\nonumber \\
~~~s.t.~~~\mathbf{B}^\intercal\mathbf{B}=\mathbf{I}_{k}.
\label{equ:equavalentproblem}
\end{align}
  The optimal solution of (\ref{equ:equavalentproblem}) is given in Theorem \ref{the}.

  It takes $\mathcal{O}\left(nN\max_i\{l_i\}(\max_i\{L_i\}+n)\right)$ time to compute the matrix $\mathbf{C}$.
  We adopt the SVD algorithm for solving $\mathbf{C}$'s eigen system, which takes $\mathcal{O}(n^3)$ time.
  Finally, computing all $\mathbf{S}_i$ takes $\mathcal{O}(nN\max_i\{l_i\}(\max_i\{L_i\}+k))$ time.
  Since the number of markers $n$ is usually small (e.g., a few tens), the dominating term in the time complexity is $\mathcal{O}(nN\max_i\{l_i\}(\max_i\{L_i\}+n+k))$.

\begin{theorem}[\cite{trace maximum}]\label{the}
  Given a symmetric matrix $\mathbf{X}$ of dimension $p\times p$ and an arbitrary orthogonal matrix \textbf{Y} of dimension $p\times q$,
  the trace of $\mathbf{Y}^T\mathbf{X}\mathbf{Y}$ is maximized when $\mathbf{Y}$ contains the $q$ eigenvectors of $\mathbf{X}$,
  which correspond to the $q$ (algebraically) largest eigenvalues.
\end{theorem}
\subsection{Quantization and Entropy Coding}

  Since the compression quality highly depends on the accuracy of the orthogonal matrix $\mathbf{B}$,
  we represent each element of $\mathbf{B}$ using 16 bits.
  For each element of $\mathbf{S}_i$, we quantize its fractional part with $Q_i$ bits per entry,
  where $Q_i$ is automatically determined by our algorithm (see Section~\ref{subsec:parameters}).
  Then we adopt the lossless coding scheme to encode the bitstream.
  Note that other advanced lossless entropy coding schemes can also be applied.
  The decoder is a straightforward process that computes decompressed clip by $\widehat{\mathbf{B}}\mathbf{\widehat{S}}_i\mathbf{D}_{r_{i}}^\intercal$,
  where $\mathbf{\widehat{B}}$ and $\mathbf{\widehat{S}}_i$ are the decoded matrices.

\section{Generalization to Mocap Database}
\label{sec:extension}

  The previous section presents our algorithm for compressing a single human motion sequence.
  In this section, we consider a general scenario of human mocap database, which may consist of a large amount of motion sequences recorded from multiple performers.
  Due to diversity, a single transformation matrix $\mathbf{B}$ may not work for all clips.
  Therefore, it is natural to consider multiple transformation matrices to improve the compression result.
  We formulate the problem as computing $K$ transform matrices $\mathbf{B}_{j}$, $j=1,\cdots,K$, so that all clips can be selectively projected on them with the least distortion
    \begin{align}
    &\min_{\mathbf{W},\{\mathbf{B}_j\}_{j=1}^K} \sum_{i=1}^{N}\sum_{j=1}^{K}W_{i,j}\left\|\mathbf{M}_i-\mathbf{B}_{j}\mathbf{S}_{i,j}\widetilde{\mathbf{D}}_{r_i}^\intercal\right\|_F^2,\nonumber\\
    &s.t.~\forall j,~\mathbf{B}_j^\intercal \mathbf{B}_j=\mathbf{I}_{k},\nonumber \\
    &~~~~~\forall i,j,~W_{i,j}\in[0, 1],\nonumber\\
    &~~~~~\forall i, \sum_{j=1}^K W_{i,j}=1,
    \label{equ:extension}
    \end{align}
  where $W_{i,j}$ is the $(i,j)$-th entry of $\mathbf{W}$ indicating whether the $i$-th clip is represented in the space spanned by the $j$-th transform matrix $\mathbf {B}_j$.
  Obviously, Eqn. (\ref{equ:construct B}) is a special case of Eqn. (\ref{equ:extension}) with $K=1$.

  Note that the optimization problem (\ref{equ:extension}) is not easy to solve due to the two types of variables involved.
  In this paper, we propose a deterministic annealing based method~\cite{multi-matrices} that iteratively solves only one variable ($\mathbf{B}_j$ or $W_{ij}$) at a time.
  In each iteration, we first update $\mathbf{S}_{i,j}$ with other variables fixed
\begin{equation}
\centering
\mathbf{S}_{i,j}=\mathbf{B}_j^\intercal\mathbf{M}_i\widetilde{\mathbf{D}}_{r_i}.
\label{equ:compute Sij}
\end{equation}
  Then we update $W_{i,j}$ using the centroid equation,
\begin{equation}
\centering
W_{i,j}=\frac{e^{-\frac{1}{t}\left\|\mathbf{M}_i-\mathbf{B}_j\mathbf{S}_{i,j}\widetilde{\mathbf{D}}_{r_i}^\intercal\right\|_F^2}}{\sum_{h=1}^K
e^{-\frac{1}{t}\left\|\mathbf{M}_i-\mathbf{B}_l\mathbf{S}_{i,h}\widetilde{\mathbf{D}}_{r_i}^\intercal\right\|_F^2}},
\label{equ:compute Wij}
\end{equation}
  where $t$ is the temperature.
  By cooling the temperature slowly, the global optimal solution is obtained~\cite{multi-matrices}.
  Next, we compute matrix $\mathbf{C}_j$
  \begin{equation}
  \centering
  \mathbf{C}_j=\sum_{i=1}^N
  W_{i,j}\mathbf{M}_i\widetilde{\mathbf{D}}_{r_i}\widetilde{\mathbf{D}}_{r_i}^\intercal\mathbf{M}_i^\intercal,
  \label{equ:compute Bj}
  \end{equation}
  and solve its eigensystem using SVD.
  Finally, we form matrix $\mathbf{B}_j$ by taking $\mathbf{C}_j$'s $k$ eigenvectors associated with the largest $k$ eigenvalues.
  We repeat the above procedures until convergence, i.e., the change of matrices $\mathbf{B}_j$ and $\mathbf{W}$ is less than a small tolerance .
  The above deterministic simulated annealing method is guaranteed to converge~\cite{multi-matrices}.
  Computational results show that it usually converges in less than 30 iterations given a tolerance of $10^{-6}$.

\begin{algorithm}[]
\caption{Deterministic annealing based alternating iteration algorithm for solving (\ref{equ:extension}).}
 \textbf{Input}: $\{\mathbf{M}_i\}_{i=1}^{N}$, $K$\\
 \textbf{Output}: $\{\mathbf{B}_j\}_{j=1}^K$ and $\mathbf{W}$\\
 \textbf{Initialization}: initialize $\mathbf{B}_j$ with random orthogonal bases and $\widetilde{\mathbf{D}}_{r_i}$ with DCT bases;
 set $W_{i,j}=1/K$ and $t>0$.
\begin{algorithmic}[1]
       \REPEAT
                \FOR {$i \leftarrow 1:N$}
                     \FOR {$j \leftarrow 1:K$}
                          \STATE $\mathbf{S}_{i,j}=\mathbf{B}_j^\intercal\mathbf{M}_i\widetilde{\mathbf{D}}_{r_i}$
                     \ENDFOR
                     \FOR {$j\leftarrow 1:K$}
                          \STATE compute $W_{i,j}$ using (\ref{equ:compute Wij})
                     \ENDFOR
                \ENDFOR
                      \FOR {$j\leftarrow 1:K$}
                           \STATE compute $\mathbf{C}_j$ using (\ref{equ:compute Bj})
                           \STATE $\mathbf{B}_j\leftarrow$ $\mathbf{C}_j$'s $k$ eigenvectors associated with the $k$ largest eigenvalues
                      \ENDFOR
           \STATE $t\leftarrow \frac{t}{2}$
       \UNTIL {convergence}
  \end{algorithmic}
  \label{Alg:solution}
\end{algorithm}

\section{Experimental Results and Discussion}
\label{sec:experiment}

\begin{table*}[]
\centering \caption{The test data is from CMU Mocap database
\cite{cmu}, where each motion sequence has 31 markers and with a
frame rate 120 fps. } \label{tab:sequences details}
\begin{tabular}{c|c|c|l }
\hline \hline Sequences& \# of frames & Size (MB)& Description  \\
\hline   \multicolumn{4}{c}{\textbf{Used in Section \ref{sec:analysis}}} \\
\hline 14\_14& 4,653 & 1.65 &jumping, jog, squats, side twists, stretches\\
\hline 15\_12 &9,086 & 3.22&wash windows, lay-up shot, pass, throw ball, dance, the dive, the twist, strew\\
\hline 38\_04 & 8,631 & 3.06&walk around, frequent turns,  cyclic walk along a line \\
\hline 83\_36  & 1,062 &0.38 &walk turn 90 degrees left walk forward\\
\hline 86\_06 & 9939& 3.63& walking, running, kicking, punching, knee kicking, and stretching\\
\hline   \multicolumn{4}{c}{\textbf{Used in Section \ref{subsec:parameters}}} \\
\hline 86\_02 & 10617  &  3.77  & walk, squats, run, stretch, jumps, punches, and drinking\\
\hline 86\_12 & 8,856 & 3.14 &walking, dragging, sweeping, dustpan, wipe window, and wipe mirror \\
\hline  \multicolumn{4}{c}{\textbf{Used in Section \ref{subsec:results}}} \\
\hline 41\_07 & 7,536 & 2.67&climb, step over, jump over, navigate around stepstool\\
\hline 56\_07 & 9,420 &3.34 &yawn, stretch, walk, run/jog,    angrily grab, jump, skip, halt \\
\hline 69\_08 &5,309  & 1.88  & walk and turn \\
\hline 86\_05 & 8,340 &2.96 &walking, jumping, jumping jacks, jumping on one foot, punching \\
\hline database\_1 &291,506 & 103.41&all sequences of subject: 31, 40  and 86\\
\hline database\_2 & 223,852 &79.41 &all sequences of subject: 105, 106, 111 and 113 \\
\hline   \multicolumn{4}{c}{\textbf{Used in Section \ref{subsec:comparison}}} \\
\hline 15\_04 &22,549 & 8.02 &wash windows, paint, hand signals, dance, the dive, the twist, boxing\\
\hline 17\_08 &  6,197  & 2.20 &muscular, heavyset person's walk \\
\hline 17\_10 &  2,783   &0.99  &boxing \\
\hline 85\_12  & 4,499 & 1.60 &jumps, flips, breakdance\\
\hline database\_3  & 341,472& 121.14&all sequences of subject: 6, 15, 16, 17,  35, 94 and 135\\
\hline database\_4 & 122,846&43.58  &all sequences of subject: 1, 2, 5, 6, 9, and 12\\
\hline \hline
\end{tabular}
\end{table*}

\subsection{Parameter Setting}\label{subsec:parameters}

  The performance of our algorithm depends on the clip lengths $L_i$, $i=1,2,\cdots, N$, which are specified by either the user or a motion segmentation algorithm, and three parameters, i.e., $k$, $l_i$, and $Q_i$, which are determined by our algorithm automatically.  For database, the user also specifies the number of transform matrices $K$.

  Throughout this paper, the compression ratio (CR) is defined as the ratio of the
  original data size to the compressed data size.
  We measure the distortion using average Euclidian distance (in cm), i.e.,
  \begin{equation}
  \centering
  \overline{d}=\frac{1}{nf}\sum_{i=1}^n \sum_{j=1}^f
  \sqrt{\left(\mathbf{p}_{i,j}-\widehat{\mathbf{p}}_{i,j}\right)\left(\mathbf{p}_{i,j}-\widehat{\mathbf{p}}_{i,j}\right)^\intercal},
  \end{equation}
  where $\mathbf{p}_{i,j}=[x_{i,j}~y_{i,j}~z_{i,j}]$ and $\widehat{\mathbf{p}}_{i,j}=[\widehat{x}_{i,j}~\widehat{y}_{i,j}~\widehat{z}_{i,j}]$ are original and decompressed 3D
  coordinates of the $i$-th marker in the $j$-th frame, respectively.

\begin{figure}[ph]
\centering \subfigure[$L$=70]{
\includegraphics[width=1.6in]{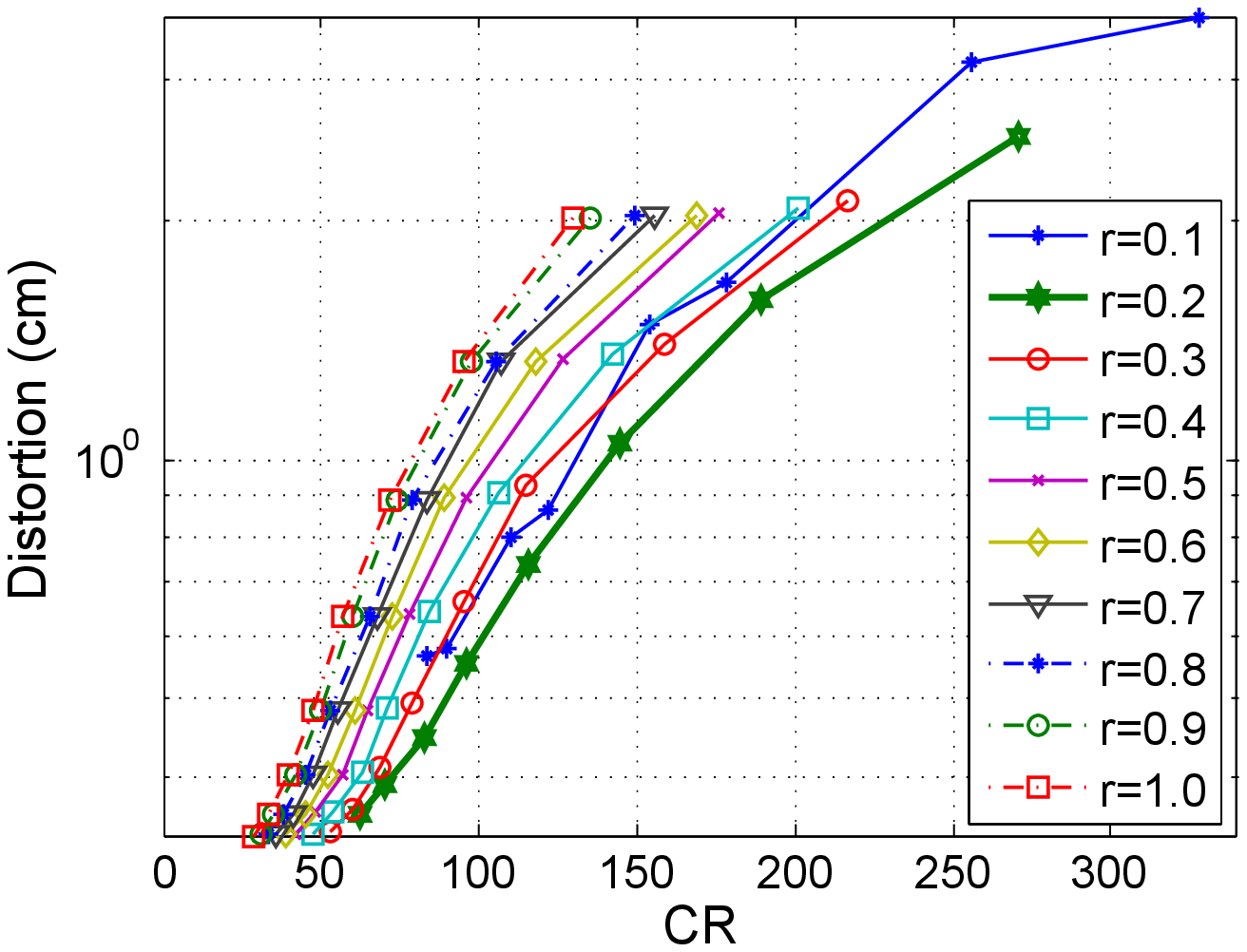}}
\subfigure[$L$=130]{
\includegraphics[width=1.6in]{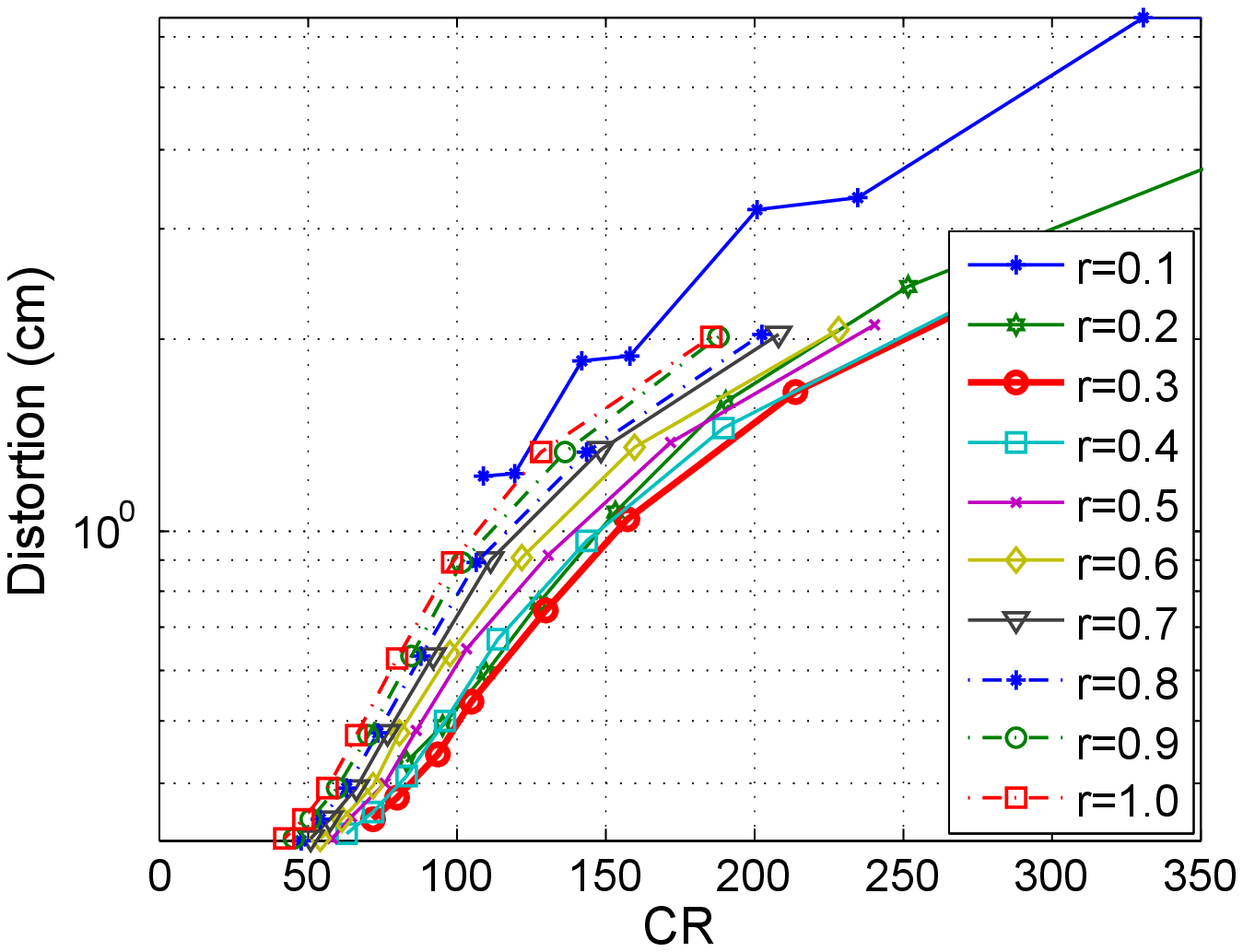}}
\subfigure[$L$=180]{
\includegraphics[width=1.6in]{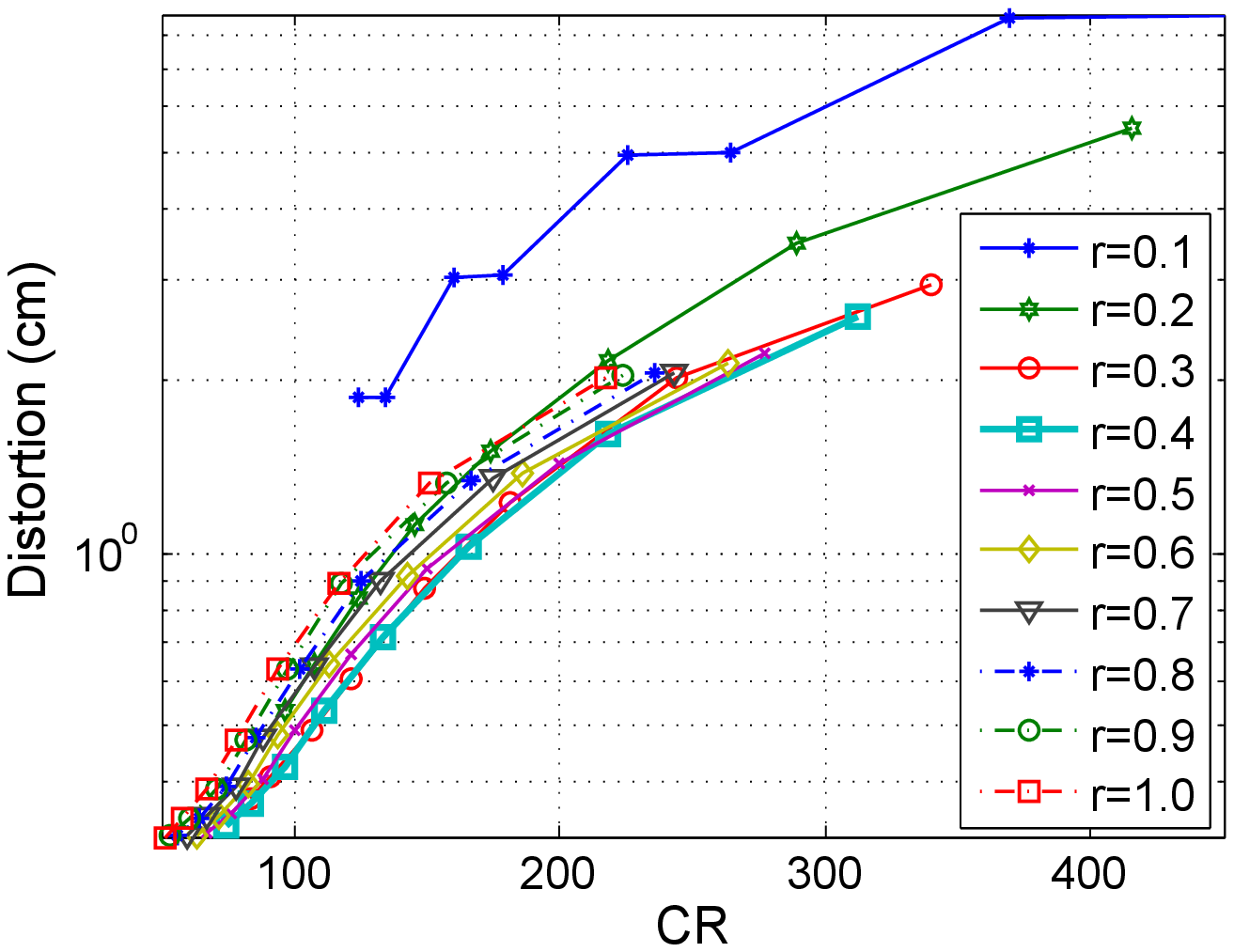}}
\subfigure[$L$=240]{
\includegraphics[width=1.6in]{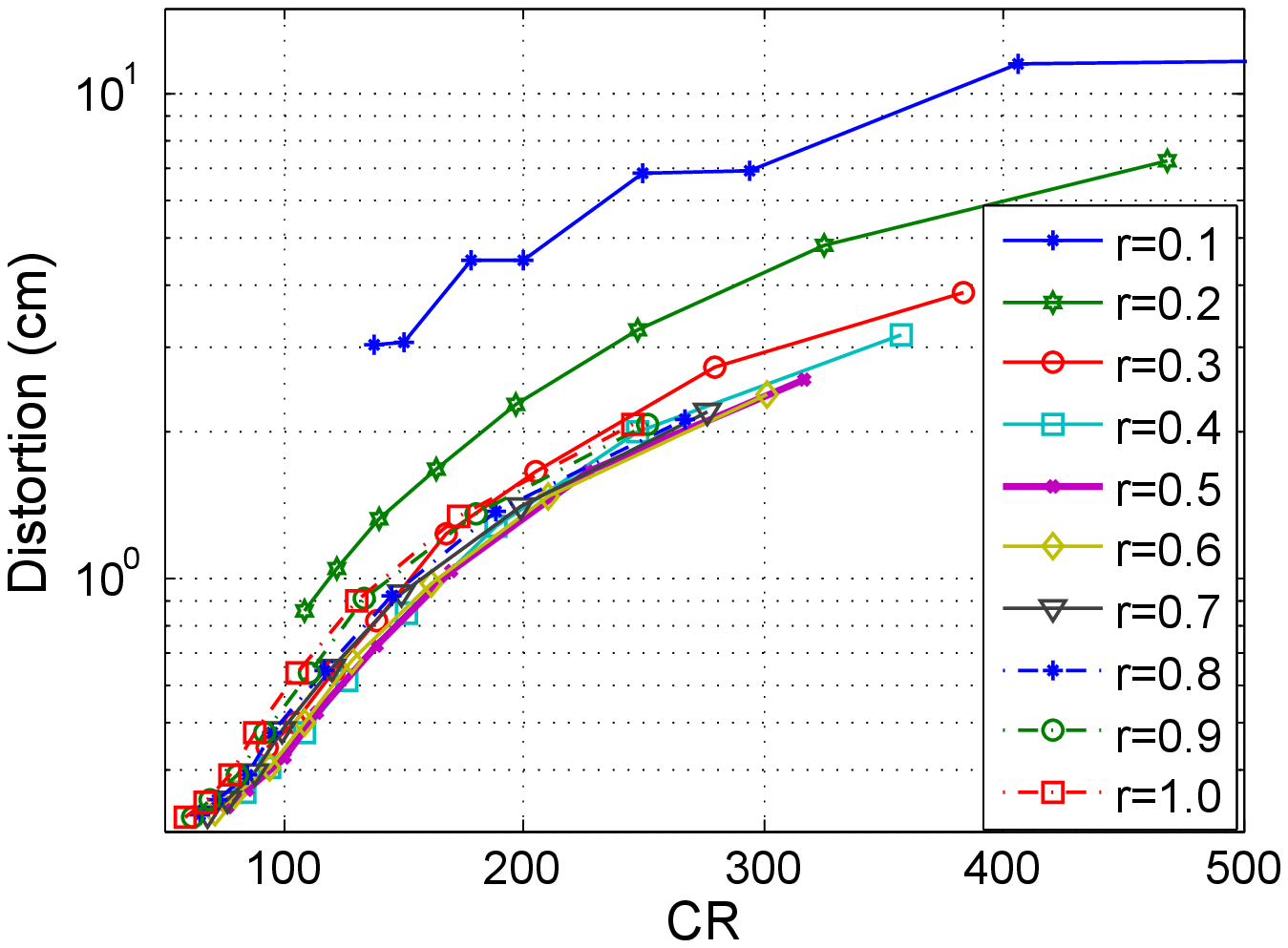}}
\subfigure[$L$=280]{
\includegraphics[width=1.6in]{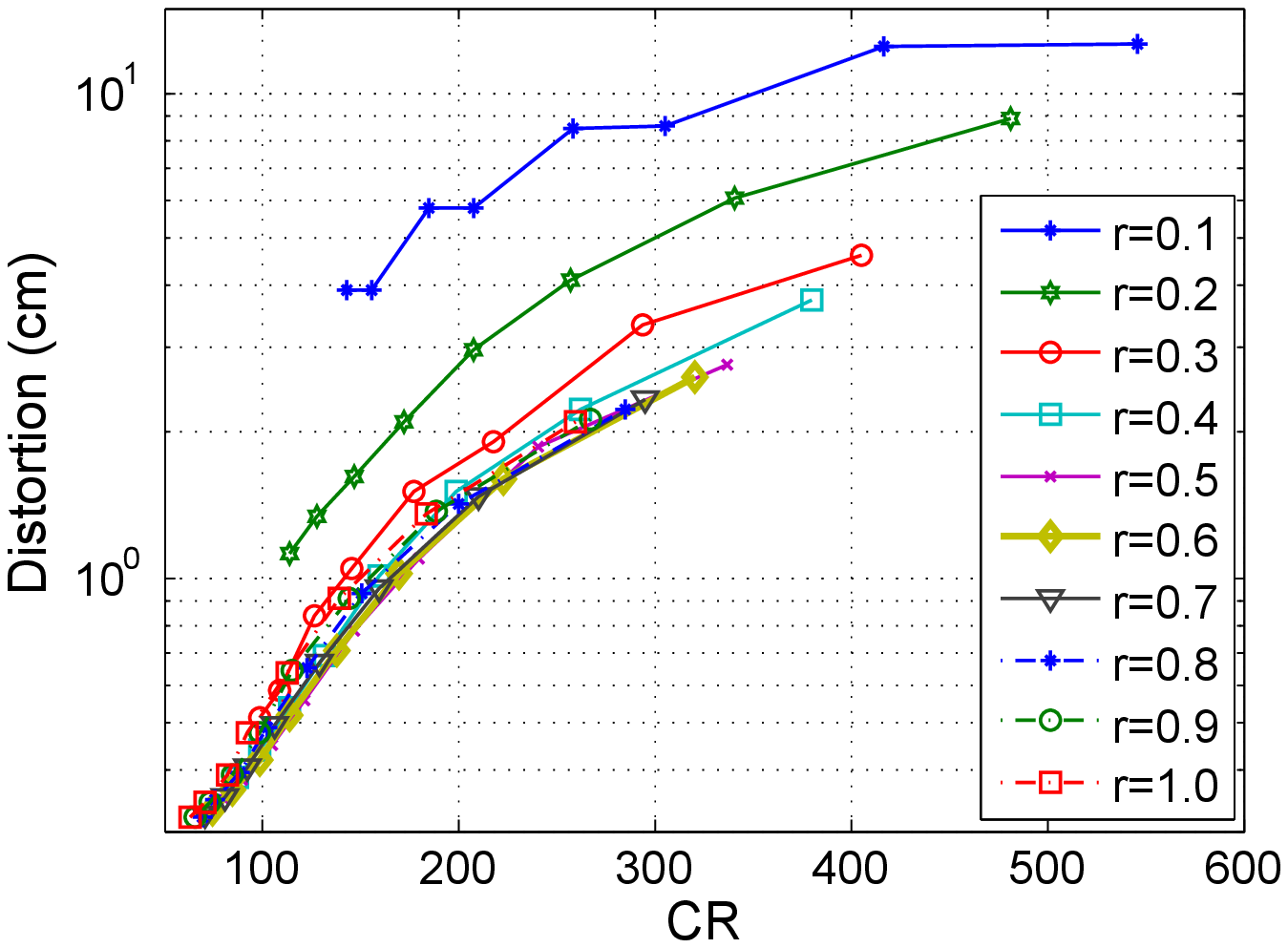}}
\subfigure[$L$=340]{
\includegraphics[width=1.6in]{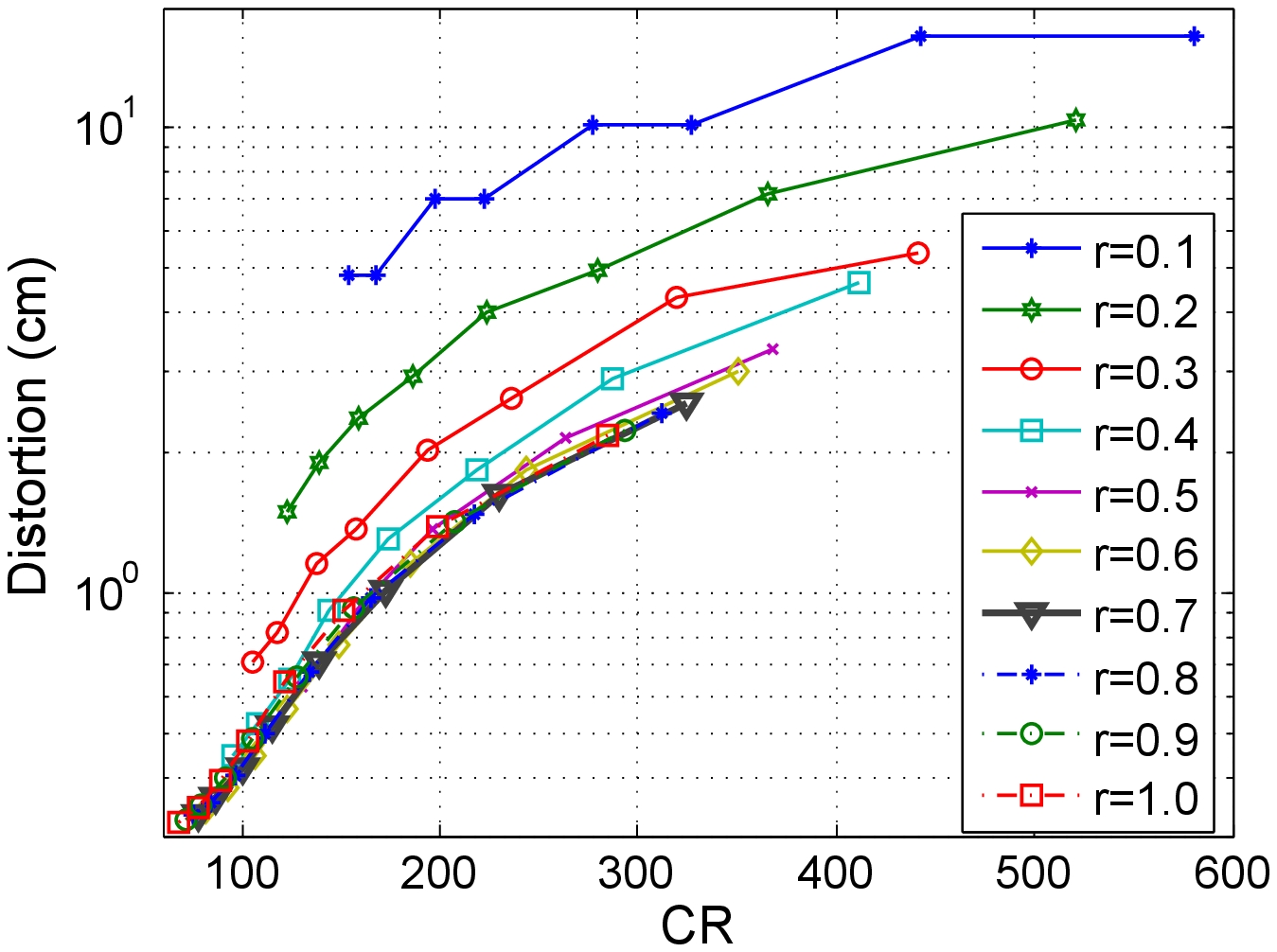}}
\caption{The impact of the ratio $r=l/k$ on compression performance
under various clip lengths on the test sequence 86\_02. In each
figure, the thick line is the optimal result.} \label{fig:optimal r}
\end{figure}

  Here, we show how the parameters $k$, $l_i$, and $Q_i$ are automatically set in our scheme. To simplify the analysis,
  we simply take all clips equally, i.e.,
  $L_i\equiv L$, $l_i\equiv l$ and $Q_i\equiv Q$ for all $i$.
  The parameters $k$ and $l$ specify the number of retained important coefficients in transform coding.
  Taking more coefficients in the direction with weaker correlation can effectively reduce
  the distortion and improve the compression performance.
  The degree of correlation within each row of $\mathbf{M}_i$ depends on the clip lengths,
  and computational results show that the ratio $r=l/k$ that leads to the least distortion is
  linearly proportional to clip length $L$ and we adopt the following empirical formula:
  \begin{equation}
  \centering r=0.1\times\left\lceil\frac{L}{50}\right\rceil, \label{equ:optimal r}
  \end{equation}
  where $\lceil\cdot\rceil$ is the ceiling function. See Fig.~\ref{fig:optimal r}.

\begin{figure}[]
\centering
\subfigure[Sequence 15\_12]{
\includegraphics[width=1.65in]{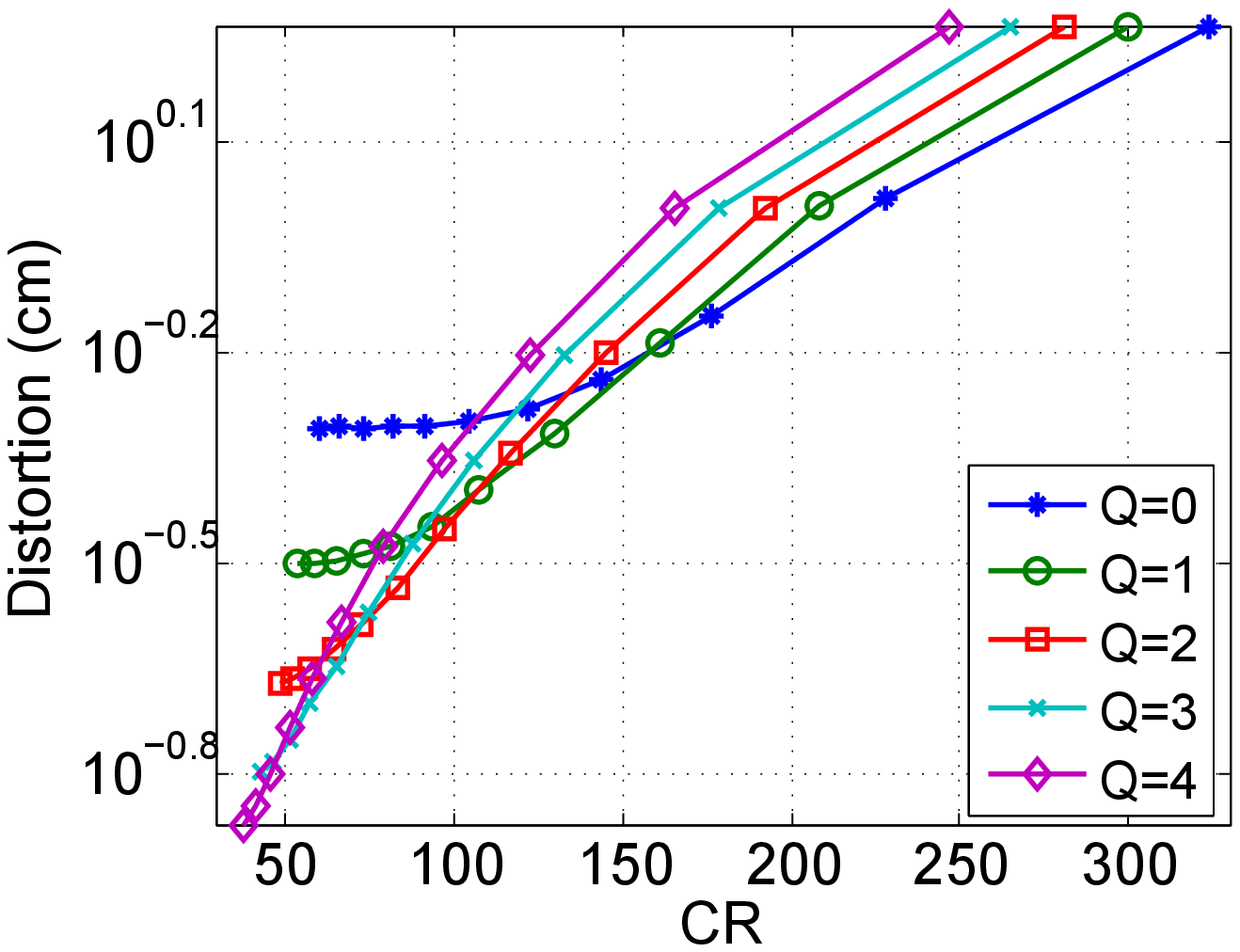}}
\subfigure[Sequence 38\_04]{
\includegraphics[width=1.65in]{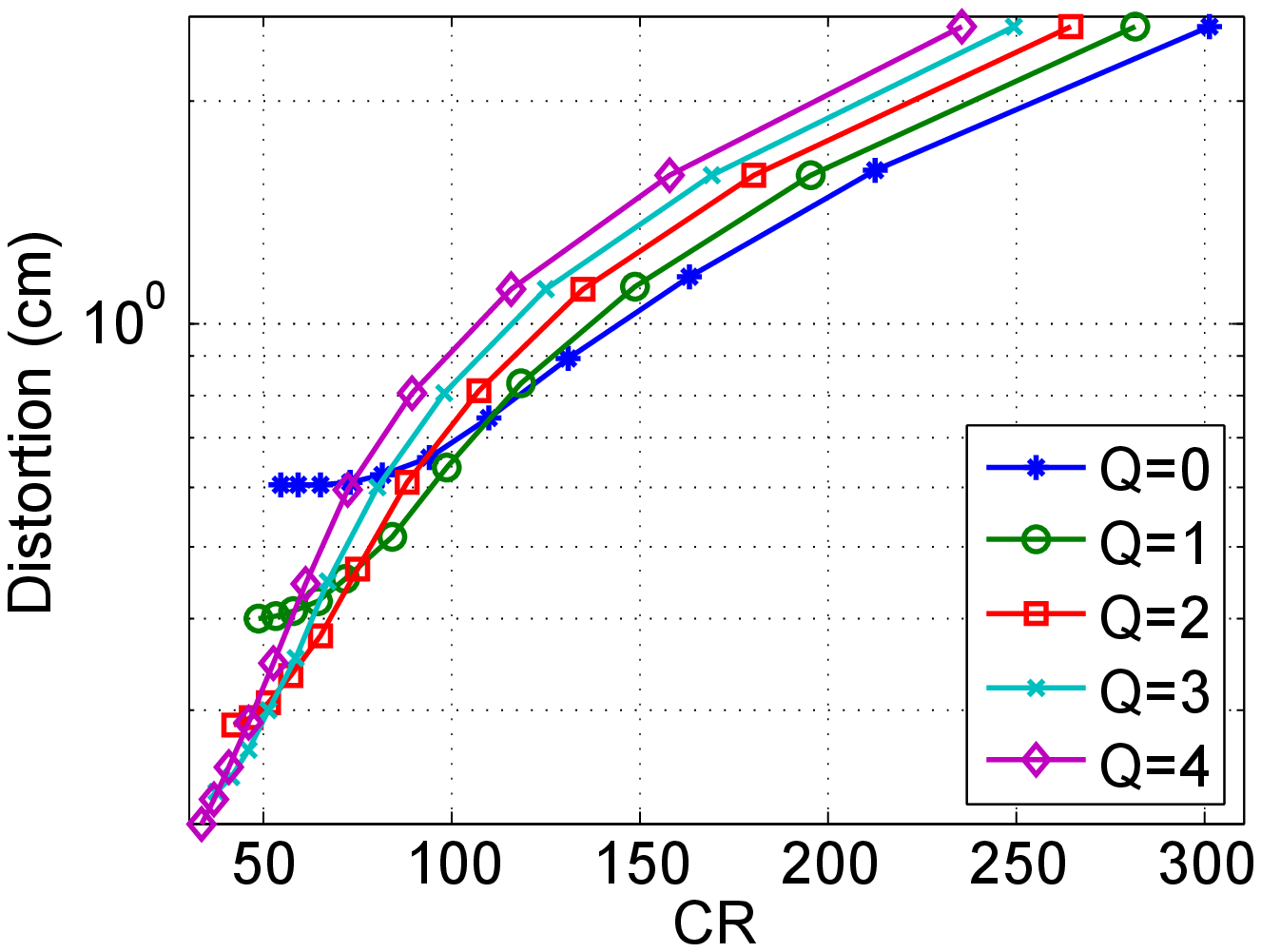}}
\subfigure[Sequence 55\_28]{
\includegraphics[width=1.65in]{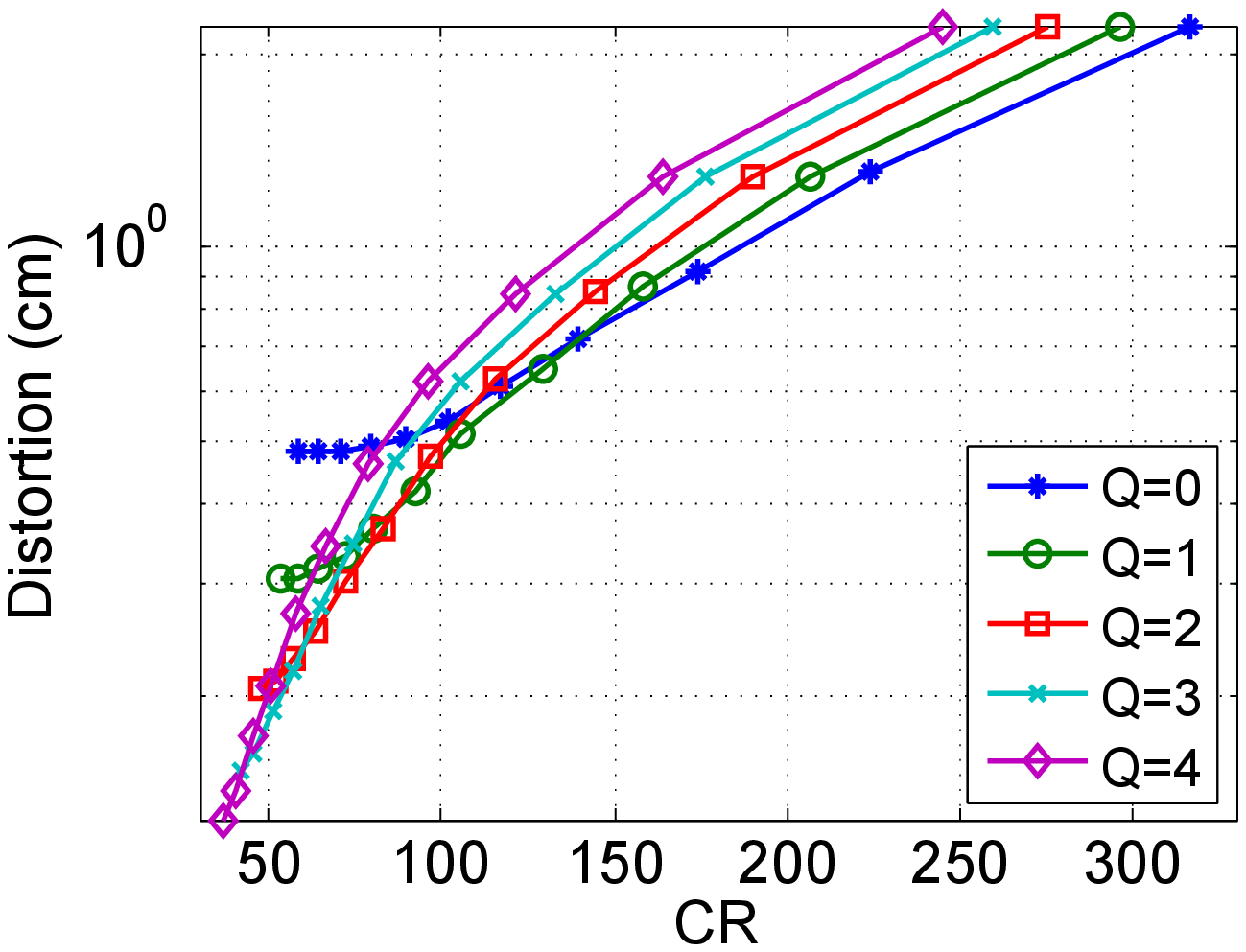}}
\subfigure[Sequence 86\_12]{
\includegraphics[width=1.65in]{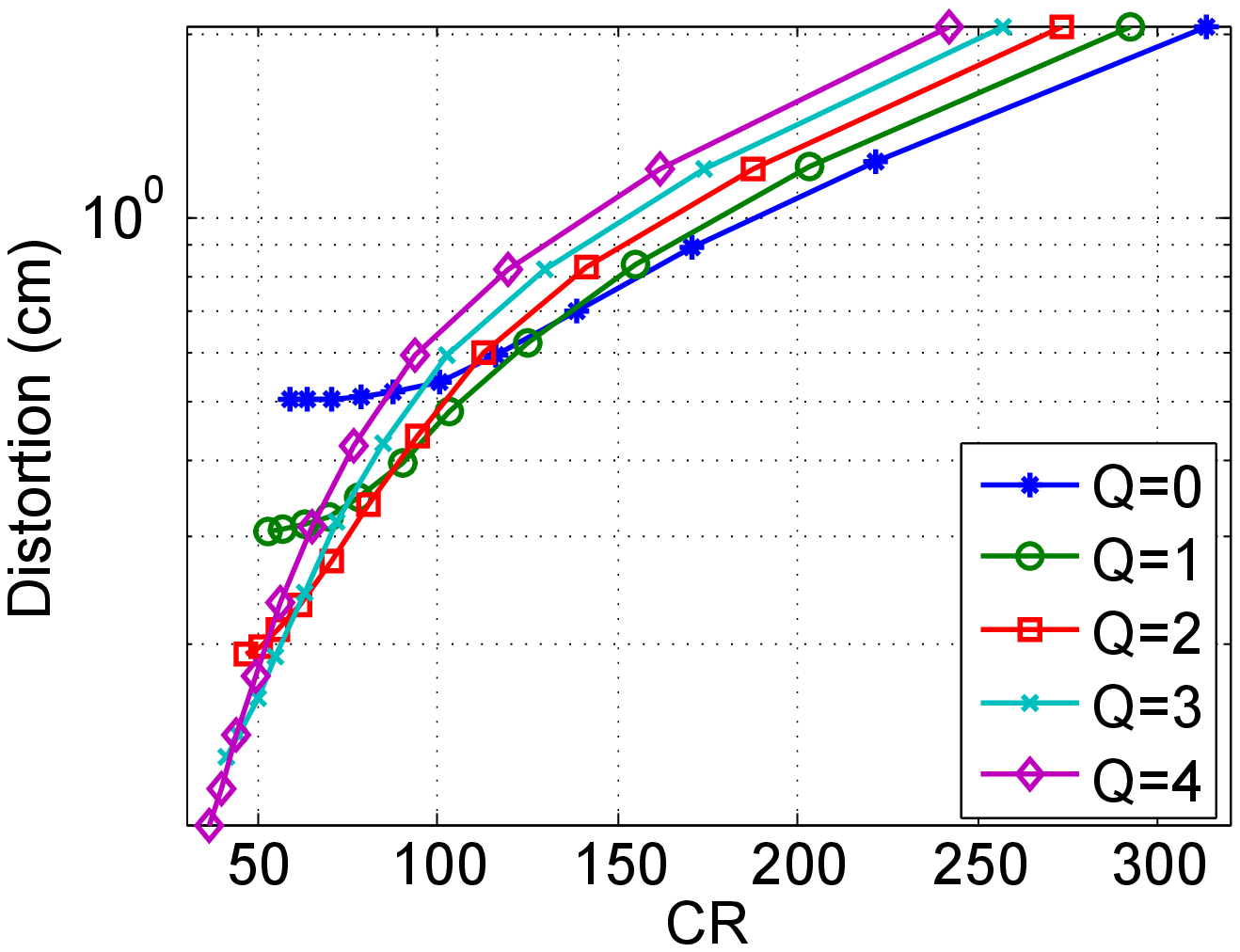}}
\caption{Compression performance and the quantization parameter $Q$. In each
curve plot, the parameter $k$ is sampled from $15$ to $65$ with a
step length $5$. The clip length $L=270$. }
\label{fig:optimal Q}
\end{figure}

  The quantization parameter $Q$ determines the size of the fractional parts of the transformed coefficients.
  Increasing $Q$ leads to more accurate results, but it reduces the compression ratio.
  Fig. \ref{fig:optimal Q} shows that smaller $Q$ produces better results at high CRs and larger $Q$ is preferred at low CRs.
  In our implementation, we adopt the following empirical formula:
  \begin{equation}
  \centering Q=\left\{
  \begin{array}{ccc}
  0&   0<k\leq 30 \\
  \\
  \left\lceil\frac{k-30}{10}\right\rceil & 30<k \leq 93.
  \end{array}
  \right.
  \label{equ:Q}
  \end{equation}

\subsection{Results} \label{subsec:results}

  We implement our algorithm in MATLAB and evaluate its performance and quality on the CMU Mocap Database \cite{cmu}.
  Table \ref{tab:sequences details} shows the details of the test data.
  We use the MATLAB toolbox \cite{matlab} to decode the .asf/.amc into 3-D marker coordinates, which are stored as 32-bit floating-points.

  We test our method under two scenarios, i.e., equal segmentation and adaptive segmentation.
  Various CRs are obtained by changing the parameter $k$.
  \begin{figure}
  \centering
  \subfigure[Sequence 41\_07]{
  \includegraphics[width=1.65in]{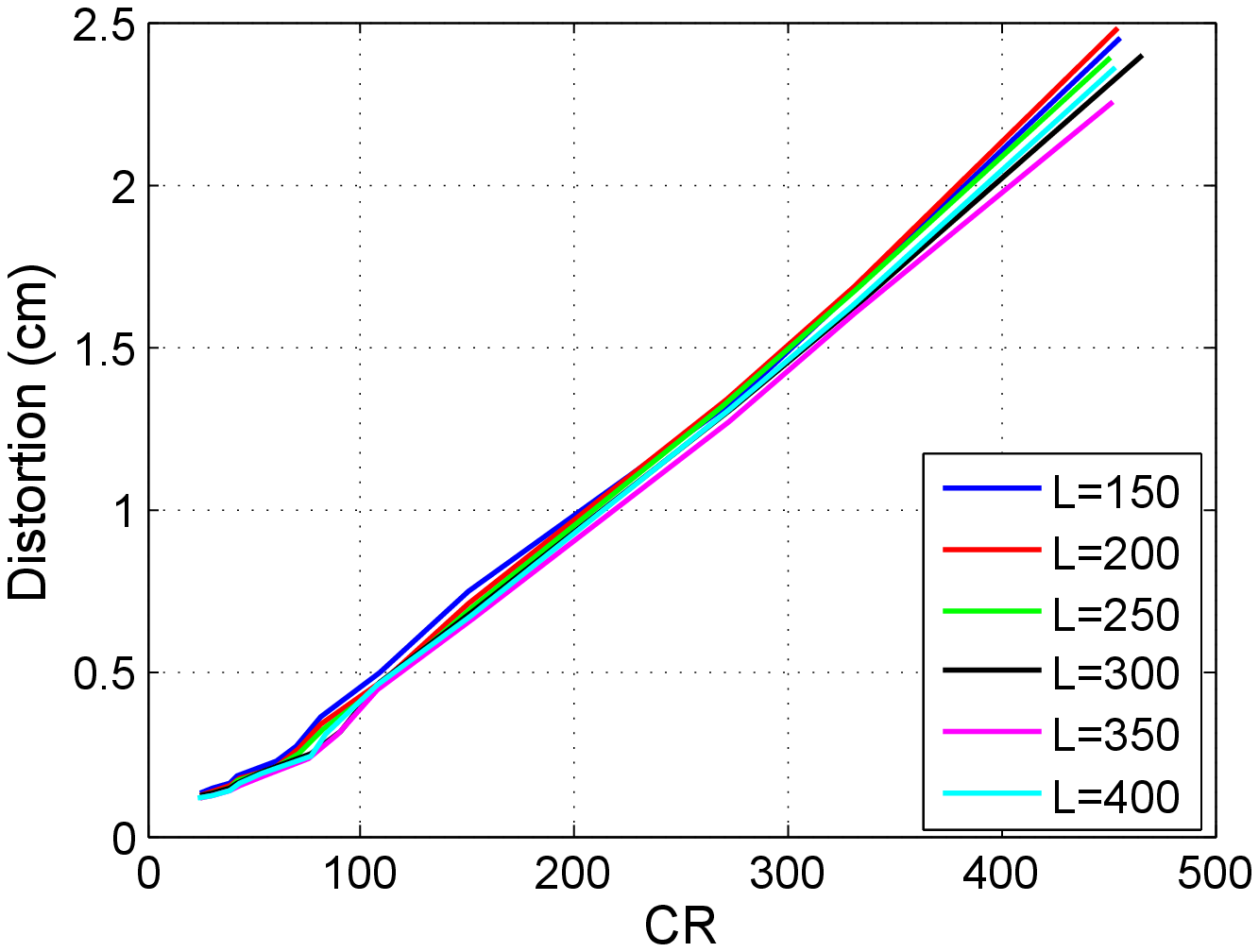}}
  \subfigure[Sequence 56\_07]{
  \includegraphics[width=1.65in]{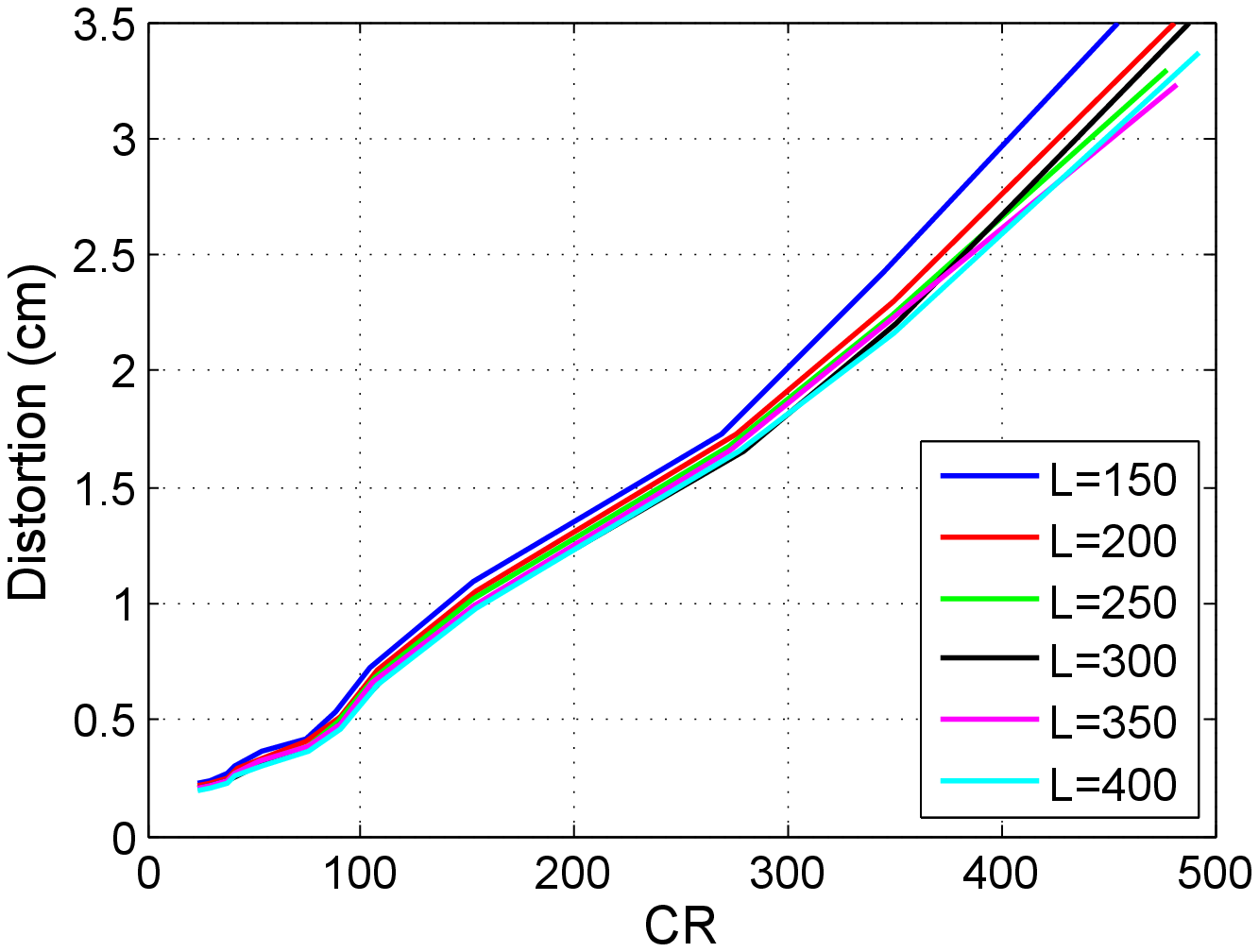}}
  \subfigure[Sequence 69\_08]{
  \includegraphics[width=1.65in]{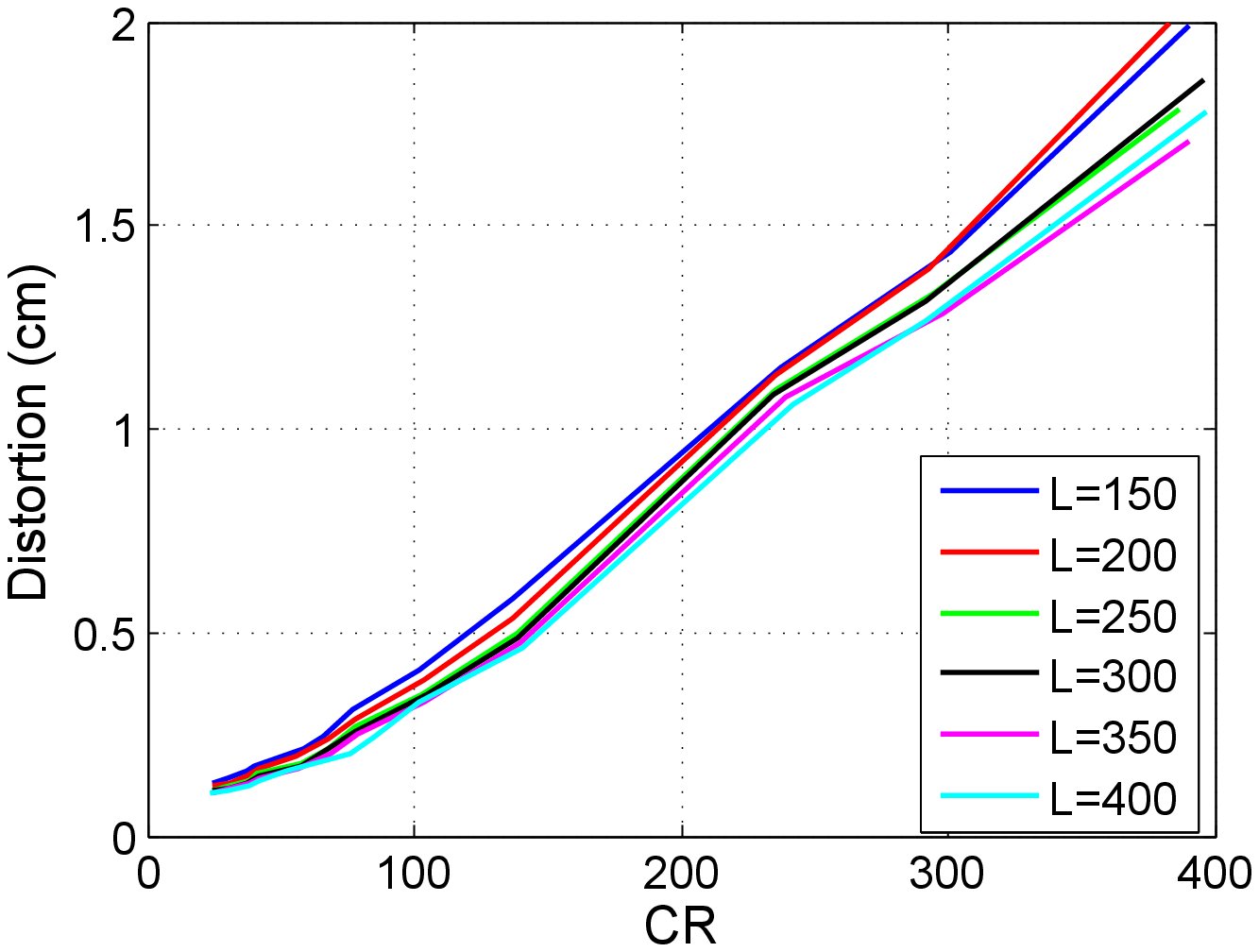}}
  \subfigure[Sequence 86\_05]{
  \includegraphics[width=1.65in]{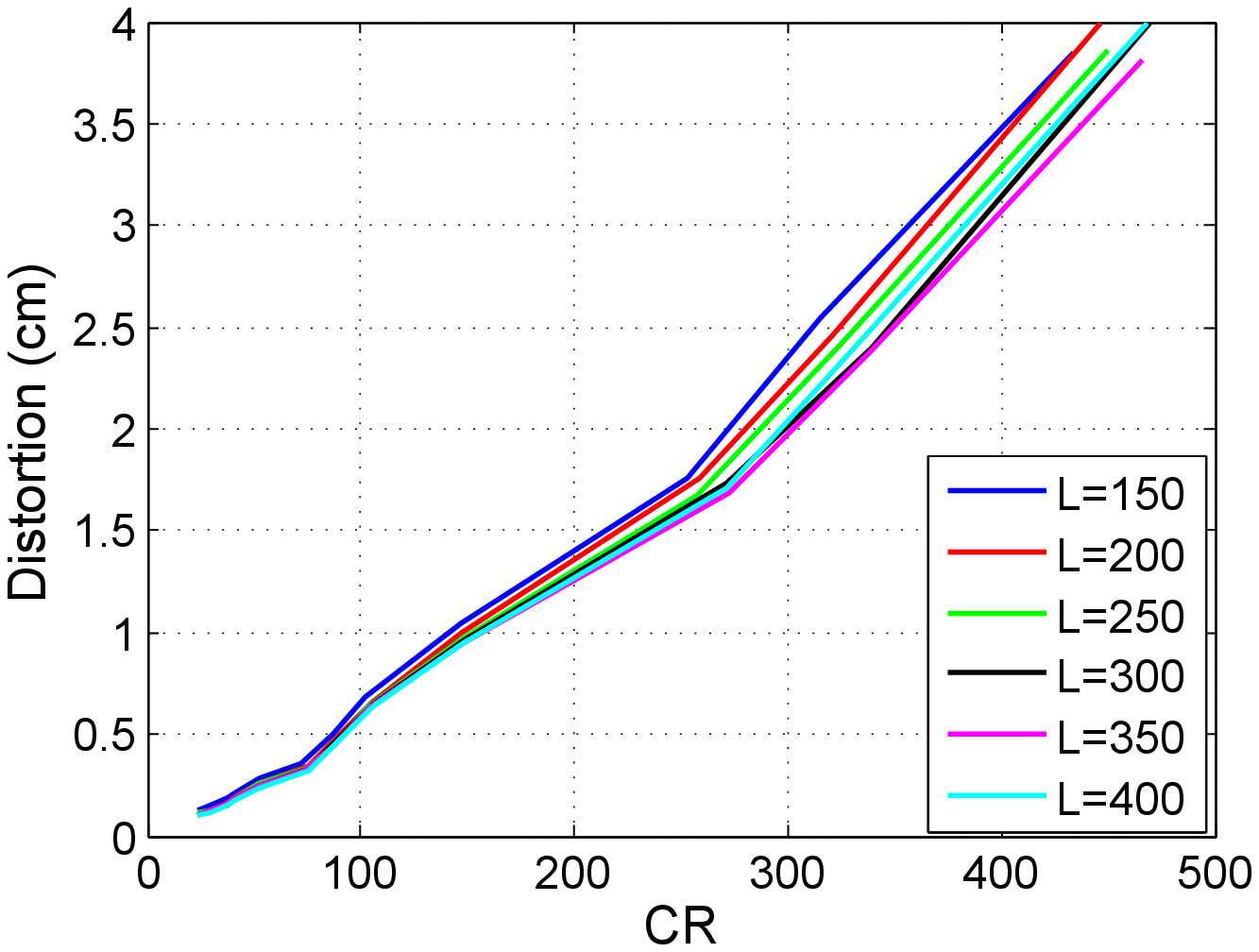}}
  \subfigure[database\_1 ($L$=280)]{
  \includegraphics[width=1.6in]{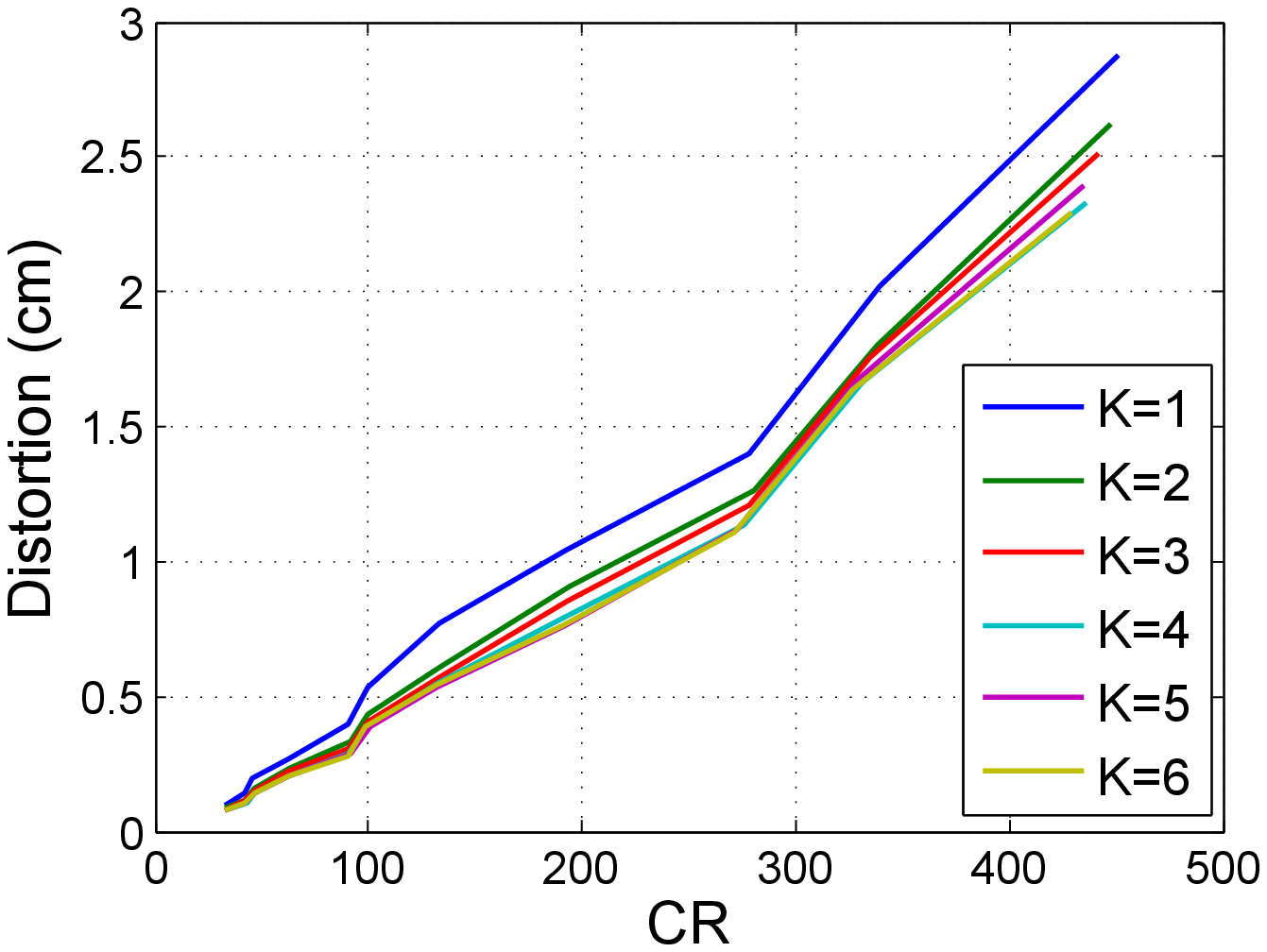}}
  \subfigure[database\_2 ($L$=280)]{
  \includegraphics[width=1.6in]{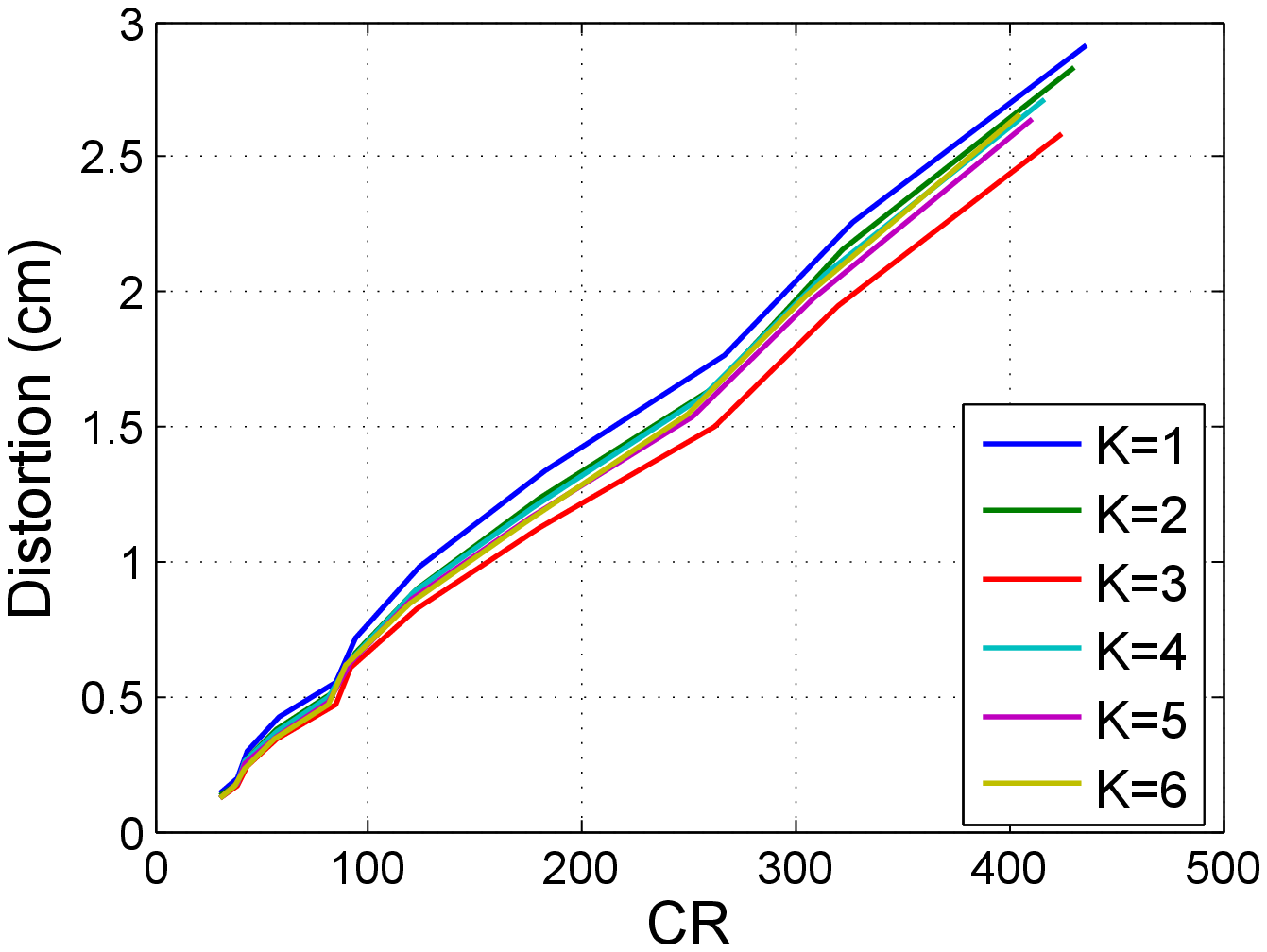}}
  \caption{CR-distortion curves for equal segmentation.}
  \label{fig:result fixed L}
  \end{figure}

\begin{table*}[]
\centering \caption{Algorithm efficiency. $E_{fps}$ and $D_{fps}$
are the encoding and decoding frame rates,
respectively.}\label{tab:averagecomplexity}
\begin{tabular}{c|c|c|c|c|c|c|c|c|c|c}
\hline \hline Data & 15\_04&17\_08 & 17\_10& 85\_12 &41\_07 & 56\_07& 69\_08& 86\_05 &database\_1 &database\_2  \\
\hline $E_{fps} (\times 10^4)$  &17.2 &7.19 &3.24 &4.90 &7.88 &9.61&5.98&8.64  &0.23 &0.21\\
\hline $D_{fps}(\times 10^4)$ &27.6 &9.65 &4.18 &6.83  &10.9 &13.5 &7.92&12.1 & 34.6 & 31.4 \\
\hline \hline
\end{tabular}
\end{table*}
\subsubsection{Equal segmentation}
In this scenario, we segment mocap data into clips with equal
length, i.e., $L_1=L_2=\cdots=L_N=L$. Equal segmentation is easy to implement and it is often
used when $L$ is specified by the user, since it is tedious for the
user to specify the length for each clip separately.

First, we test our method on mocap sequences by setting various
$L$s. We observe that increasing $L$ can improve the compression
performance, but the improvement becomes less significant when
$L\geq 350$. As Figs. \ref{fig:result fixed L} and \ref{fig:visual
results} show, the distortion is less than 1cm when CR$\geq 100$.
See also the accompanying video for more results.

Then, we apply our method to mocap databases. We specify one more
parameter $K$ for the number of transform matrices. Intuitively, the
more the data-dependent transform matrices, the less the distortion.
However, since the orthogonal matrices $\mathbf{B}_i$ are difficult
to compress, more spaces are required to store them, which reduces
the compression ratio. Fig. \ref{fig:result fixed L} shows that we
achieve the optimal compression performance for database\_1 and
database\_2 when $K$ is equal to 4 and 3, respectively.

Table \ref{tab:averagecomplexity} shows the performance statistics
of our method. Our unoptimized and serial MATLAB code is able to
encode more than 10,000 frames per second for a single mocap
sequence. It can also encode 2,000 fps for mocap databases. The
decoding speed is even faster than encoding. It is worth noting that
our method can be further sped up by parallel computation. For
example, computing $\mathbf{S}_i$ can be implemented in parallel on
the GPUs.

\begin{figure*}
\centering
\includegraphics[width=2.2in]{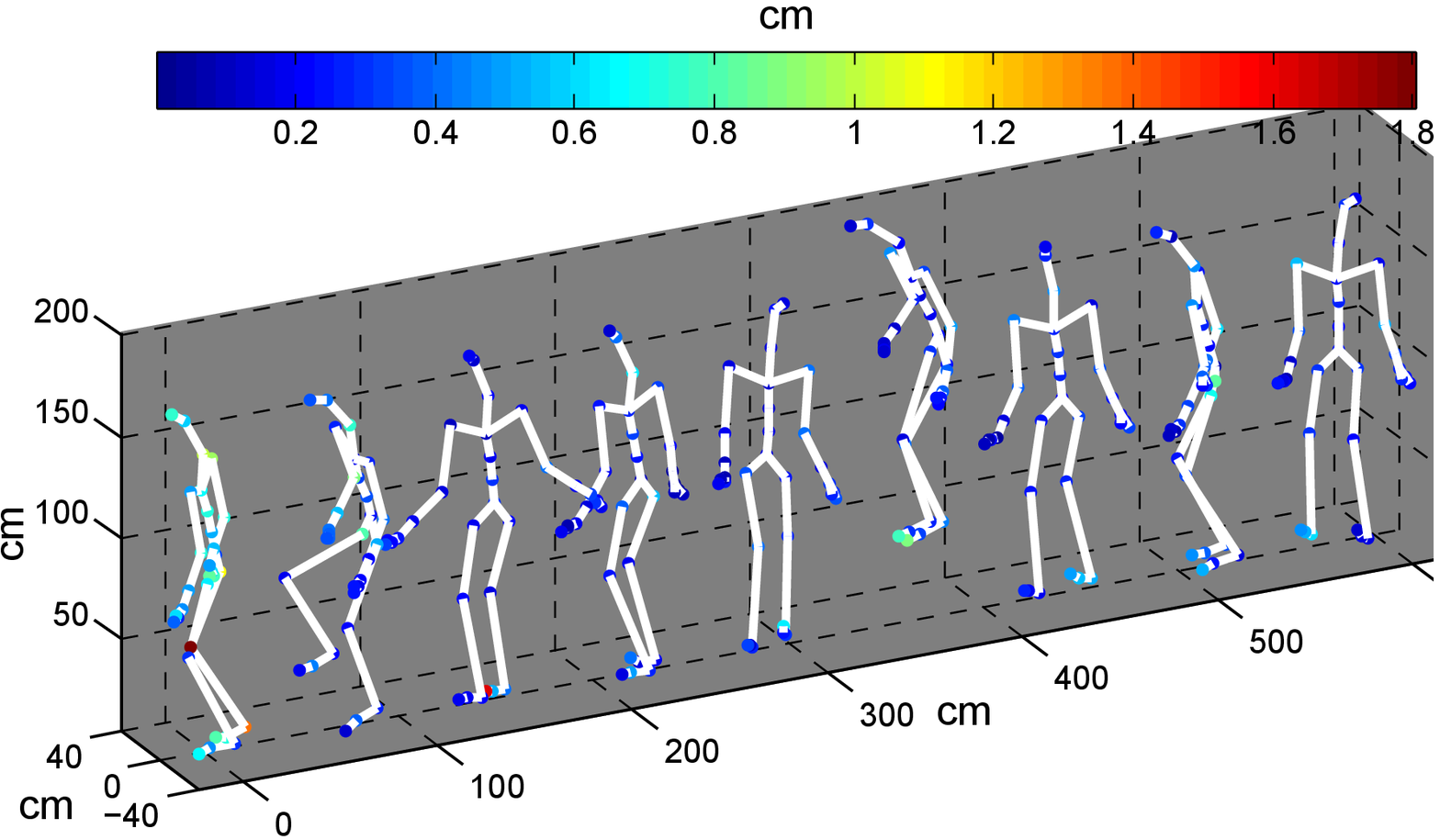}
\includegraphics[width=2.2in]{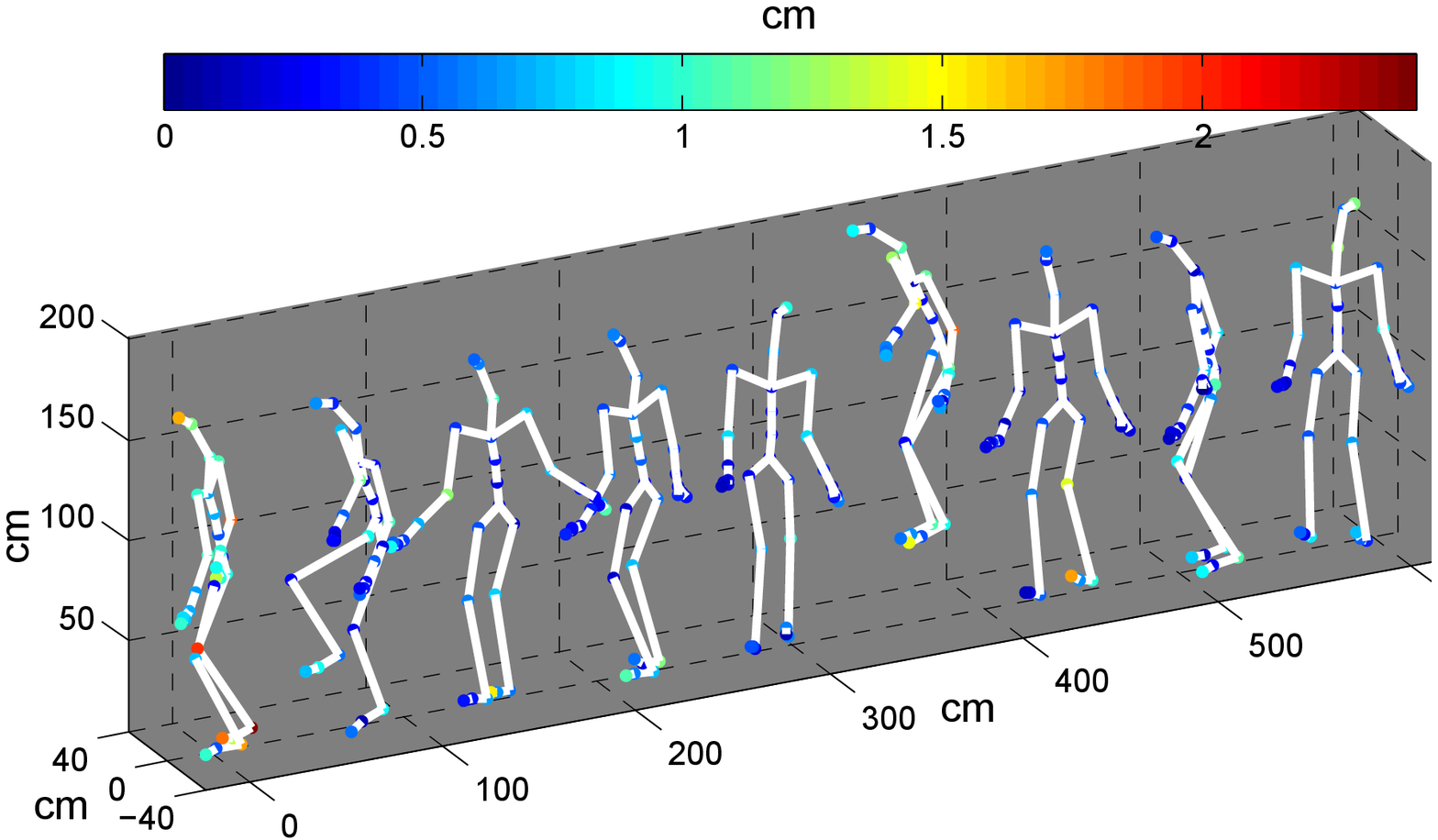}
\includegraphics[width=2.2in]{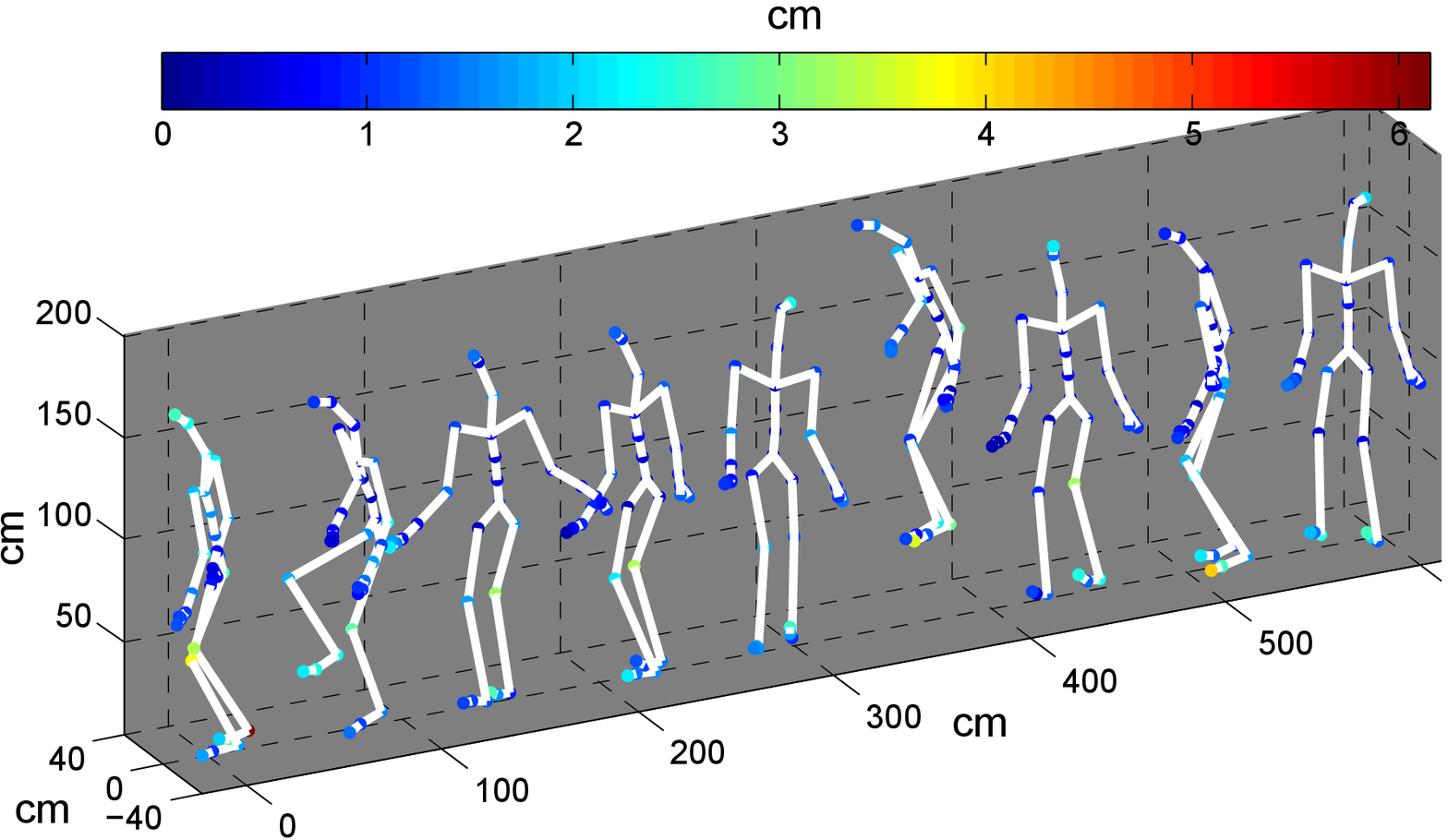}
\makebox[2.2in]{\footnotesize CR=97.9}\makebox[2.2in]{\footnotesize
CR=146.8}\makebox[2.2in]{\footnotesize CR=281.3}\\
\makebox[4.4in]{(a) Sequence 41\_07}\\
\includegraphics[width=2.2in]{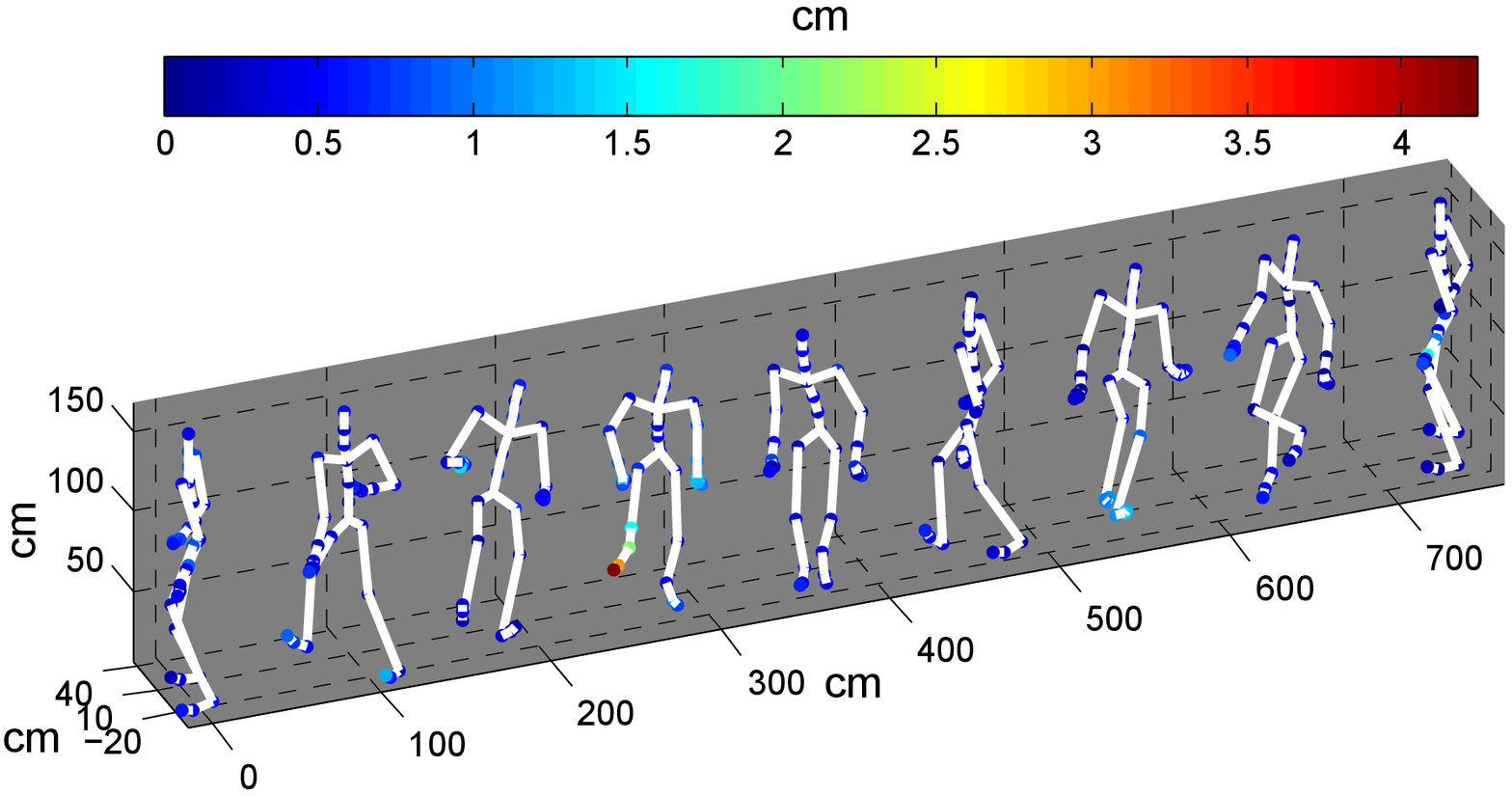}
\includegraphics[width=2.2in]{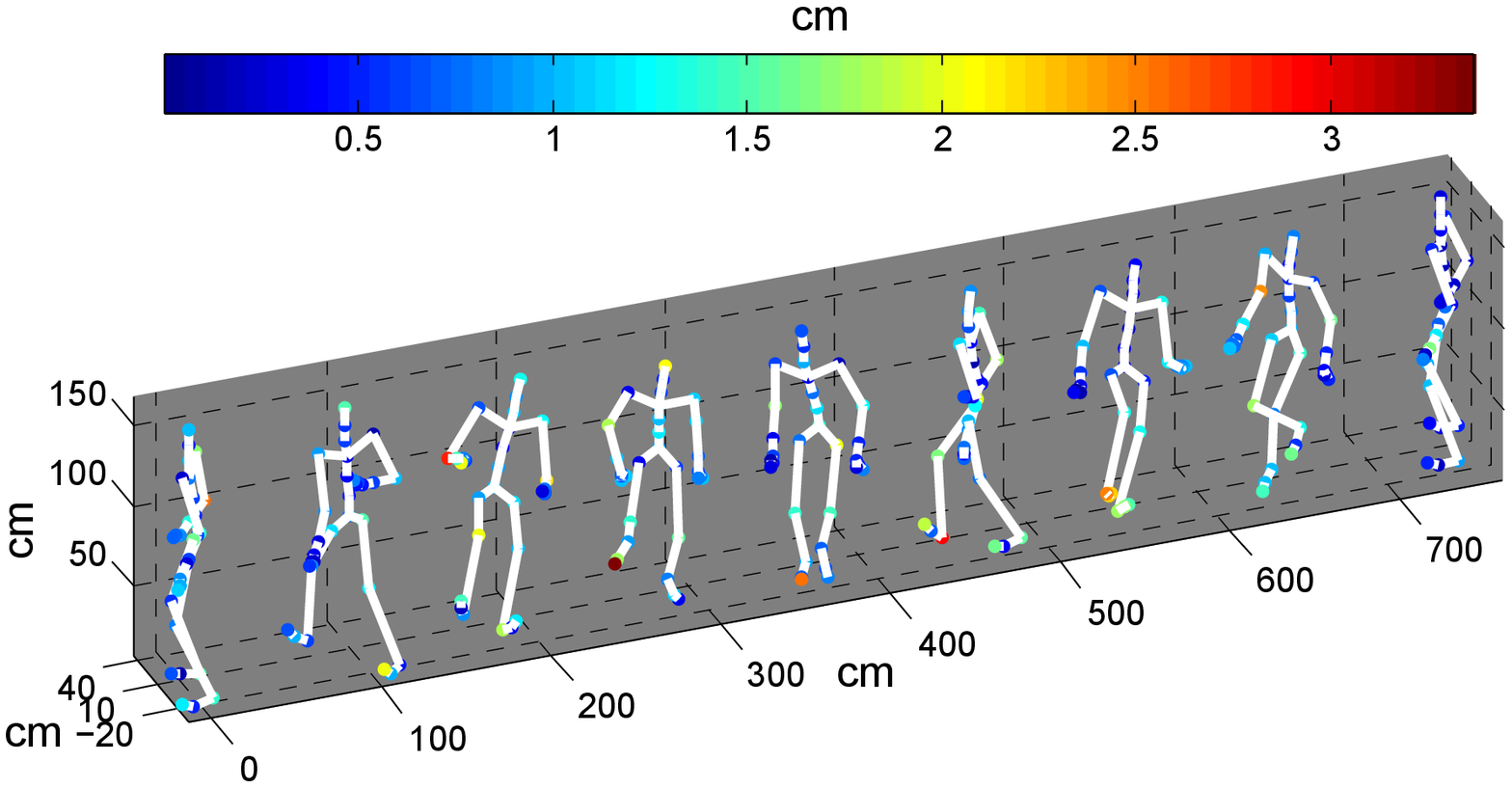}
\includegraphics[width=2.2in]{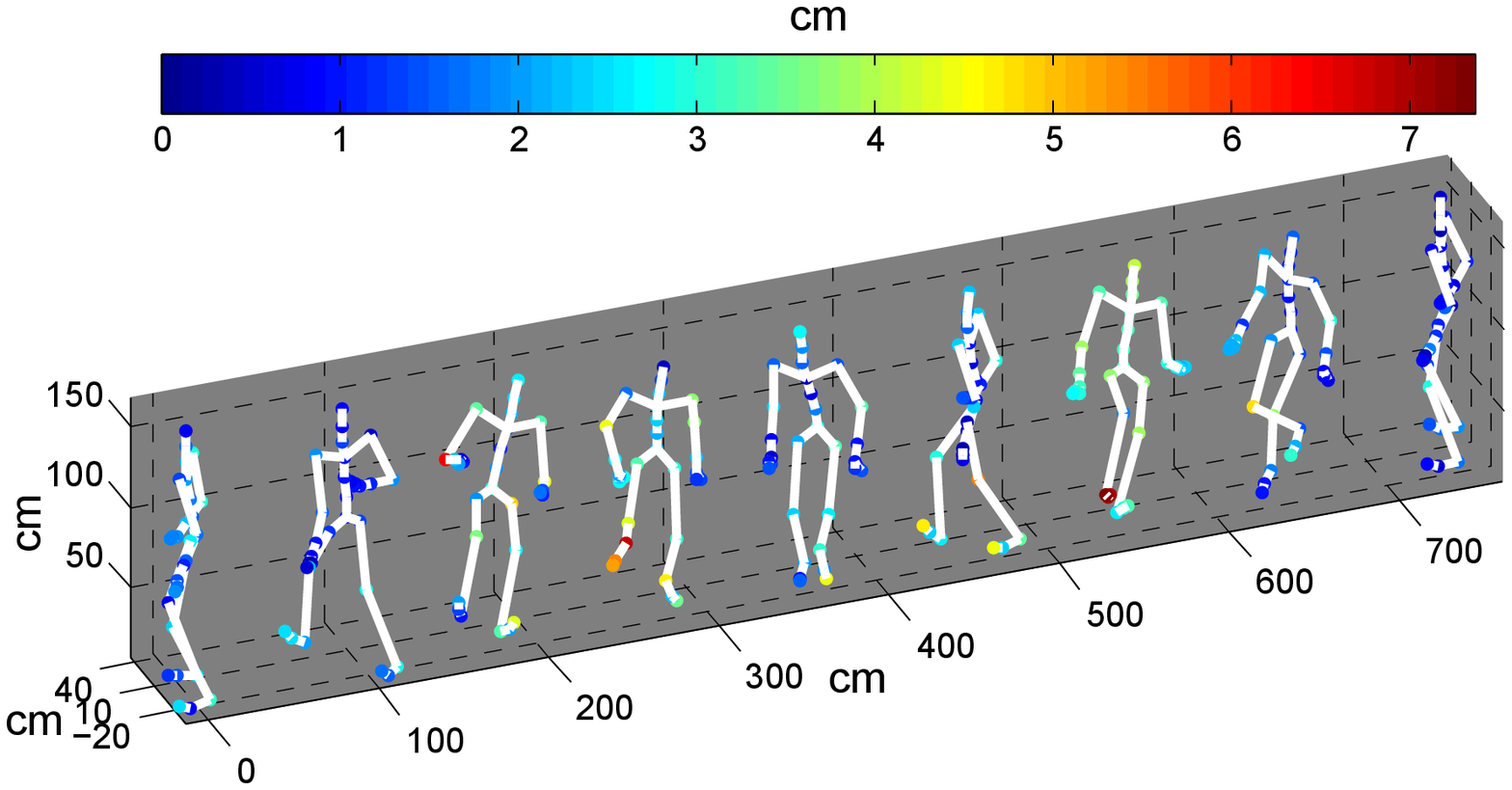}
\makebox[2.2in]{\footnotesize CR=86.9}\makebox[2.2in]{\footnotesize
CR=148.6}\makebox[2.2in]{\footnotesize CR=289.9}\\
\makebox[4.4in]{(b) Sequence 56\_07}\\
\includegraphics[width=2.2in]{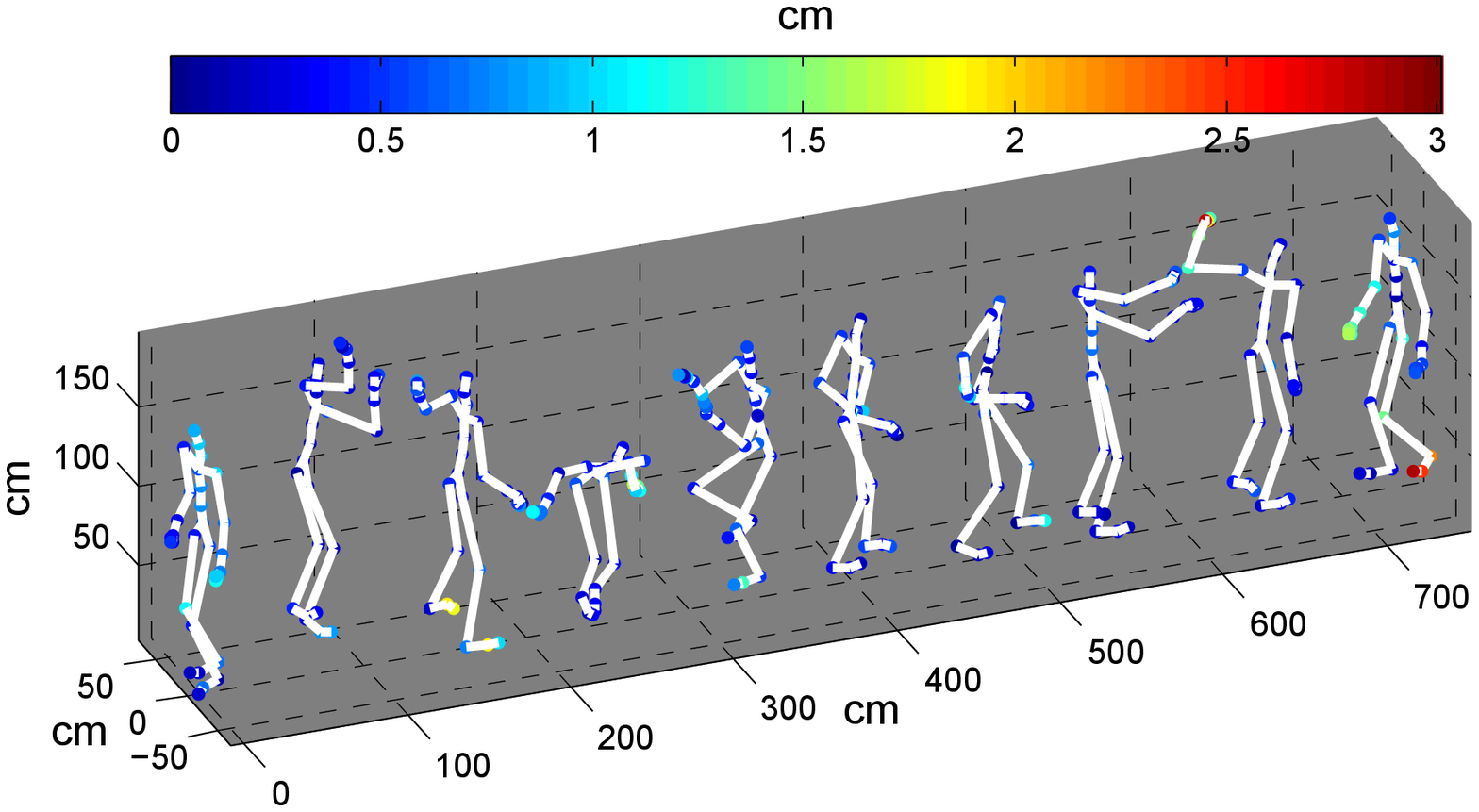}
\includegraphics[width=2.2in]{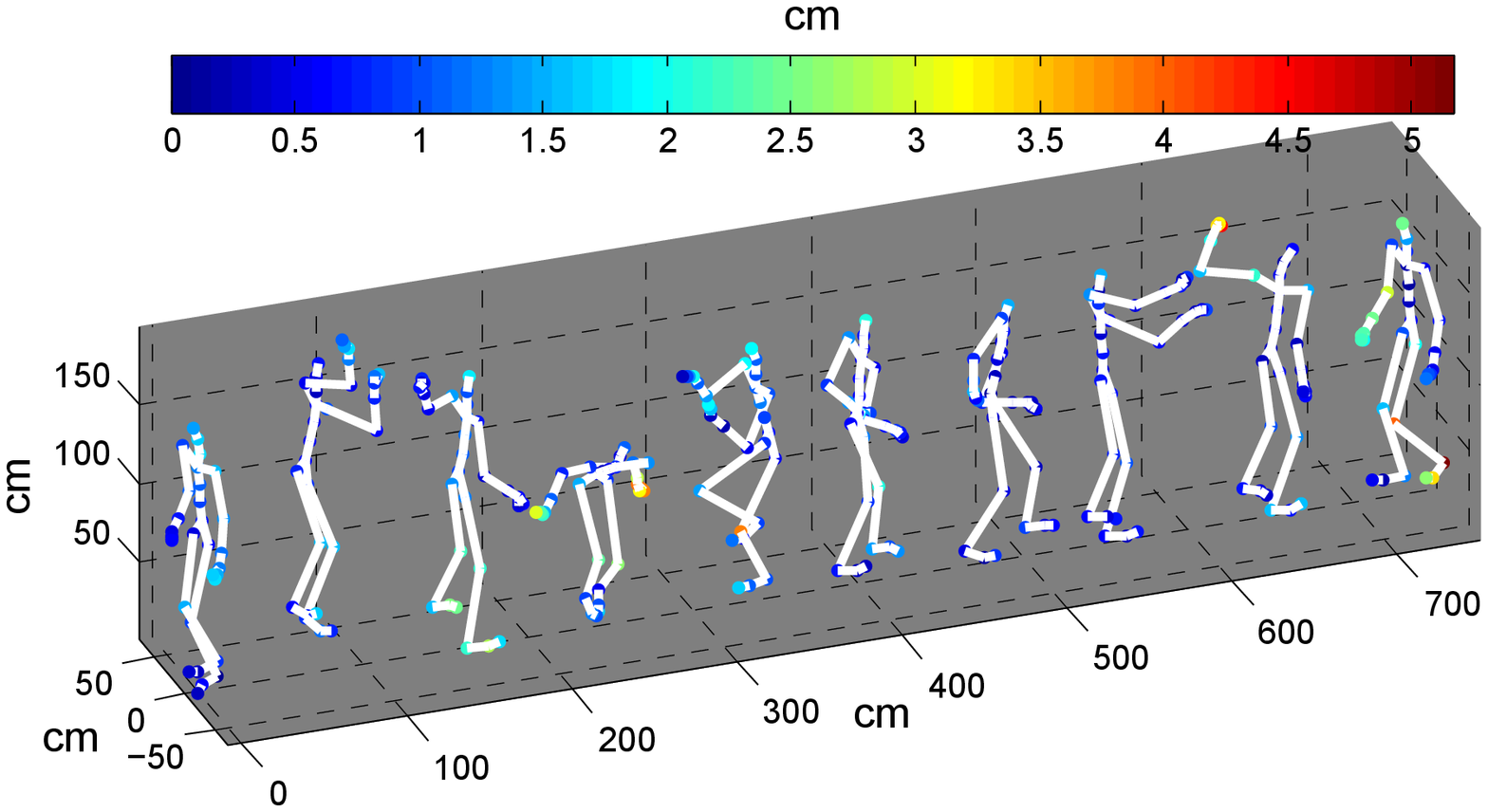}
\includegraphics[width=2.2in]{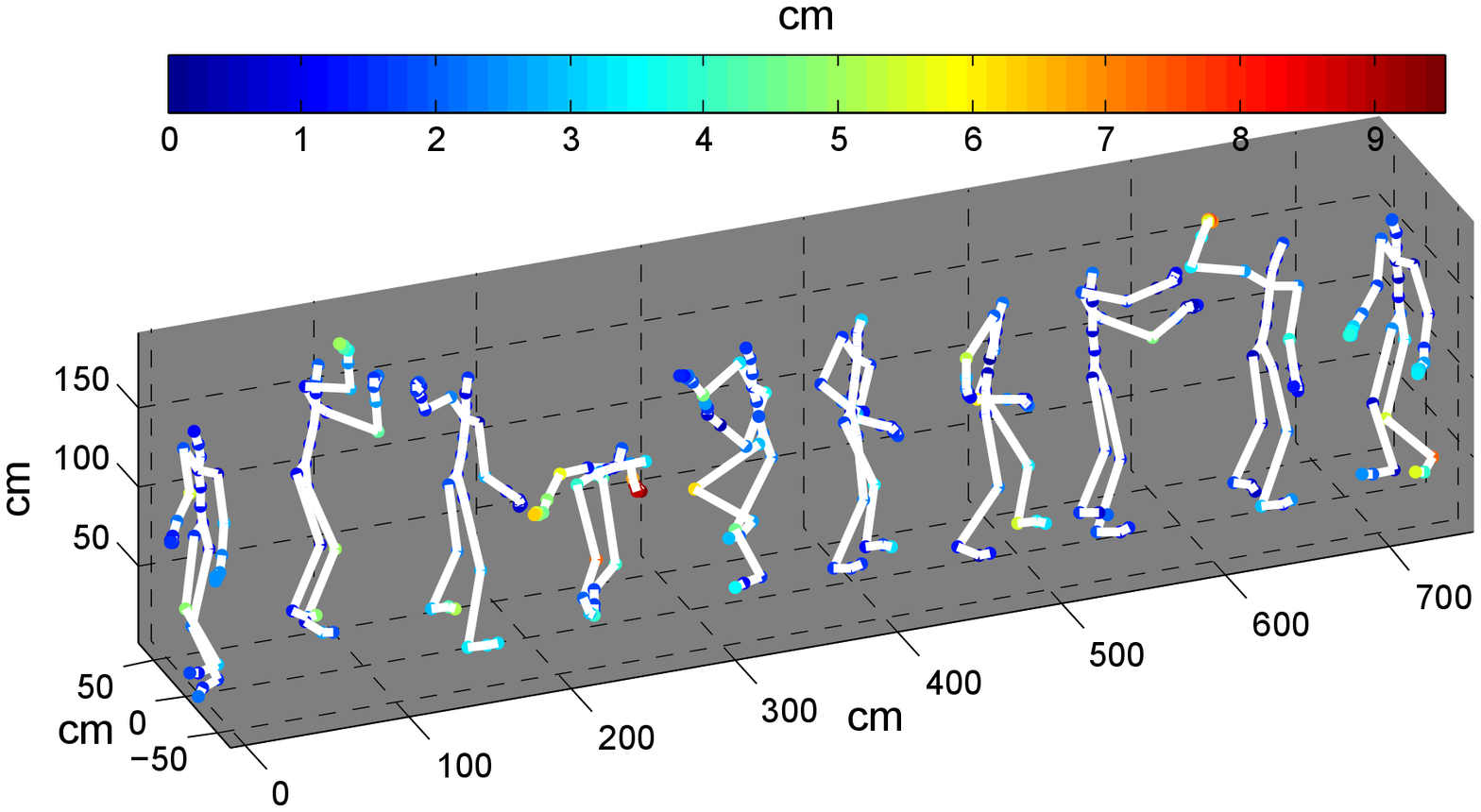}
\makebox[2.2in]{\footnotesize CR=79.1}\makebox[2.2in]{\footnotesize
CR=142.5}\makebox[2.2in]{\footnotesize CR=271.9}\\
\makebox[4.4in]{(c) Sequence 86\_05} \caption{Visual results for equal clip segmentation ($L$=280). The joint distortions are colored in heatmap.} \label{fig:visual results}
\end{figure*}

\begin{figure}
  \centering
  \subfigure[Sequence 41\_07]{
  \includegraphics[width=1.65in]{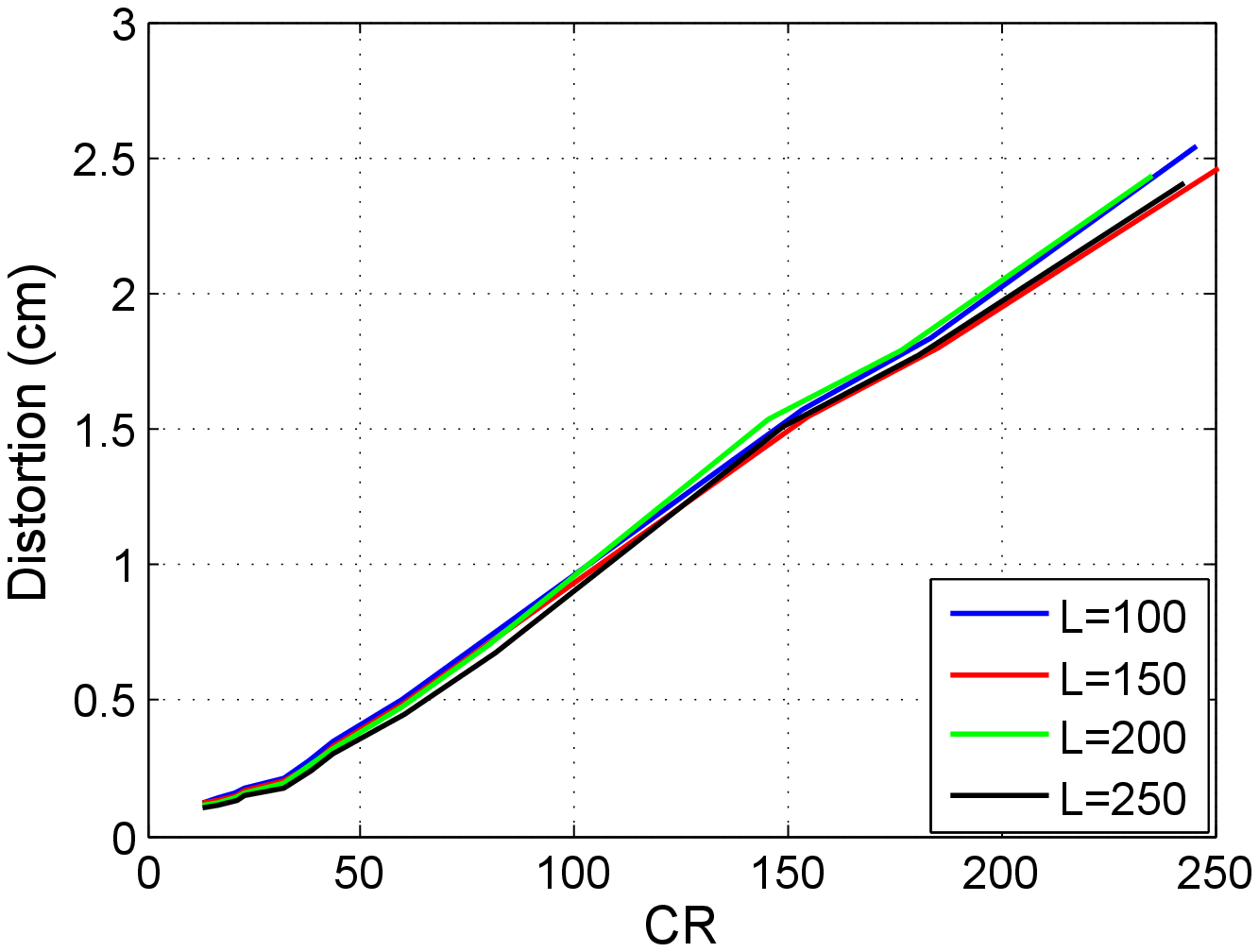}}
  \subfigure[Sequence 56\_07]{
  \includegraphics[width=1.65in]{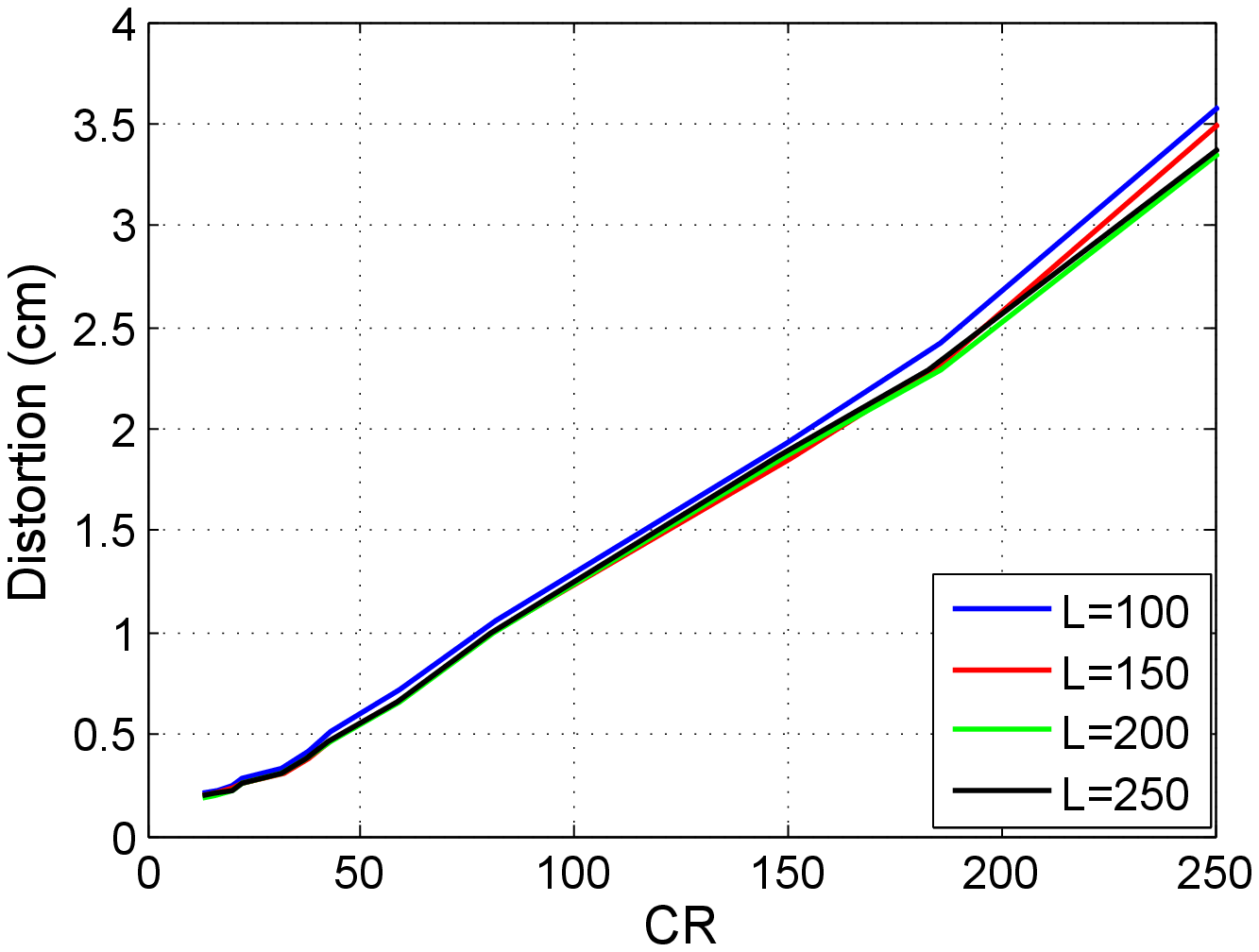}}
  \subfigure[Sequence 69\_08]{
  \includegraphics[width=1.65in]{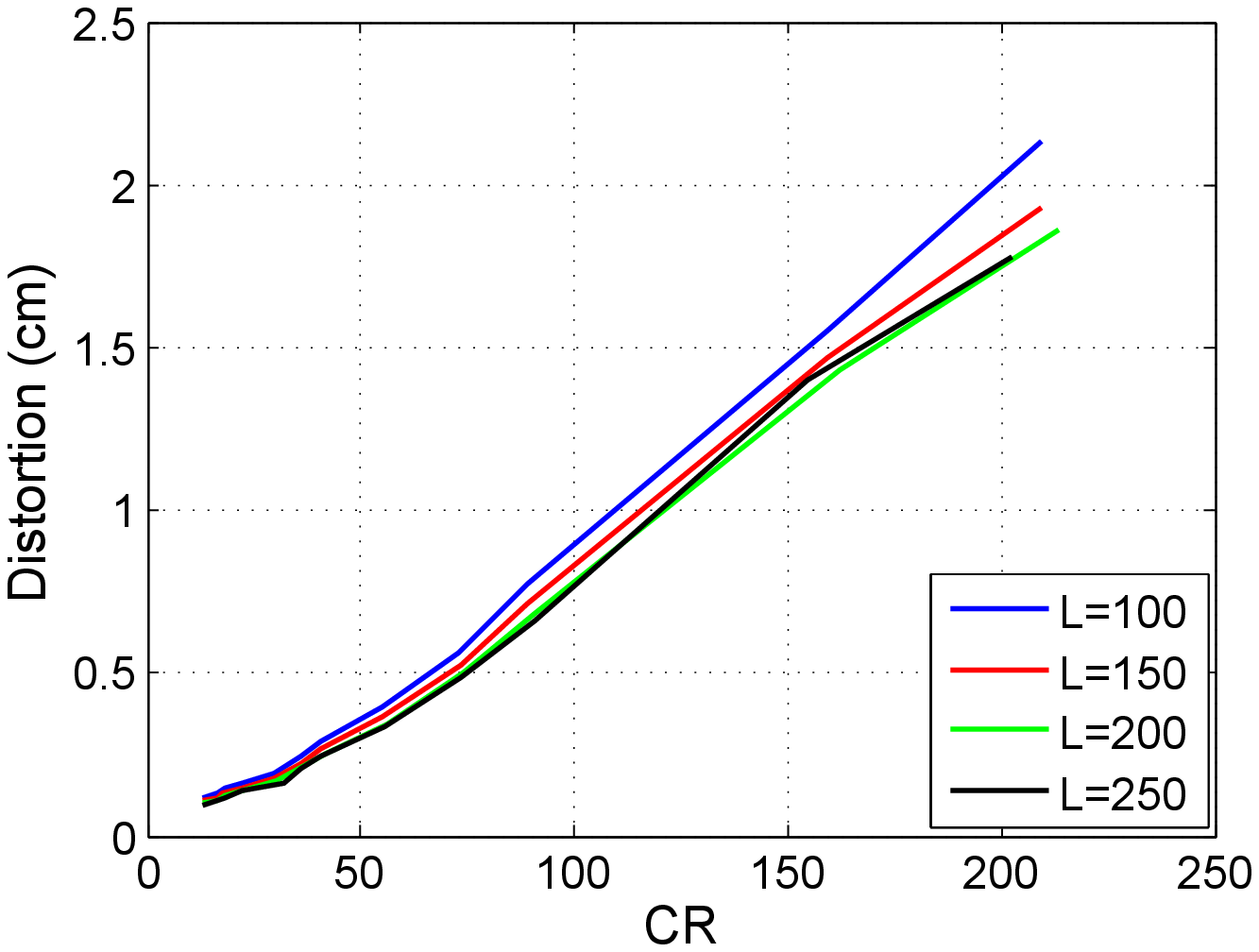}}
  \subfigure[Sequence 86\_05]{
  \includegraphics[width=1.65in]{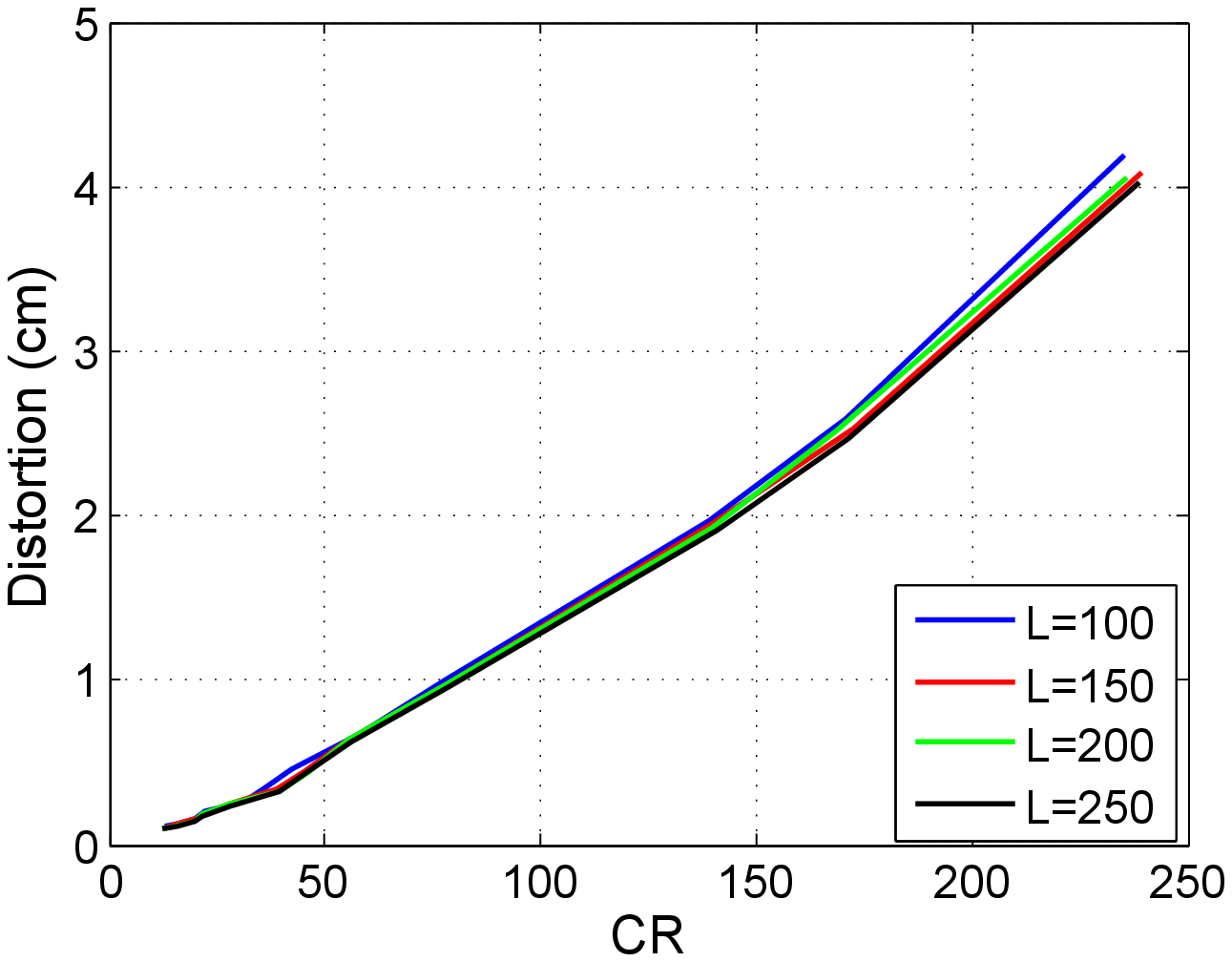}}
  \caption{Our results are not sensitive the frame rate.
  The original test data were obtained at 120 fps and we downsampled each sequence to 60fps and
  segmented them into clips of equal length. With doubled compression ratio, the distortion curves are similar to Figure~\ref{fig:result fixed L}.
  }
  \label{fig:result fixed L lfr}
  \end{figure}

To demonstrate the robustness of our method, we also test our method
on sequences with low frame rates, which have weaker data
correlation than the ones at full frame rates. Given the original
sequences with 120 fps, we uniformly downsampled them to 60 fps.
Correspondingly, their data size and number of frames halved. As
Fig. \ref{fig:result fixed L lfr} shows, our method still works very
well, i.e., under the same distortion, the CRs is approximate 50\%
of those in Fig. \ref{fig:result fixed L}. We also observe that the
optimal range $L\in[150,200]$ is approximately half of the one in
Fig. \ref{fig:result fixed L}.

\subsubsection{Adaptive segmentation}
\begin{figure}
  \centering
  \subfigure[Sequence 41\_07]{
  \includegraphics[width=1.6in]{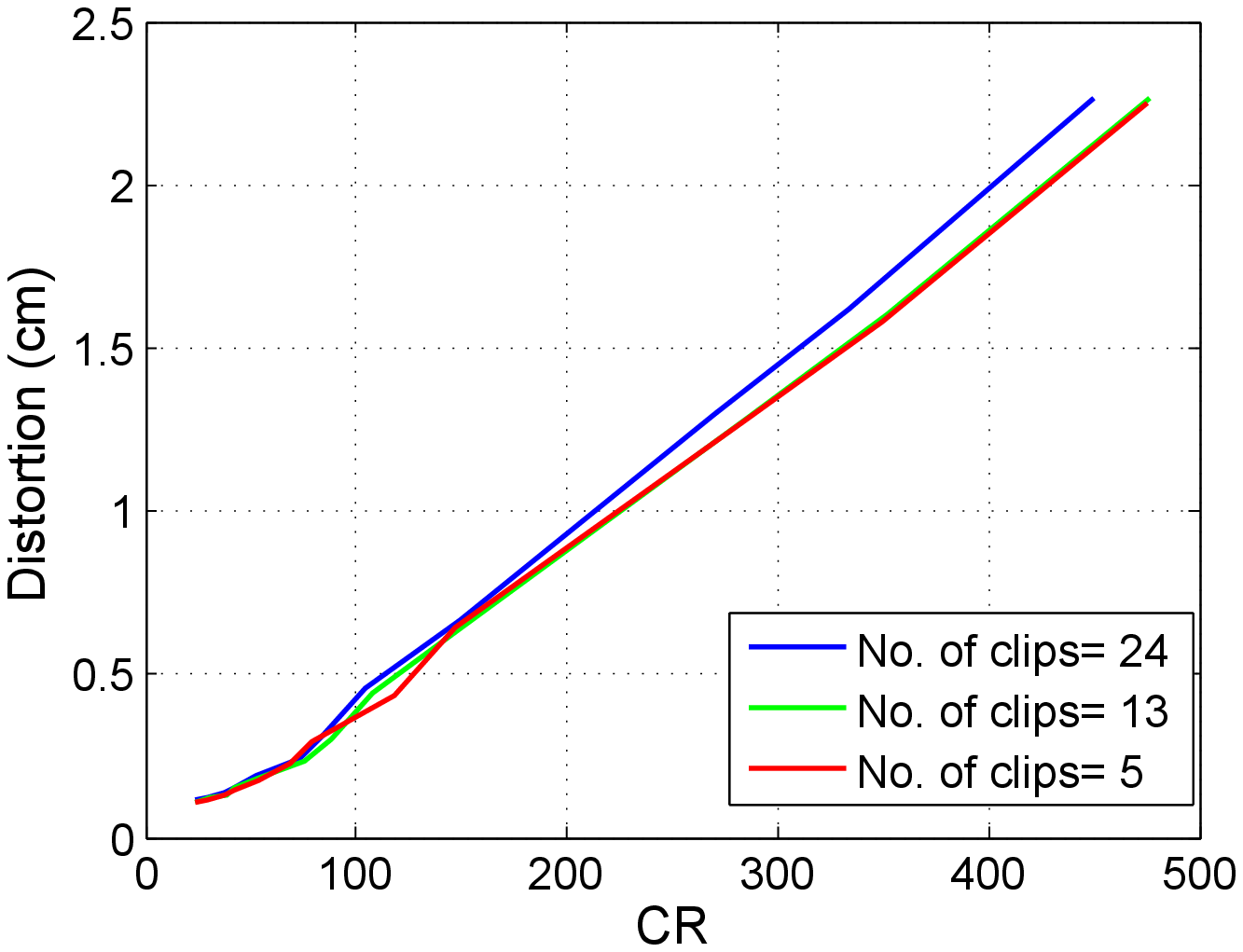}}
  \subfigure[Sequence 56\_07]{
  \includegraphics[width=1.6in]{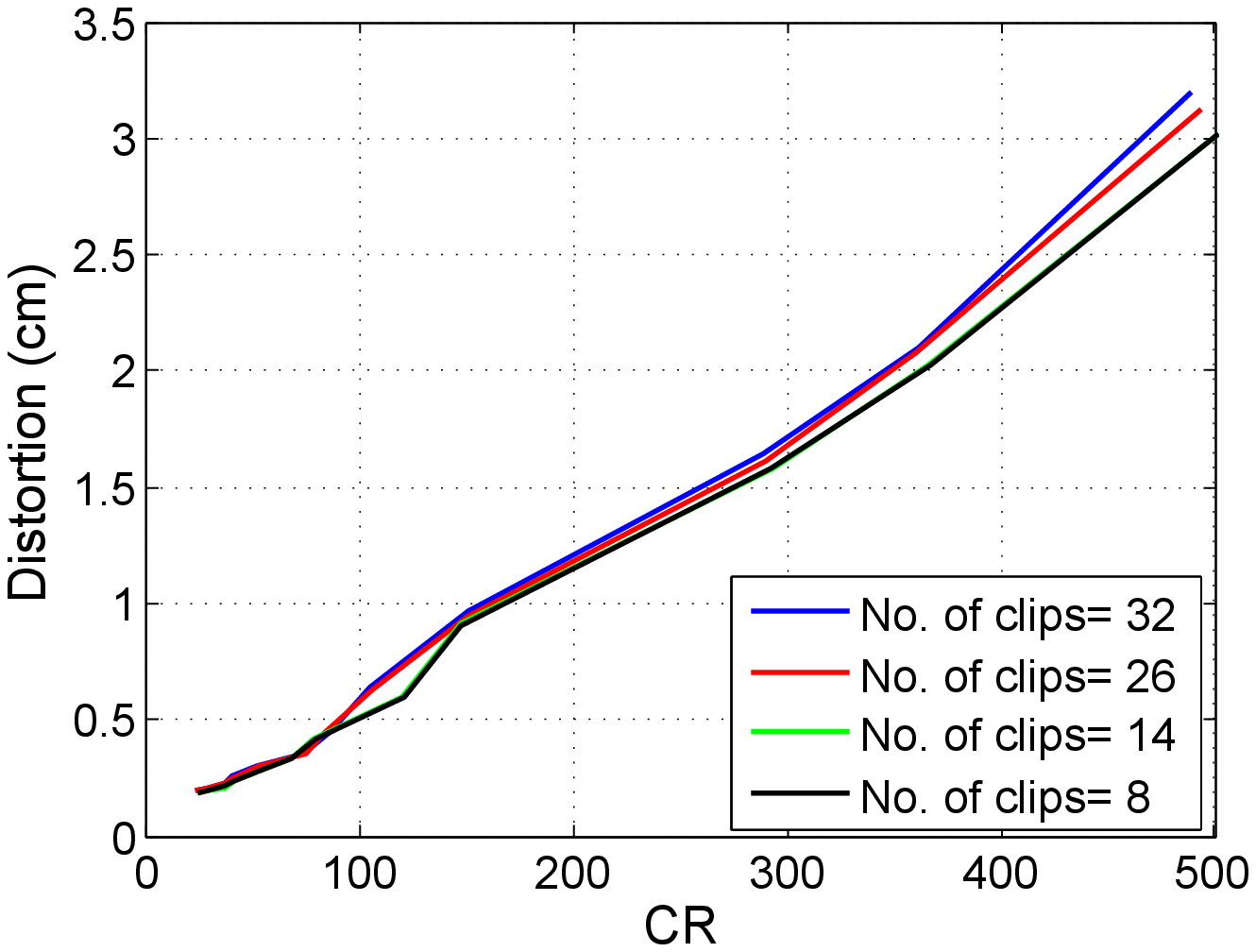}}
  \subfigure[Sequence 69\_08]{
  \includegraphics[width=1.6in]{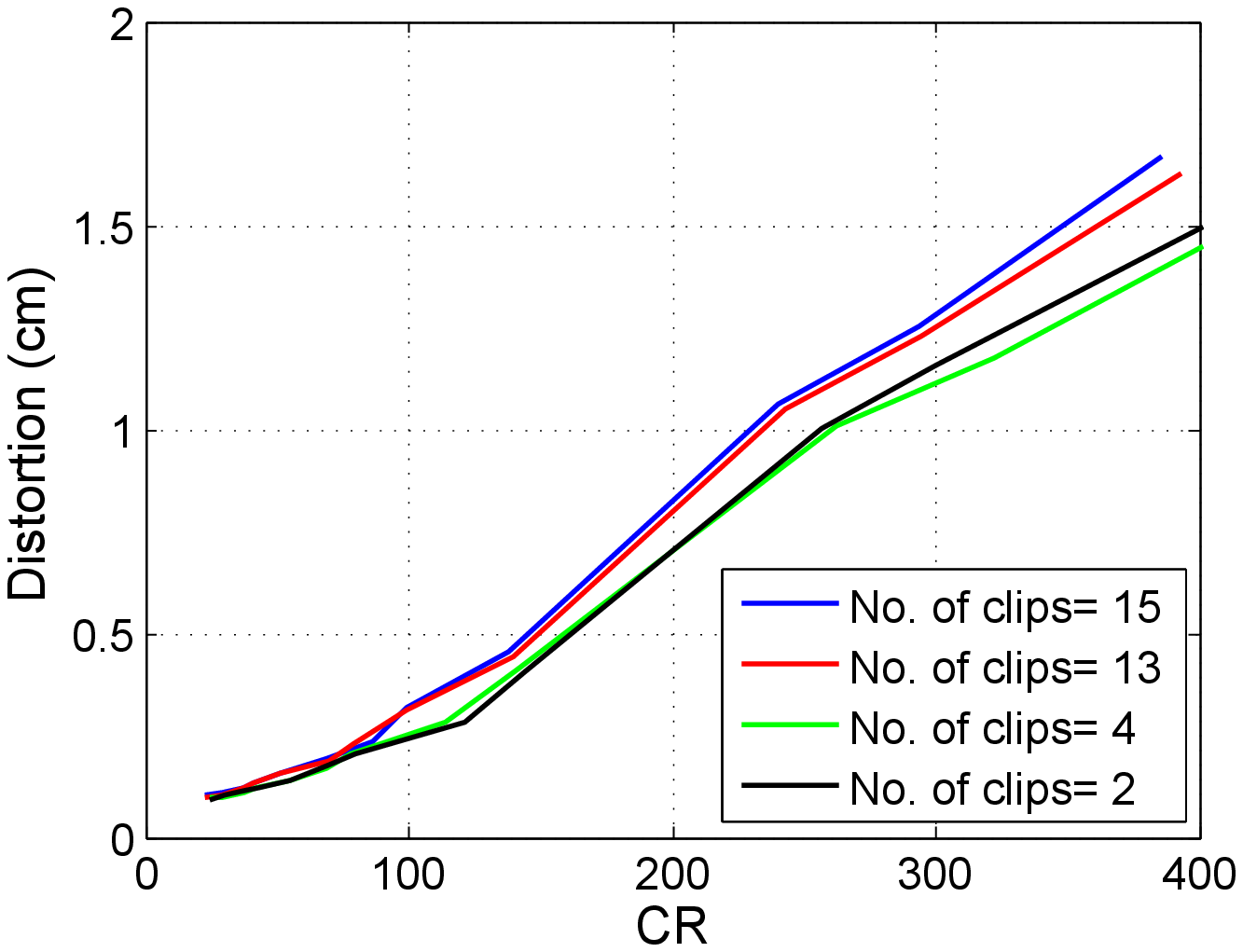}}
  \subfigure[Sequence 86\_05]{
  \includegraphics[width=1.6in]{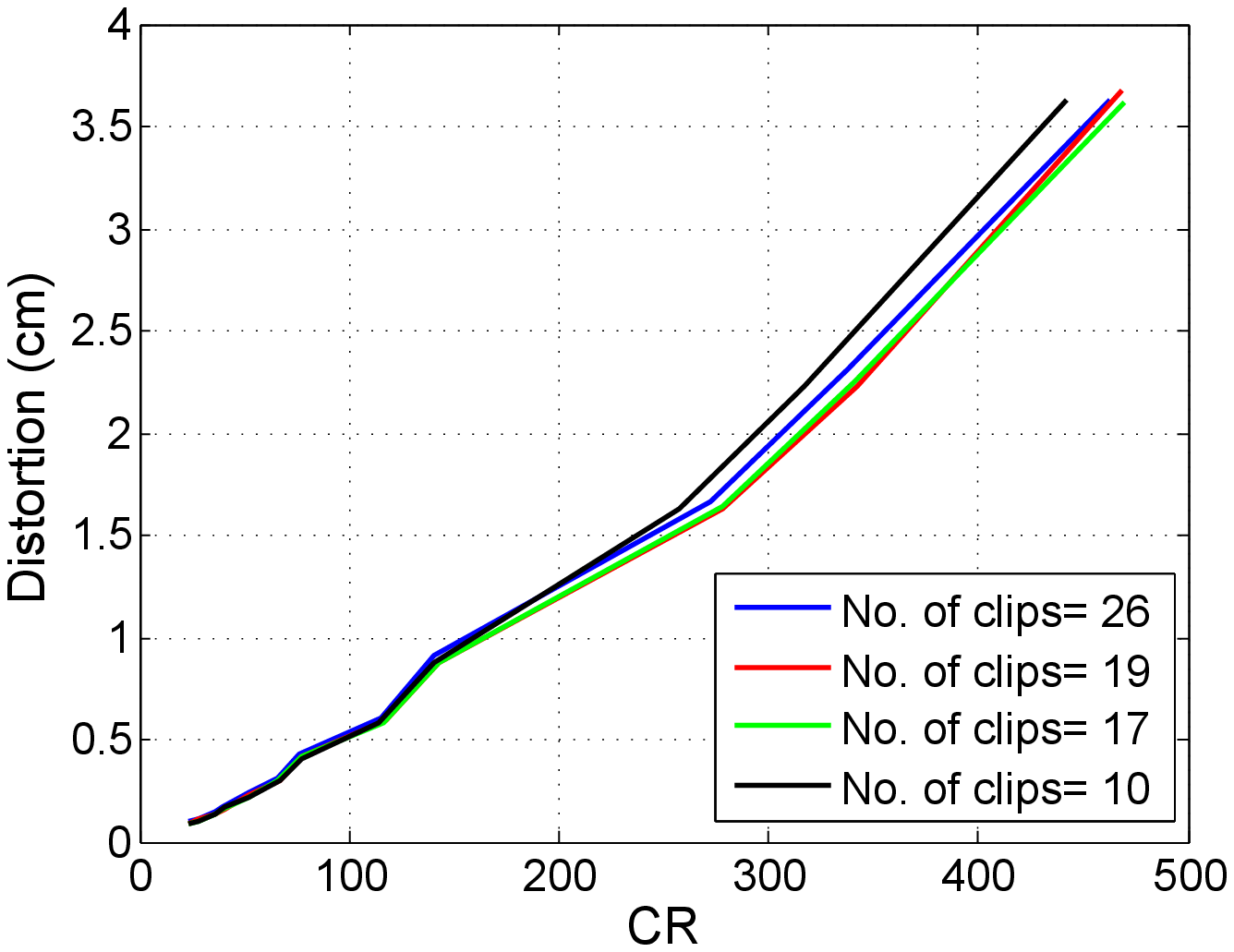}}
  \caption{CR-distortion curves for adaptive
  segmentation with various clip numbers.
  }
  \label{fig:result ada L}
  \end{figure}

When advanced motion analysis tool (e.g., \cite{PPCA},
\cite{PAMI-segment}, \cite{segmenSCA2004}, \cite{segmenSCA2014}) is
available, one can segment the input data to clips of physically
meaningful actions, which have stronger correlation than
fixed-length clips. In our implementation, we adopt the
probabilistic principal component analysis (PPCA)-based segmentation
algorithm \cite{PPCA} for its simplicity and effectiveness, in which
a cut is made when the distribution of human poses changes. Other
advanced and complicated motion segmentation algorithms, e.g.,
\cite{PAMI-segment} and \cite{segmenSCA2014}, can also be employed.
The parameters of the $i$-th clip, i.e., $l_i$ and $Q_i$, are
obtained by Eqns. (\ref{equ:optimal r}) and (\ref{equ:Q}) according
to the clip length $L_i$. We also set various numbers of clips
during the segmentation to investigate how it affects the
compression performance. As shown in Fig. \ref{fig:result ada L},
the optimal number of clips (or segments) depend on the motion
characteristics. For example, the sequence 69\_08 contains only
walking and turning, so a small number of clips works well. In
contrast, the sequences 56\_07 and 86\_05 contain complicated
motions, such as climbing punching. As a result, a large number of
clips produces smaller distortion. As expected, at relatively high
CRs, adaptive segmentation can achieve higher compression
performance than equal segmentation.

\subsection{Comparison} \label{subsec:comparison}

\begin{figure}
\centering
\subfigure[Sequence 15\_04]{
\includegraphics[width=1.65in]{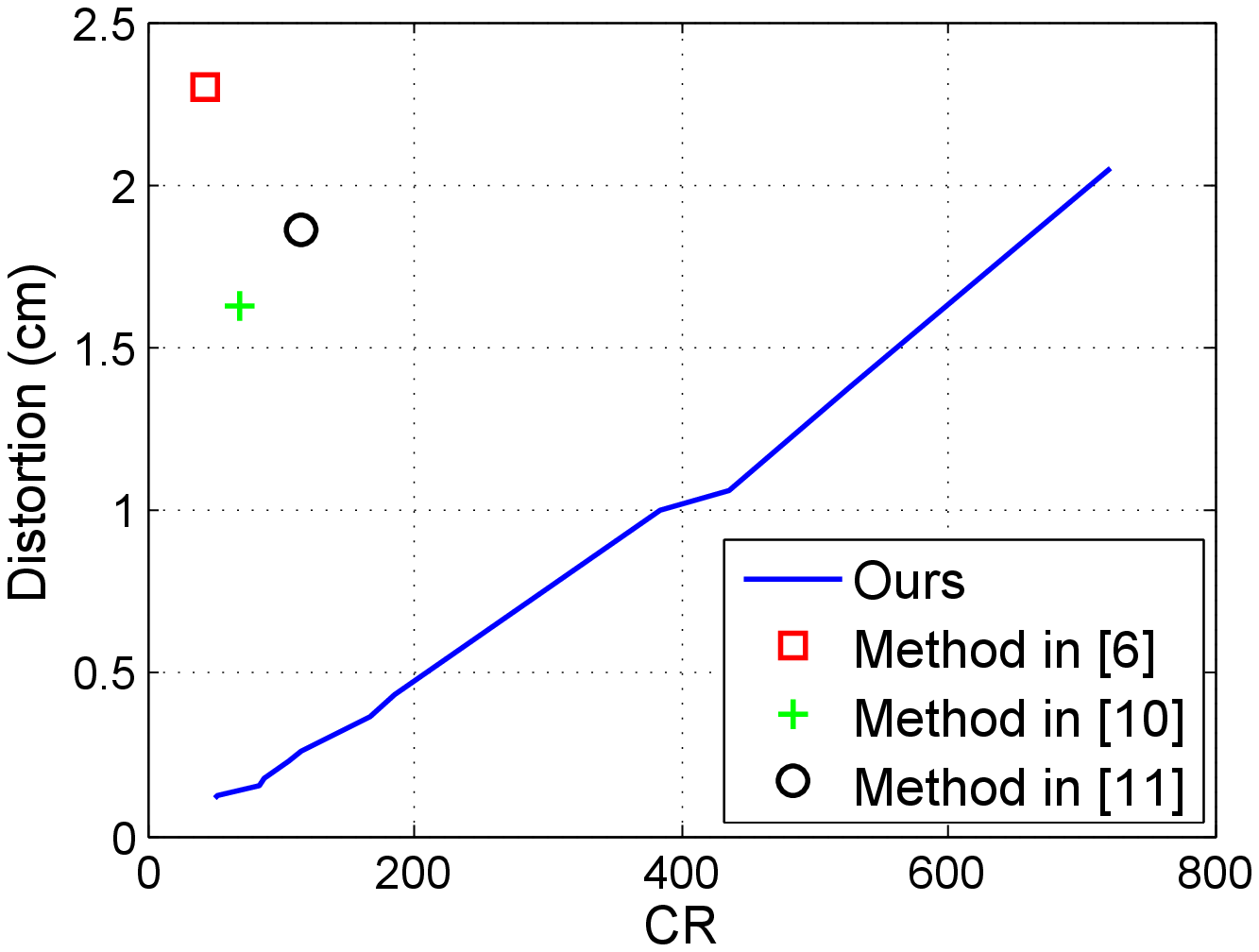}}
\subfigure[Sequence 17\_08]{
\includegraphics[width=1.65in]{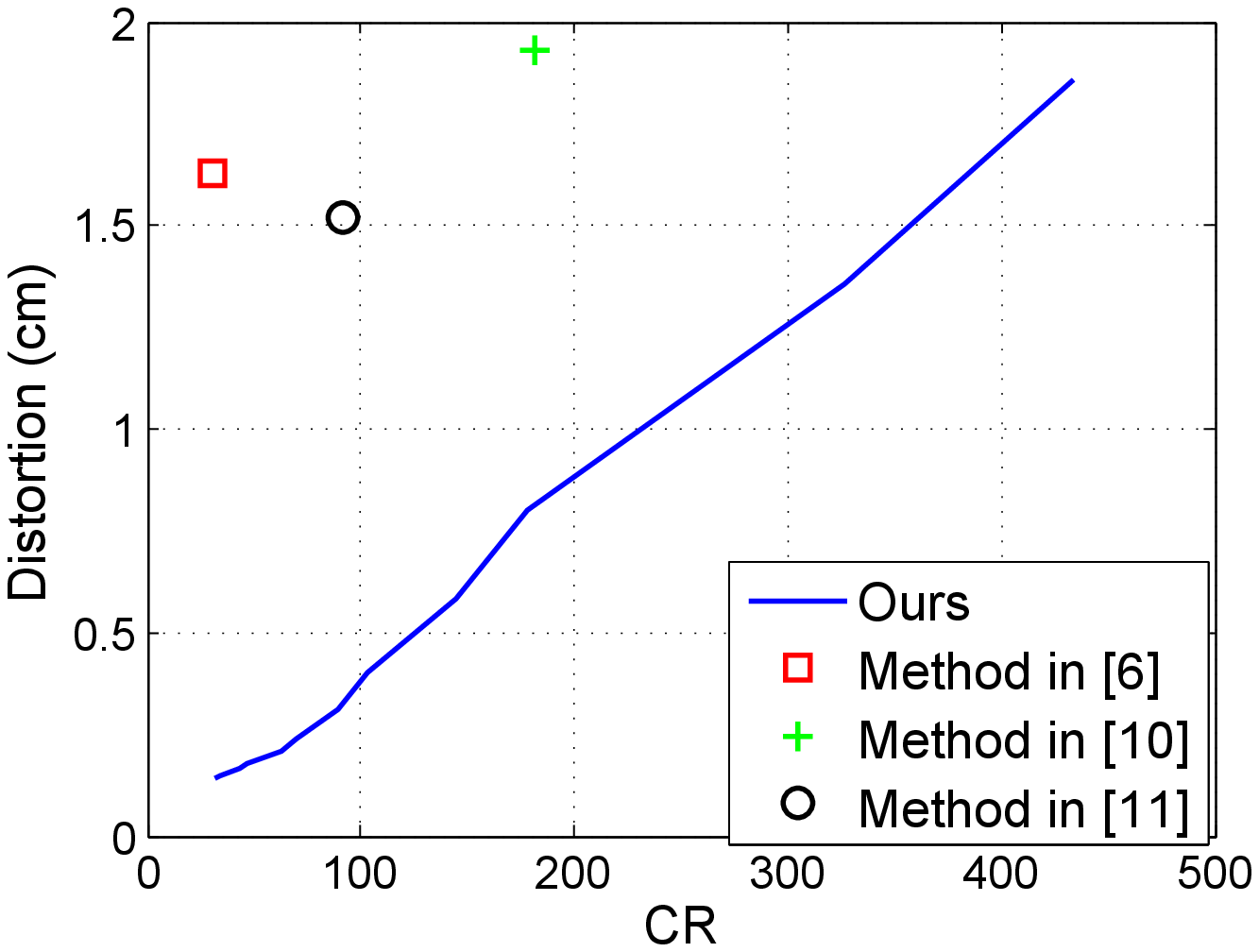}}
\subfigure[Sequence 17\_10]{
\includegraphics[width=1.65in]{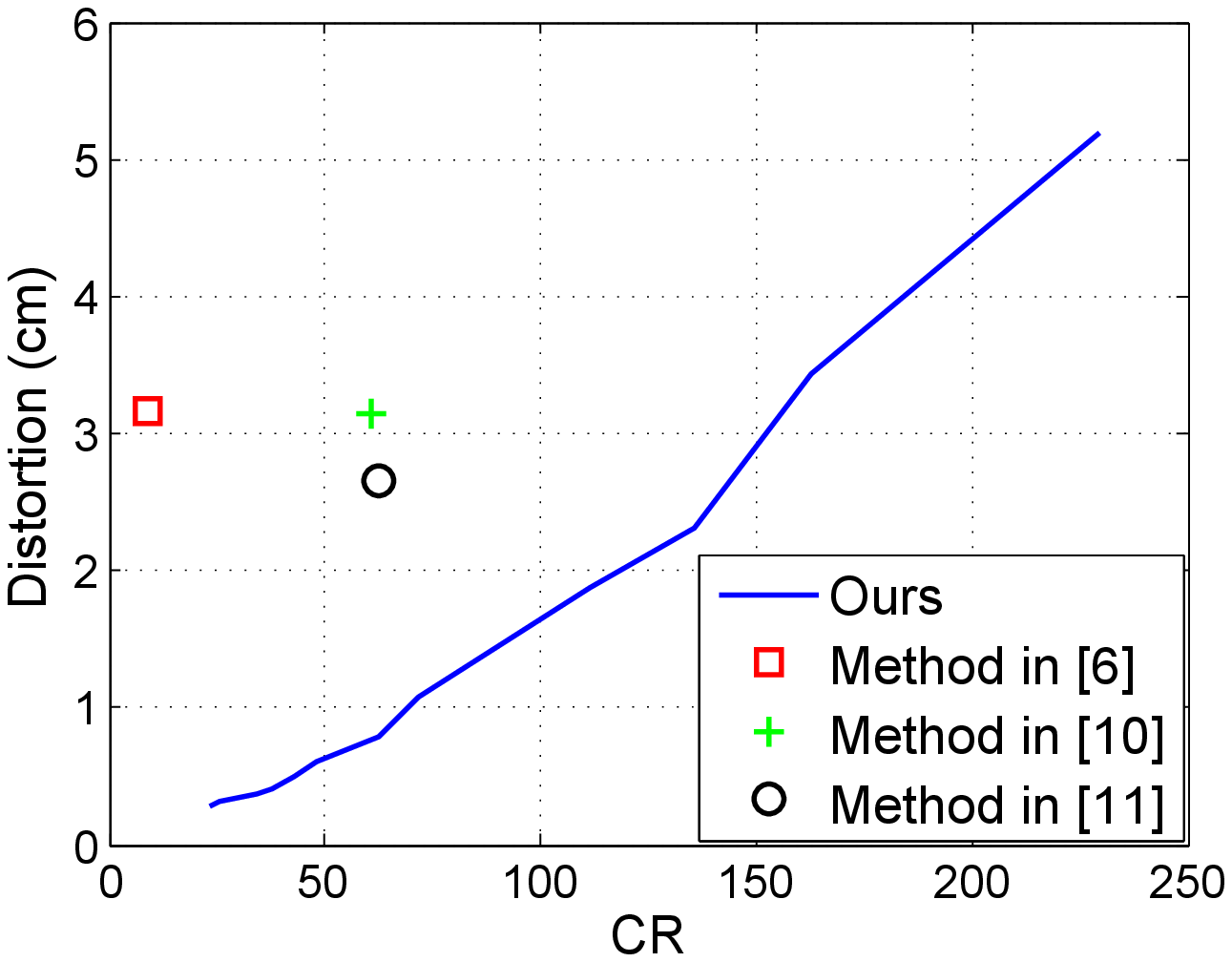}}
\subfigure[Sequence 85\_12]{
\includegraphics[width=1.65in]{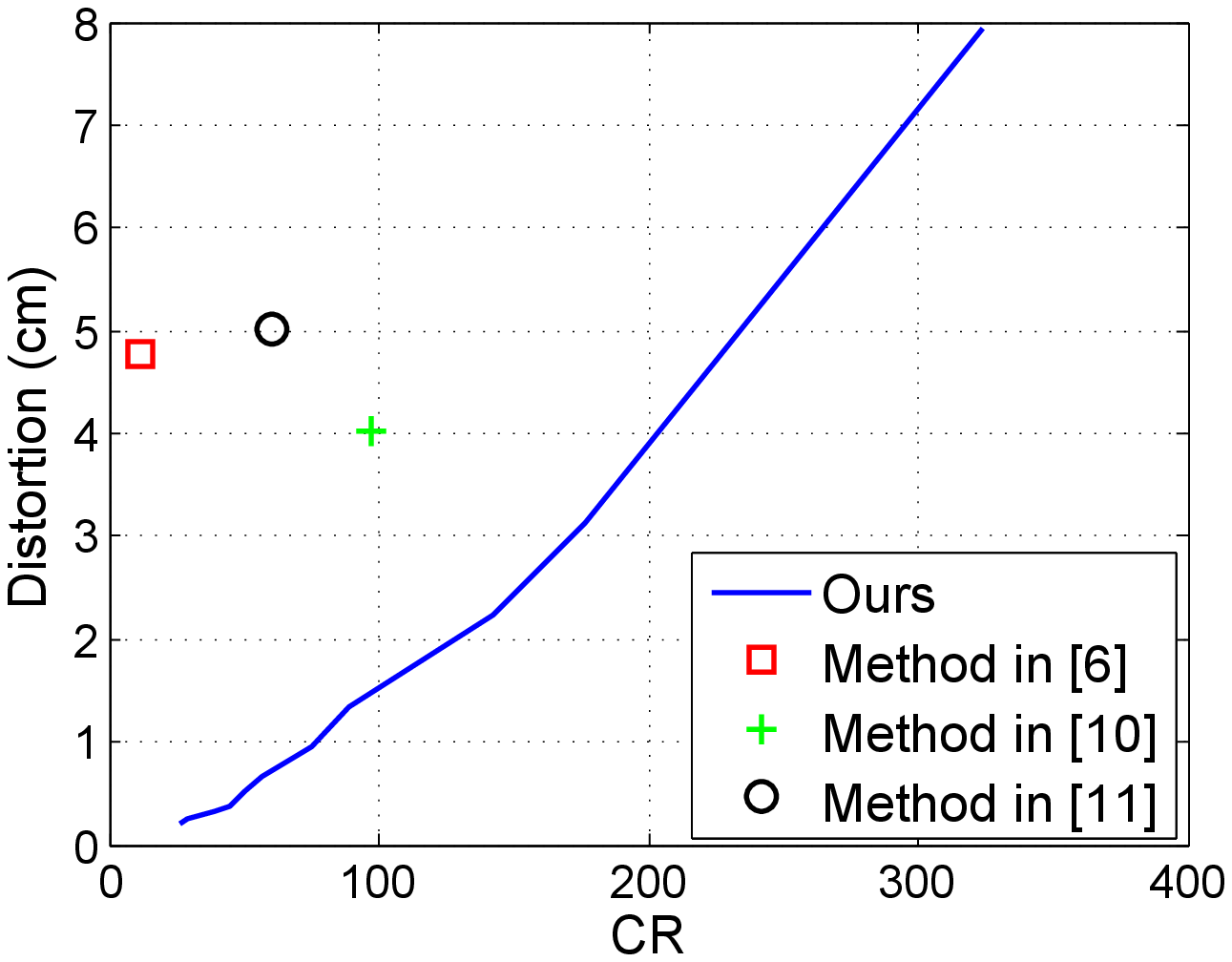}}
\subfigure[database\_3]{
\includegraphics[width=1.6in]{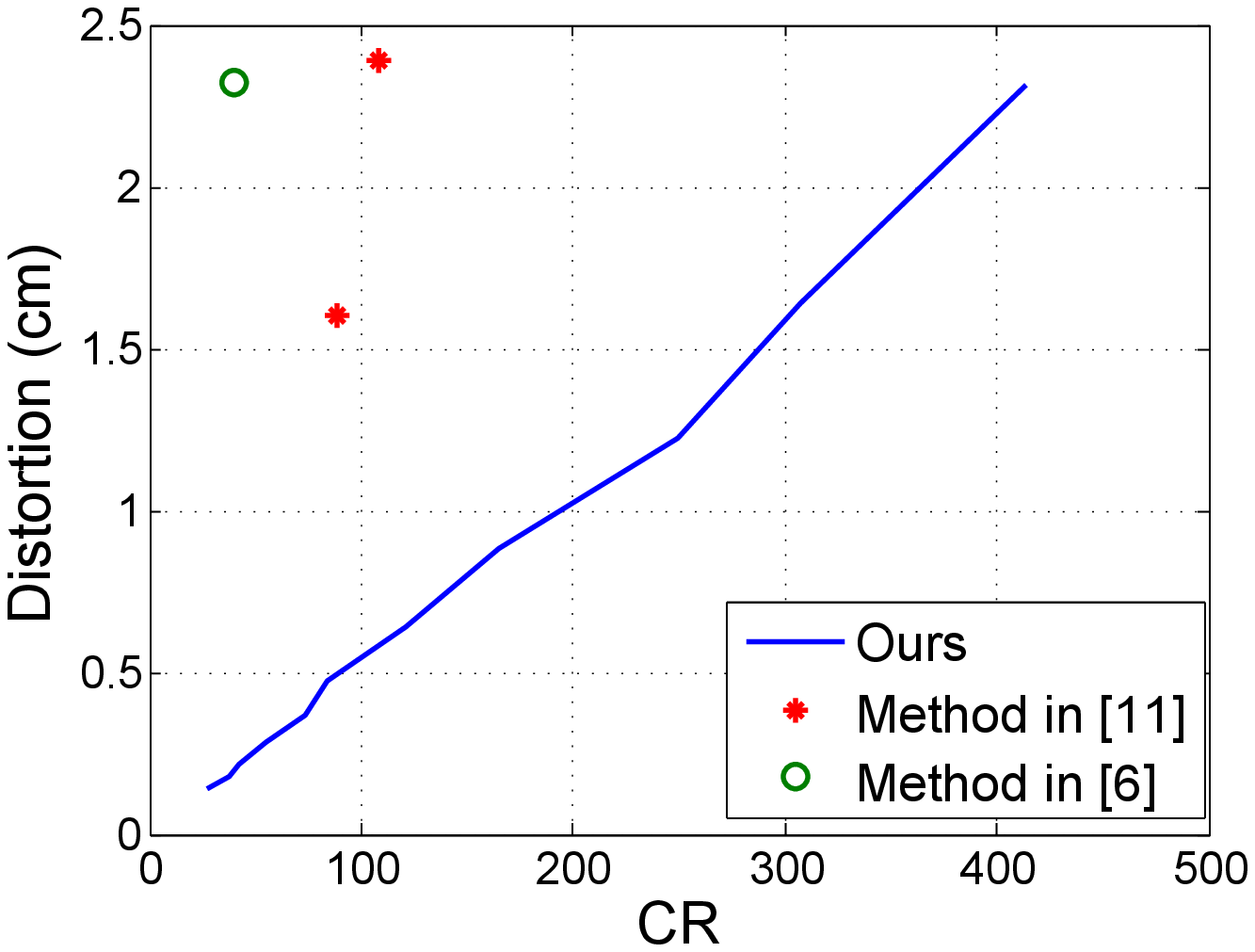}}
\subfigure[database\_4]{
\includegraphics[width=1.6in]{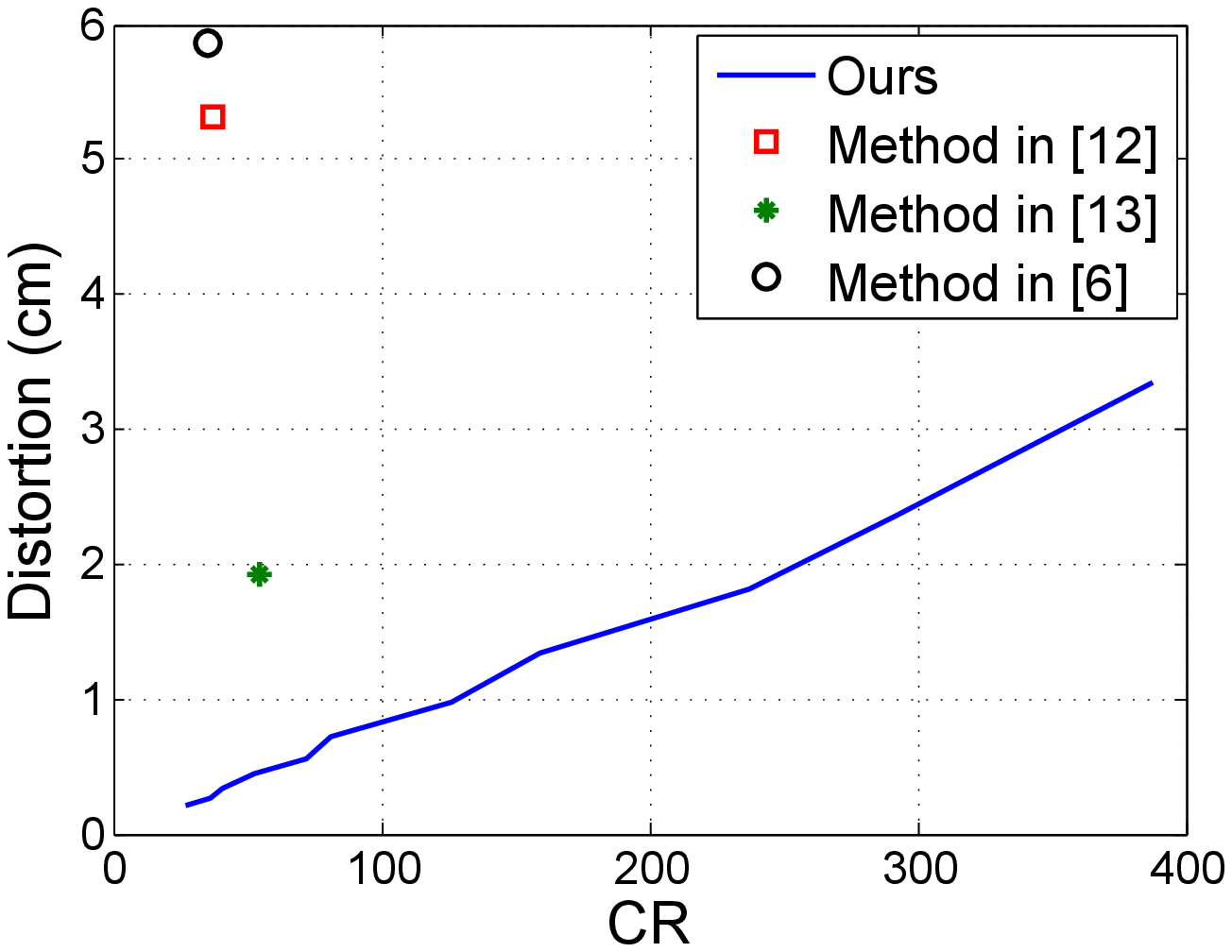}}
\caption{Comparison in terms of compression performance. Our method
is performed under equal segmentation with $L=280$. In (a), (b), (c)
and (d), the results of~\cite{siggraph}, \cite{Priciple geodesic},
\cite{repeate} are taken from \cite{repeate}, where only a single CR
was provided. For database\_3, $K=4$ and the results of
\cite{repeate} and \cite{siggraph} are taken from \cite{repeate}.
For database\_4, $K=3$ and the results of \cite{siggraph},
\cite{pattern} and \cite{Quaternion} are taken from
\cite{Quaternion}.} \label{fig:CR-D}
\end{figure}

  We compare our method with several state-of-the-art techniques~\cite{siggraph},~\cite{Priciple geodesic},~\cite{repeate},~\cite{pattern},~\cite{Quaternion}
  in terms of compression performance and efficiency.
  The results of~\cite{siggraph},~\cite{Priciple geodesic}, and ~\cite{repeate} are taken from~\cite{repeate},
   and the results of \cite{pattern} and \cite{Quaternion} are taken from \cite{Quaternion}.
   Note that only results with a single compression ratio were provided in both \cite{repeate} and \cite{Quaternion}.
   For fair comparison,  our scheme is performed under the equal segmentation scenario.
  Fig. \ref{fig:CR-D} shows our CR-distortion curves on some test sequences and mocap databases.
  We observe that our CR-distortion curves are consistently below the other methods, indicating
  that our method outperforms them in terms of compression performance.
  Table~\ref{tab:comparison} shows that our method is also much more efficient than the existing methods.

\begin{table*}[]
  \centering \caption{Efficiency comparison. The results of Arikan~\cite{siggraph} and Lin \textit{et al.}~\cite{repeate} are taken from \cite{repeate}, which were measured on an Intel 3.0GHz CPU.
  We evaluated our algorithm on an Intel 3.1GHz CPU. Computational results show that our method is significantly faster than the existing methods and also produces results of better
  quality. $\overline{d}$ is the distortion (in cm), $E_{fps}$ and $D_{fps}$
are the encoding and decoding frame rates, respectively.}
  \label{tab:complexity1}
 \begin{scriptsize}
  \begin{tabular}{c|c|c|c|c|c|c|c|c|c|c|c|c|c|c|c|c}
  \hline \hline & \multicolumn{4}{c|}{Arikan~\cite{siggraph}}& \multicolumn{4}{c|}{Ours}&\multicolumn{4}{c|}{Lin \textit{et al}~\cite{repeate}} & \multicolumn{4}{c}{Ours}\\
  \hline Seq. & CR & $\overline{d}$ & $E_{fps}$ &$D_{fps}$ & CR & $\overline{d}$ & $E_{fps}$ & $D_{fps}$ & CR & $\overline{d}$ & $E_{fps}$ &$D_{fps}$ &CR & $\overline{d}$ & $E_{fps}$ &$D_{fps}$ \\
  \hline 15\_04  &42.5& 2.30 &705.6 & 2693.7 & 45.7&0.12 &125,934 &261,563&114.7 &1.86 &219.1& 2530.8 & 114.1& 0.29&169,567&281,478 \\
  \hline 17\_08 &30.0 & 1.63&778.8 &2979.8 &35.0 &0.15 &58,576 &88,967&91.2 & 1.52&539.1& 2945.2 &93.8 &0.37 &69,856&91,729\\
  \hline 17\_10 &9.2 &3.16 &376.7 & 4596.8&14.2 & 0.22& 27,748 &39,254& 62.6& 2.65&486.6& 4288.1&63.2 & 1.07&32,348& 40,239\\
  \hline  85\_12 &11.1& 4.78&585.7&3903.1  &16.1 & 0.12& 43,173&69,187&60.1&5.01 &506.4&3096.4 &59.3 &0.84 &52,269&68,546\\
  \hline database\_3 &39.8 & 2.32& 530.5& 4902.3& 40.6& 0.20&1,805 &294,536&88.7 &1.60 &372.4 &4286.8 &83.7 & 0.47&1,798 &293,947\\
  \hline \hline
  \end{tabular}
  \end{scriptsize}
  \label{tab:comparison}
\end{table*}

\subsection{Limitation}
 The high compression performance of our method comes by significantly exploring the temporal correlation, that is, it requires the entire sequence or at least a sequence of
  sufficiently long period to compute the left transform matrix using Eq. (\ref{equ:construct B}).
  This will cause the latency problem, which also occurs in other approaches, e.g.,~\cite{segmentpca,siggraph,tmm,tmm-perceptual,repeate}.
  Thus, it is not suitable for applications where data acquisition and compression have to be done simultaneously or within a very small time interval.


\section{Conclusion and Discussion}
\label{sec:conclusion}

  This paper has presented a mocap-tailored transform coding for compressing human mocap data.
  Our method is highly effective in that it takes advantage of the features of mocap data.
  Furthermore, it has low computational cost and can be easily extended to compress mocap
  databases.
  Unlike existing approaches, it requires neither training nor complicated parameter setting.
  Experimental results demonstrate that the proposed scheme significantly outperforms
  state-of-the-art algorithms in terms of compression performance and
  speed.

  Owning to its high performance, we believe our method has potential to many mocap applications, such as  retrieval~\cite{Muller,Muller2,Retrieval1,Retrieval2}, where the clip dimensionality can be significantly reduced
  without comprising data quality, leading to low computational complexity.
  Besides mocap data, our method can be extended for compressing dynamic meshes~\cite{cpca,Vasa}, in which each vertex is considered as a marker.

  Our method can also inspire interesting future work.
\begin{enumerate}[1)]
\item
We observed that the compression performance depends on the compressed transform matrices. Therefore,
advanced matrix compression algorithms, e.g., sparse and redundancy representation, can improve our results.
\item
Perception-based compression schemes~\cite{SSIM rate,CSVT
perceptual} have proven to be effective to natural images and
videos, since they are adaptive to human vision system. It would be
interesting to investigate the perceived-based methods for mocap
data compression. For example, optimally allocating  bits to clips
guided by a perceptual distortion metric, that is, clips which can
be significantly compressed with little perceptual distortion should
be assigned fewer bits, and vice versa, so that the overall
compression performance is higher.
\item
In our implementation, we focus on human motion capture data. It is
also interesting to check its performance on other types of mocap
data, e.g., facial expressions.
\item
A possible solution to solve the latency problem is to apply machine
learning techniques to pre-compute some left transform matrices from
learning datasets. As a result, the compression can be carried out
whenever a clip with small length is available. 
\item
Last but not least, it is desirable to develop a fully automatic
algorithm for determining the optimal (or near-optimal) parameter,
i.e., the clip lengths for equal segmentation and the number of
clips for adaptive segmentation.
\end{enumerate}

\section*{Acknowledgments}
The test data is courtesy of the CMU
Motion Capture Database~\cite{cmu}.


\begin{thebibliography}{1}
\bibitem{segmentpca}
G. Liu and L. McMillan, ``Segment-based human motion compression,"
\emph{Proc. ACM SIGGRAPH/Eurographics Symposium on Computer
Animation}, pp. 127-135, 2006.
\bibitem{Power Aware Approach}
S. Chattopadhyay, S. M. Bhandarkar, and K. Li, ``Human motion
capture data compression by model-based indexing: a power aware
approach," \emph{IEEE Trans. Visualization and Computer Graphics},
vol. 13 ,no. 1, pp. 5-14, 2007.
\bibitem{wavelet compression}
P. Beaudoin, P. Poulin, and Michiel van de Panne, ``Adapting wavelet
compression to human motion capture clips," \emph{Proc. Graphics
Interface}, pp. 313-318, 2007.
\bibitem{keyframe2}
I.S. Lim and D. Thalmann, ``Key posture extraction out of human
motion data by curve simplification", \emph{23rd Annual
International Conference IEEE Engineering in Medicine and Biology
Society 2001}, pp. 1167-1169, 2001.
\bibitem{cgi}
J. Xiao, Y. Zhuang, T. Yang, and F. Wu, ``An efficient keyframe
extraction from motion capture data," \emph{Proc. Computer Graphics
International Conference}, pp. 494-501, 2006.
\bibitem{siggraph}
O. Arikan, ``Compression of Motion Capture database," \emph{ACM
Trans. on Graphics}, vol. 25, no. 3, pp. 890-897, 2006.
\bibitem{keyframe1}
M.-H Kim, L.-P Chau, and W.-C. Siu, ``Keyframe selection for motion
capture using motion activity analysis," in \emph{Proc. IEEE
International Symposium on Circuits and Systems}, pp. 612-615, 2012.
\bibitem{sparse}
Y. Li, C. Fermuller, Y. Aloimonos, and H. Ji,  ``Learning
shift-invariant sparse representation of actions," in \emph{Proc.
IEEE Conference on Computer Vision and Pattern Recognition}, pp.
2630-2637, 2010.
\bibitem{tmm}
B.-S. Chew, L.-P. Chau, K.-H. Yap, ``A fuzzy clustering algorithm
for virtual character animation representation," \emph{IEEE Trans.
Multimedia}, vol. 13, no. 1, pp. 40-49, 2011.
\bibitem{Priciple geodesic}
M. Tournier, X. Wu, N. Courty, E. Arnaud, and L. Reveret, ``Motion
compression using principal geodesics analysis," \emph{Computer
Graphics Forum}, vol. 28, no. 2, pp. 355-364, 2009.
\bibitem{repeate}
I-C. Lin, J.-Y. Peng, C.-C. Lin, and M.-H. Tsai, ``Adaptive motion
data representaion with repeated motion analysis," \emph{IEEE Trans.
Visualization and Computer Graphics}, vol. 17, no. 4, pp. 527-538,
2011.
\bibitem{pattern}
Q. Gu, J. Peng, and Z. Deng, ``Compression of human motion capture
data using motion pattern indexing," \emph{Computer Graphics Forum},
vol. 28, no. 1, pp. 1-12, 2009.
\bibitem{Quaternion}
M. Zhu, H. Sun, Z. Deng, ``Quaternion space sparse decomposition for
motion compression and retrieval," \emph{Proc. ACM
SIGGRAPH/Eurographics Symposium on Computer Animation}, pp. 183-192,
2012.
\bibitem{phdthesis}
L. Ren, ``Statistical analysis of natural human motion for
animation," Ph.D. dissertation, Pittsburgh, PA, USA, 2006.
\bibitem{BBA}
M. Preda and F. Preteux, ``Virtual character within MPEG-4 animation
framework eXtension," \emph{IEEE Trans. Circuits Syst. Video
Technol.}, vol. 14, no. 7, pp. 975-988, 2004.
\bibitem{marker optimization}
B. Le, M. Zhu, and Z. Deng, ``Marker optimization for facial motion
acquisition and deformation," \emph{IEEE Trans. Visualization and
Computer Graphics}, vol. 19, no. 11, pp. 1859-1871, 2012.
\bibitem{ICME}
J. Hou, Z.-P. Bian, L.-P. Chau, N. Magnenat-Thalmann, and Y. He,
``Restoring corrupted motion capture data via jointly low-rank
matrix completion," \emph{Proc. IEEE International Conference on
Multimedia \& Expo (ICME)}, pp. 1-6, 2014.
\bibitem{hosvd}
G. Bergqvist and E.G. Larsson, ``The higher-order singular value
decomposition: theory and an application," \emph{IEEE Signal
Processing Magazine}, vol. 27, no. 3, pp. 151-154, 2010.
\bibitem{trace maximum}
E. Kokiopoulou, J Chen, and Y. Saad, ``Trace optimization and
eigenproblems in dimension reduction methods," \emph{Numerical
Linear Algebra with Applications}, vol. 18, no. 3, pp. 565-602,
2011.
\bibitem{multi-matrices}
T. Hofmann and J. M. Buhmann, ``Pairwise data clustering by
deterministic annealing," \emph{IEEE Trans. Pattern Analysis and
Machine Intelligence}, vol. 19, no. 1, pp. 1-14, 1997.
\bibitem{tmm-perceptual}
A. Firouzmanesh, I. Cheng, and A. Basu, ``Perceptually guided fast
compression of 3-d motion capture data,"  \emph{IEEE Trans.
Multimedia}, vol. 13, no. 4, pp. 829-834, 2011.
\bibitem{tensor}
J. Hou, L.-P. Chau, N. Magnenat-Thalmann, and Y. He, ``Scalable and
compact representation for motion capture data using tensor
decomposition," \emph{IEEE Signal Processing Letters}, vol. 21, no.
3, pp. 255-259, 2014.
\bibitem{PAMI-segment}
F. Zhou, F. De La Torre, and J. Hodgins, ``Hierarchical aligned
cluster analysis for temporal clustering of human motion,"
\emph{IEEE Trans. Pattern Analysis and Machine Intelligence}, vol.
35, no. 3, pp. 582-596, 2013.
\bibitem{segmenSCA2004}
Y. Sakamoto, S. Kuriyama, and T. Kaneko, ``Motion map: image-based
retrieval and segmentation of motion data," \emph{Proc. ACM
SIGGRAPH/Eurographics Symposium on Computer Animation}, pp. 259-266,
2004.
\bibitem{PPCA}
J. Barbi{\v{c}}, A. Safonova, J.-Y. Pan, C. Faloutsos, J. Hodgins
and N. Pollard, ``Segmenting motion capture data into distinct
behaviors," \emph{Proc. of Graphics Interface}, pp. 185-194, 2004.
\bibitem{Muller}
M. M{\"u}ller, T. R{\"o}der, and M. Clausen, ``Efficient
content-based retrieval of motion capture data," \emph{ACM Trans. on
Graphics}, vol. 24, no. 3, pp. 677-685, 2005.
\bibitem{Muller2}
M. M{\"u}ller and T. R{\"o}der, ``Motion templates for automatic
classification and retrieval of motion capture data," \emph{Proc.
ACM SIGGRAPH/Eurographics Symposium on Computer Animation}, pp.
137-146, 2006.
\bibitem{segmenSCA2014}
A. V{\"o}gele, B. Kr{\"u}ger, and R. Klein, ``Efficient unsupervised
temporal segmentation of human motion," \emph{Proc. ACM
SIGGRAPH/Eurographics Symposium on Computer Animation}, pp. 167-176,
2014.
\bibitem{Retrieval1}
B. Kr{\"u}ger, J. Tautges, A. Weber, and A. Zinke, ``Fast local and
global similarity searches in large motion capture databases,"
\emph{Proc. ACM SIGGRAPH/Eurographics Symposium on Computer
Animation}, pp. 1-10, 2010.
\bibitem{Retrieval2}
M. Kapadia, I-K. Chiang, T. Thomas, N. Badler,  and J. Kider Jr
``Efficient motion retrieval in large motion databases," \emph{Proc.
ACM SIGGRAPH Symposium on Interactive 3D Graphics and Games}, pp.
19-28, 2013.
\bibitem{cpca}
M. Sattler R. Sarlette, and R. Klein, ``Simple and efficient
compression of animation sequences," \emph{Proc. ACM
SIGGRAPH/Eurographics Symposium on Computer Animation}, pp. 209-217,
2005.
\bibitem{Vasa}
L. V\'{a}\v{s}a, S. Marras, K. Hormann, and G. Brunnett,
``Compressing dynamic meshes with geometric laplacians,"
\emph{Computer Graphics Forum}, vol. 33, no. 2, pp. 145-154, 2014.
\bibitem{SSIM rate}
T.-S. Ou, Y.-H. Huang, and H.H. Chen, ``SSIM-based perceptual rate
control for video coding," \emph{IEEE Trans. Circuits and Systems
for Video Technology}, vol. 21, no. 5, pp. 682-691, 2011.
\bibitem{CSVT perceptual}
M. Naccari and F. Pereira, ``Advanced H.264/AVC-based perceptual
video coding: architecture, tools, and assessment," \emph{IEEE
Trans. Circuits and Systems for Video Technology}, vol. 21, no. 6,
pp. 766-782,  2011.
\bibitem{cmu}
CMU Mocap Database, [Online] Available:
\url{http://mocap.cs.cmu.edu}.
\bibitem{matlab}
Neil Lawrence, ``Mocap toolbox for matlab," [Online] Available:
\url{http://www. cs. man. ac. uk/neill/mocap}.
\end{thebibliography}
\end{document}